\documentclass[a4paper,11pt]{article}
\usepackage{jheppub_personal} 
\usepackage{fontawesome}

\usepackage{enumerate}

\renewcommand{\thesection}{\arabic{section}}

\usepackage{hyperref}
\usepackage{placeins}
\usepackage{amsfonts}
\usepackage{amsmath}
\usepackage{relsize}
\usepackage{mathrsfs}
\usepackage{epsfig}
\usepackage{graphicx}
\usepackage{wasysym}
\usepackage{url}
\usepackage[dvipsnames]{xcolor}
\usepackage{hyperref}
\usepackage{float}
\usepackage{color}
\usepackage{ dsfont }
\usepackage{comment}
\usepackage{tcolorbox}

\usepackage{graphicx}
\usepackage{epsf}

\definecolor{peera_col}{RGB}{240, 94, 28}
\definecolor{blue_col}{RGB}{0,92,175}
\definecolor{red_col}{RGB}{203,64,66}

\usepackage{hyperref}
\hypersetup{linktocpage,
    colorlinks=true,
    linkcolor=red_col,
    filecolor=Mahogany,      
    urlcolor=blue_col,
    citecolor=blue_col,
}

\usepackage{slashed}

\def\be{\begin{equation}}
\def\ee{\end{equation}}

\usepackage{footmisc}
\usepackage{setspace}

\preprint{}

\title{Cosmic string gravitational wave backgrounds at LISA \\\Large{\,I. Signal survey, template reconstruction, and model comparison}}

\author[1]{Androniki Dimitriou,}

\author[1]{Daniel G. Figueroa,}

\author[1,2]{Peera Simakachorn}

\author[1]{and Bryan Zaldívar}

\affiliation[1]{Instituto de F\'{i}sica Corpuscular (IFIC), CSIC‐Universitat de Val\`{e}ncia,\\
Parc Científic UV, c/ Catedr\'{a}tico Jos\'{e} Beltrán, 2, E-46980 Paterna (Val\`{e}ncia), Spain}
\affiliation[2]{Khon Kaen Particle Physics and Cosmology Theory Group (KKPaCT),
Department of Physics, Faculty of Science, Khon Kaen University,
123 Mitraphap Rd., Khon Kaen, 40002, Thailand}

\emailAdd{androniki.dimitriou@ific.uv.es, daniel.figueroa@ific.uv.es, peera.sima@gmail.com, bryan.zaldivar@ific.uv.es}

\abstract{We present a catalog of gravitational wave background (GWB) signal templates from cosmic-string networks, based on relevant 
models proposed in the literature. We classify templates 
as {\bf conventional}, 
based on standard cosmology and Nambu-Goto results (VOS and BOS), and {\bf beyond conventional}, based on 
modifications of {\bf a)} the loop number density (LRS, super,  metastable, current-carrying strings),  {\bf b)} the expansion history (non-standard cosmologies, extra degrees of freedom, either thermal or secluded), 
or {\bf c)} the loop properties (birth length, power emission). 
Using 
the SBI package
\href{https://github.com/AndronikiDimitriou/GWBackFinder}{\faGithub\,{\tt GWBackFinder}}, we quantify 
the reconstruction precision of each signal by LISA, scanning over their parameter space, and performing model comparisons. For conventional signals, LISA reconstructs the tension $G\mu$ with an error $\lesssim 10\%$ for $G\mu \gtrsim 5\cdot 10^{-15}$, which decreases 
down to  $2-3\%$ for $G\mu \gtrsim 10^{-12}$. BOS and VOS modelings 
become distinguishable confidently for $G\mu \gtrsim 5\cdot 10^{-13}$. For beyond-conventional signals, we 
identify SNR and error-threshold intervals for each parameter, and determine (for few examples) the regions where they can be distinguished from conventional signals. Analogous quality reconstruction studies of cosmic-string GWBs,  superimposed over leading astrophysical foregrounds in the LISA window, will be presented in a series of upcoming papers.}

\begin{document}
 \maketitle
\flushbottom

\section{Introduction}
\label{sec:intro}

A new era of cosmic exploration has been ushered in since 2015, thanks to the direct detection of gravitational wave (GW) signals by the LIGO/Virgo/KAGRA (LVK) collaboration in the $\sim 10-100$ Hz frequency band~\cite{Abbott:2016nmj,Abbott:2017vtc,Abbott:2017gyy,Monitor:2017mdv,Abbott:2017oio,TheLIGOScientific:2017qsa,LIGOScientific:2018mvr,LIGOScientific:2021usb,LIGOScientific:2021djp}.  
Furthermore, different pulsar timing array (PTA) collaborations  
announced in 2023 compelling evidence for a GW background (GWB) around the $\sim 10^{-9}$ Hz frequency region~\cite{NANOGrav:2023gor,Antoniadis:2023ott,Reardon:2023gzh, Xu:2023wog}. On top of this, a plethora of new detectors are expected to become operative in the 2030's, improving the sensitivity of currently explored frequency bands, and extending the search for GWs to new frequencies. Upcoming observatories include next generation ground based detectors, like e.g.~the Einstein Telescope (ET)~\cite{Hild:2010id,Punturo:2010zz,Abac:2025saz} or Cosmic Explorer (CE)~\cite{LIGOScientific:2016wof,Reitze:2019iox}, and space-based missions, like the Laser Interferometer Space Antenna (LISA)~\cite{LISA:2017pwj,LISACosmologyWorkingGroup:2022jok}, which 
will search for GWs around mHz frequencies, sitting in between PTA's and terrestrial observatories. 

The great potential of GW astronomy as a new probe of our Universe remains yet to be fully unfold, as plenty of new GW signals are expected to be potentially detectable. 
In the case of GW signals originated in the early Universe, these can only be in the form of backgrounds, commonly referred to as cosmological backgrounds. 
The Universe might be permeated by a large variety of these, ranging from signals originated during inflation~\cite{Grishchuk:1974ny,Starobinsky:1979ty, Rubakov:1982df,Fabbri:1983us,Anber:2006xt,Sorbo:2011rz,Pajer:2013fsa,Adshead:2013qp,Adshead:2013nka,Maleknejad:2016qjz,Dimastrogiovanni:2016fuu,Namba:2015gja,Ferreira:2015omg,Peloso:2016gqs,Domcke:2016bkh,Caldwell:2017chz,Guzzetti:2016mkm,Bartolo:2016ami,DAmico:2021zdd,DAmico:2021vka,Fumagalli:2020nvq,Fumagalli:2021mpc}, to backgrounds possibly generated after inflation, due to e.g.~non-perturbative particle production~\cite{Easther:2006gt,GarciaBellido:2007dg,GarciaBellido:2007af,Dufaux:2007pt,Dufaux:2008dn,Dufaux:2010cf,Bethke:2013aba,Bethke:2013vca,Figueroa:2017vfa,Adshead:2018doq,Adshead:2019lbr,Adshead:2019igv}, kination-domination~\cite{Giovannini:1998bp,Giovannini:1999bh,Boyle:2007zx,Li:2016mmc,Li:2021htg,Figueroa:2018twl,Figueroa:2019paj,Li:2021htg,Gouttenoire:2021wzu,Co:2021lkc,Gouttenoire:2021jhk,Oikonomou:2023qfz}, thermal plasma motions~\cite{Ghiglieri:2015nfa, Ghiglieri:2020mhm, Ringwald:2020ist, Ghiglieri:2022rfp}, oscillon dynamics~\cite{Zhou:2013tsa,Antusch:2016con,Antusch:2017vga,Liu:2017hua,Amin:2018xfe}, first order phase transitions~\cite{Kamionkowski:1993fg,Caprini:2007xq,Huber:2008hg,Hindmarsh:2013xza,Hindmarsh:2015qta,Caprini:2015zlo,Hindmarsh:2017gnf,Cutting:2018tjt,Cutting:2018tjt,Cutting:2019zws,Pol:2019yex,Caprini:2019egz,Cutting:2020nla,Han:2023olf,Ashoorioon:2022raz,Athron:2023mer,Li:2023yaj}, cosmic defects~\cite{Vachaspati:1984gt,Sakellariadou:1990ne,Damour:2000wa,Damour:2001bk,Damour:2004kw,Fenu:2009qf,Figueroa:2012kw,Hiramatsu:2013qaa,Blanco-Pillado:2017oxo,Auclair:2019wcv,Gouttenoire:2019kij,Figueroa:2020lvo,Gorghetto:2021fsn,Chang:2021afa,Yamada:2022aax,Yamada:2022imq,Kitajima:2023cek}, large scalar fluctuations~\cite{Matarrese:1992rp,Matarrese:1993zf, Matarrese:1997ay,Nakamura:2004rm,Ananda:2006af,Baumann:2007zm,Domenech:2021ztg, Dandoy:2023jot}, or others, see~\cite{Caprini:2018mtu} for a comprehensive review. 
While the PTA signal is likely due to supermassive black hole binaries (SMBHBs) \cite{Kelley:2017lek,NANOGrav:2023hfp,Antoniadis:2023xlr}, cosmological backgrounds also represent a viable explanation~\cite{NANOGrav:2023hvm,Antoniadis:2023xlr,Figueroa:2023zhu}.

The detection of any cosmological background will open a new window into the early Universe, probing the high energy physics -- beyond the Standard Model (BSM) -- that generate these signals, typically at much higher energies than those accessible by terrestrial means. In this work, we focus on the GWB emitted by a network of \emph{cosmic strings}, i.e.~one-dimensional field theory topological configurations~\cite{Kibble:1976sj}, arising in BSM particle physics scenarios with certain pattern of 
spontaneous symmetry breaking~\cite{Hindmarsh:1994re,Vilenkin:2000jqa,Jeannerot:2003qv}. 
Alternatively, cosmic strings can also be  fundamental strings of String Theory, stretched out to cosmological scales~\cite{Dvali:2003zj,Copeland:2003bj}. Independently of their origin, a network of cosmic strings consist of infinitely `long' strings stretching across the observable universe, and string loops. 
Most importantly, in the Nambu-Goto (NG) approximation of infinitely thin strings, loops decay solely into GWs, leading to a potentially observable GWB, see e.g.~\cite{Vilenkin:1981bx,Hogan:1984is,Vachaspati:1984gt,Damour:2000wa,Damour:2001bk,Damour:2004kw,Figueroa:2012kw,Sousa:2013aaa,Blanco-Pillado:2017rnf,Blanco-Pillado:2017oxo,Auclair:2019wcv,Gouttenoire:2019kij,Figueroa:2020lvo,Sousa:2020sxs,Gorghetto:2021fsn,Chang:2019mza,Chang:2021afa,Gouttenoire:2021jhk,Yamada:2022aax,Yamada:2022imq,Servant:2023mwt,Servant:2023tua,Blanco-Pillado:2024aca,Wachter:2024zly,Schmitz:2024gds,Avgoustidis:2025svu}. The amplitude of the signal is controlled by the string tension $\mu$, which is determined by the energy scale at which the network forms. Typically, the tension is indicated by the dimensionless fraction $G\mu$, with $G$ Newton's constant.

Due to the variety of cosmic string modeling and details in the GW calculation techniques, the corresponding GWB spectra can exhibit, however, different shapes, resulting in various signal templates, even for the same underlying particle physics parameters. 
Furthermore, as cosmic strings are primarily field theory objects, a natural 
decay channel -- ignored in the NG picture --, is particle emission~\cite{Hindmarsh:2017qff,Matsunami:2019fss,Auclair:2019jip,Saurabh:2020pqe,Hindmarsh:2021mnl,Auclair:2021jud,Baeza-Ballesteros:2023say,Baeza-Ballesteros:2024otj}. Based on this, Ref.~\cite{Baeza-Ballesteros:2024otj} has shown recently that the primary decay route for local field theory string loops formed after a phase transition, is particle production, which implies a suppression of the GWB emitted by a network. 
Cosmic string networks can be still probed in an almost model-independent way, through the imprints in the Cosmic Microwave Background (CMB) of the long strings, via 2- and 3-point functions~\cite{Garcia-Bellido:2010qjz,Fenu:2013tea,Ade:2013xla,Durrer:2014raa,Lizarraga:2014xza,Charnock:2016nzm,Lizarraga:2016onn,Lopez-Eiguren:2017dmc,Figueroa:2010zx,Ringeval:2010ca,Regan:2014vha}. The stringent constraint on the strings tension, arises however from the direct searches of the GWB, which do depend strongly on the modeling assumptions. For instance, fitting PTA data to a standard signal NG cosmic string templates leads to the tight constraint $G\mu \lesssim 10^{-10}$~\cite{NANOGrav:2023hvm,Antoniadis:2023xlr,Figueroa:2023zhu}, whereas the constraint loosens to 
$G \mu \lesssim 10^{-7}$ when fitting the same data to a field theory network that allows for particle production~\cite{Kume:2024adn}. 

In the case of direct detection experiments, the GWB from cosmic strings can be confused with the detector's noise (as any GWB is {\it de facto} another noise), leading to a challenge in data extraction from a real data set. Furthermore, astrophysical backgrounds may actually act as `contaminating' foregrounds that will diminish an experiment's ability to extract a potentially buried cosmological signal in the data. Using a naive method based on achieving certain signal-to-noise-ratio levels, theorists have assessed the detectability of a background in the past, based on comparison of the GWB spectrum against a curve called \emph{power-law sensitivity} (PLS)~\cite{Thrane:2013oya}, built from the noise characterization of a given experiment. In order to improve the reliability of detection claims/expectations, i.e.~to properly quantify the parameter space of a concrete GWB that can be probed by some experiment, it is however 
more appropriate to utilize proper statistical data-analysis methods. 

Several techniques have been recently laid out to quantify the ability of direct detection experiments to reconstruct GWBs, where both signal-agnostic and (spectrum) template-dependent  approaches are used, to reconstruct the spectral shape or parameter space of a GWB, see e.g.~Refs.~\cite{Karnesis:2019mph,Caprini:2019pxz,Flauger:2020qyi,Baghi:2023qnq,Pozzoli:2023lgz,Dimitriou:2023knw,Alvey:2023npw}. 
Recently, the novel technique of the \emph{simulated-based inference} (SBI) for reconstructing GWBs has been pushed forward~\cite{Dimitriou:2023knw, Alvey:2023npw}, and has been proven to be as accurate as the Monte Carlo Markov Chain (MCMC) technique, but much faster. The SBI techniques allow to explore a broad range of BSM parameter space and address the question of how \emph{good} a corresponding GWB signal can be reconstructed (or detected) at future detectors. Furthermore, these techniques are presented in freely available packages that any one can use, removing the need to belong to specific experimental collaborations to have access to signal-detection packages.

The principal aim of our work is to study the ability of LISA to measure the GWB from a cosmic string network, using the package \href{https://github.com/AndronikiDimitriou/GWBackFinder}{\faGithub\,{\tt GWBackFinder}}, based on SBI techniques and publicly free. We aim to investigate cosmic string modelings and circumstances proposed in the literature, including Nambu-Goto's {\tt conventional} templates, such as semi-analytical~\cite{Martins:1995tg,Martins:1996jp,Martins:2000cs,Sousa:2013aaa} and simulation-based~\cite{Blanco-Pillado:2013qja,Blanco-Pillado:2017rnf} modelings, but also other circumstances which we group under the umbrella of 
{\tt beyond conventional} templates, considering: {\it\bf a)} {\it modifications of the loop number density}, as in the LRS modeling~\cite{Lorenz:2010sm}, metastable strings~\cite{Leblond:2009fq,Buchmuller:2021mbb}, current carrying strings~\cite{Auclair:2022ylu} and String theory cases~\cite{Dvali:2003zj,Copeland:2003bj}; 
{\it\bf b)} {\it non-standard cosmologies}~\cite{Cui:2017ufi,Cui:2018rwi,Gouttenoire:2019kij}, including non-standard Hubble rates~\cite{Cui:2017ufi,Cui:2018rwi}, extra thermal and dark sector degrees of freedom ({\it dof})~\cite{Gouttenoire:2019kij}; 
and {\it\bf c)} {\it loop assumptions}, considering atypical loop sizes~\cite{Sousa:2014gka} and GW emission power~\cite{Vilenkin:2000jqa}. 
By analyzing all these signals in the LISA window, {we aim to address the following questions:}

\begin{enumerate}[\it I)]

\item {\it Figure(s) of merit}. How good the signals from each model can be reconstructed at LISA? In order to address this question, we quantify the precision of reconstruction in a case-by-case basis, and present easy-to-interpret figures of merit for each modeling. 
    
\item {\it Parameter space exploration}. Which part of the parameter space of each modeling can be reconstructed and claimed to be detectable at LISA? In order to address this point, we scan over the vast parameter space of all modelings considered.

\item {\it Model comparison}. How confident can one claim that a detected signal comes from one model and not others? In order to address this aspect, we present a systematic model comparison based on the average over sampled datasets of the log-Bayes factor. 

\item {\it Impact of astrophysical foregrounds}. How will astrophysical backgrounds affect the above points ? In order to address this relevant aspect, we will re-consider our analysis of points $I), II), III)$, but in the presence of expected astrophysical foregrounds.
    
\end{enumerate}

\begin{figure}[t!]
    \centering
    \includegraphics[width=0.85\linewidth]{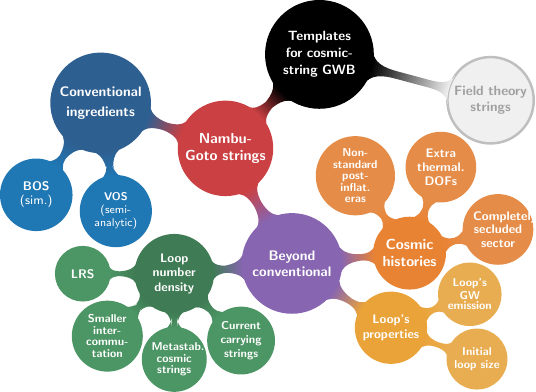}\\[-1em]
\caption{List of  leading templates of cosmic-string GWB signals. Conventional templates (BOS and VOS) rely on standard cosmology and Nambu-Goto results, which are often discussed in literature, 
see Sect.~\ref{sec:conventional}. Beyond-conventional templates are classified into different classes, depending on the modified ingredient in the GWB calculation: {\bf a)} the loop-number density, { \bf b)} the cosmic history, or { \bf c)} the loop properties, see Sect.~\ref{sec:beyondConventional}. Except LRS, all beyond-conventional templates in this work are based on VOS model. Frequency-spectra of all templates are available at the time of writing (April 2025) at \href{https://github.com/peerasima/cosmic-strings-GWB}{\faGithub~\tt repository}, except for the field-theory case (which might eventually become another class of template}, but presently constitutes only work in progress).
    \label{fig:template_summary}
\end{figure}

Addressing all these points constitutes an ambitious program for quantifying LISA's ability to measure a GWB signal from a cosmic string network. We plan to present the completion of these goals in a series of papers, in which we will progressively incorporate more complexity. The present paper is thus intended only as the first one in the series, in which we address points {\it I)}, {\it II)} and {\it III)} in the absence of astrophysical foregrounds, i.e.~ignoring yet point {\it IV)}. We highlight that for Nambu-Goto's conventional templates, points {\it I)} \& {\it II)} have been already addressed by the LISA collaboration~\cite{Blanco-Pillado:2024aca}, both in the absence and presence of foregrounds. In this first paper we re-evaluate points {\it I)} \& {\it II)} for conventional templates in the absence of foregrounds, using SBI methods instead, so we can compare techniques as a starting point. Most relevantly, in this paper we extend the analysis of points {\it I)} \& {\it II)} to beyond conventional templates, which constitute a much larger pool of cases than the conventional templates, with higher parameter complexity dependencies in most cases. Furthermore, here we also include 
a model comparison analysis, i.e.~point {\it III)}, for both conventional and non-conventional templates, yet in the absence of foregrounds. Actually, points {\it I)}-{\it II)} for non-conventional templates, and point {\it III)} for either conventional or non-conventional cases, require a scan over a broad range of parameter spaces, which is only made feasible precisely because of the use of SBI techniques. Figure~\ref{fig:template_summary} and Table~\ref{tab:summary_templates_beyond} indicate all templates, both conventional and non-conventional, that constitute our catalog of cosmic-string GWB signals.
All templates in the present work are calculated with Nambu-Goto string frameworks. VOS and BOS templates assume that all three ingredients (loop number density, loop's properties, and cosmic history) are conventional. Except for the LRS template, which relies on its own framework, all beyond-conventional templates are based on the VOS framework, with one of the three ingredients altered. Note that one can always modify these ingredients in any calculation framework if they are allowed to be changed.

We leave the study of point {\it IV)} for upcoming papers in the series, where we will study cosmic string signals in the presence of astrophysical foregrounds in the LISA window. 
We note that there have several studies on the impact of foregrounds on the template reconstruction, e.g., \cite{Baghi:2023qnq,Muratore:2023gxh,Dimitriou:2023knw,Kume:2024xvh} and especially on the reconstruction of conventional cosmic-string templates~\cite{Blanco-Pillado:2024aca}.
In the next papers of our series we will study the detectability of template signals over relevant astrophysical foregrounds, including those from extragalactic black hole and neutron star binaries, white dwarfs in our galaxy, and extragalactic white dwarfs. The latter source, which was not considered in the analysis of Ref.~\cite{Blanco-Pillado:2024aca}, leads however to the dominant expected  foreground over the entire frequency window of LISA~\cite{Staelens:2023xjn,Hofman:2024xar}. In particular, our second paper of the series will quantity the significant 
impact of such dominant foreground on the detectability of conventional signals, providing a realistic assessment on the parameter space detectable by LISA of these signals, when considering them superimposed on top of all leading foregrounds expected in the LISA window. 

We will also consider an analogous study of the detectability of non-conventional signals over relevant astrophysical foregrounds in the LISA window. The inclusion of the foregrounds will have again a significant impact on the parameter space of such signals. However, given the complexity of such multi-parameter analysis, we will separate it from the conventional signals over foregrounds analysis, and present the details in separate papers of the series. Finally, we also note that we leave open the possibility to add even more papers into the series, e.g.~considering new signal templates, or potential improvements in the data analysis pipeline and/or the noise modeling of LISA.
\newpage

\vspace*{-0.5cm}

\begin{tcolorbox}[width=\textwidth]
\begin{center}
    \underline{\bf Series of Papers -- Guide}
\end{center}

{\bf 1st Paper (this one)}. We quantify the ability of LISA to probe a catalog of cosmic-string GWBs, from conventional to beyond-conventional templates, in the absence of astrophysical foregrounds. We present figures of merit, parameter space exploration, and model comparison analysis.\vspace*{0.25cm} 

\noindent {\bf 2nd Paper.} We study the ability of LISA to probe conventional cosmic-string signal templates in the presence of expected astrophysical foregrounds in the LISA window, including extragalactic black-hole and neutron star binaries, intragalactic and extragalactic white dwarf binaries, and possibly others.\vspace*{0.25cm} 

\noindent {\bf 3rd Paper.} We analyse the ability of LISA to probe non-conventional cosmic-string signal templates in the presence of expected astrophysical foregrounds in the LISA window, including extragalactic black-hole and neutron star binaries, intragalactic and extragalactic white dwarf binaries, and possibly others.

\vspace*{0.25cm} 

{{\noindent {\bf Further Papers.} Possible further papers in the series, e.g. considering new signal
templates or potential improvements in the data analysis pipeline and/or LISA
noise modeling.}}

\end{tcolorbox}

\newpage

\begin{center}
    {\bf -- Paper outline --}
\end{center}\vspace*{-0.3cm}

Section~\ref{sec:GWB_recap} provides a brief review of cosmic strings and of the calculation of its associated GWB, based on specifying the loop number density, the cosmic history, and certain properties of the loops. We present templates for conventional signals in Sect.~\ref{sec:conventional}, where we discuss both semi-analytical and simulation-based templates, and the uncertainty in the energy-running of the Standard Model effective degrees of freedom. In Section~\ref{sec:beyondConventional} we introduce the templates for non-conventional signals, based on varying either of the following ingredients: the loop number density (Sect.~\ref{subsec:LoopNumberDensityBSM}), the cosmic history (Sect.~\ref{sec:GWB_cosmic_history}), and/or the loop properties (Sect.~\ref{subsec:loopProperties}). In Section~\ref{sec:reconstruction}, we discuss the goodness of reconstruction of conventional templates (Sect.~\ref{sec:recon_result_conven}), and also quantify the ability of LISA to differentiate between conventional modelings (Sect.~\ref{subsec:ModelComparisonVOSvsBOSvsDOF}), given a data set. In Sect.~\ref{sec:recons_beyond} we present an analogous analysis on the goodness of reconstruction by LISA of beyond-conventional templates, considering separately models that modify beyond conventional templates the loop number density (Sect.~\ref{subsec:recon_loop_number}), the cosmic history (Sect.~\ref{subsec:recon_explore_cosmic_hist}), and the loop properties (Sect.~\ref{subsec:recons_loopProperties}). We present again a study on model comparison (Sect.~\ref{sec:model_comparison_beyond_conventional}), though reducing the discussion to few representative beyond-conventional scenarios vs conventional signals. We summarize our results and discuss future perspectives/improvements in Section~\ref{sec:conclude}. In App.~\ref{app:GWB_derivation} we provide a detailed  derivation of the master formula for a GWB from a cosmic string-network, as well as technicalities for speeding up computations. App.~\ref{app:generalized_VOS_current} recaps the generalized equations describing the evolution of string network.  App.~\ref{app:diff_model_calculation} discuss some subtleties in the calculation of  conventional BOS template. 
Brief summaries on the LISA noise model (including data generation method), the SBI technique, the MCMC method, and the Bayesian model comparison are provided in Apps.~\ref{app:LISA_noise}, \ref{app:SBI_summary}, \ref{app:mcmc}, and \ref{app:model_comparison}, respectively.

\section{The GW background from a cosmic string network}
\label{sec:GWB_recap}

Cosmic strings are line-like objects arising naturally in beyond the Standard Model (BSM) field theories with spontaneous symmetry-breaking, where the resulting vacuum manifold has non-trivial first homotopy group (i.e.~it corresponds to a non-shrinkable loop). The simplest case is the spontaneous breaking of a $U(1)$ symmetry, which leads to global or local strings, depending on whether the broken group is global or gauge. After the symmetry-breaking process, a network of cosmic strings emerges in the universe, characterized by an \emph{energy density per unit length} or \emph{string tension}\footnote{The string tension $T$ can differ from the energy density per unit length $\mu$ in the case of wiggly strings~\cite{Vachaspati:1991sy}. We assume in this work that strings do not possess wiggles and $\mu = T$.} $\mu$, determined by the energy scale of the symmetry breaking $\eta$. As $\mu$ is of mass dimension $+2$, it is often expressed through the dimensionless ratio
\begin{align}
    G \mu \equiv \left(\frac{\eta}{M_{\rm Pl}}\right)^2 \simeq 6.7 \times 10^{-11} \left(\frac{\eta}{10^{14}~ {\rm GeV}}\right)^2,
    \label{eq:tension}
\end{align}
where $G = 1/M_{\rm Pl}^2$ is the gravitational constant, and $M_{\rm Pl} \simeq 1.22 \times 10^{19} ~ {\rm GeV}$ is the Planck mass scale. The strings of the network subsequently evolve, \emph{intercommuting} with themselves and chopping off small loops that later decay into particles and GWs\footnote{The long strings generate GWs as well but with an emission power much smaller than that of loops~\cite{Figueroa:2012kw,Figueroa:2020lvo,CamargoNevesdaCunha:2022mvg}.}. The resulting network thus consists of {\it long} strings, stretching across the horizon, and {\it loops} of sub-horizon size, which continuously shrink till their eventual decay. This process leads to a continuous energy loss of the network, which enters into a \emph{scaling} regime, where the network's energy density becomes proportional to the background energy density of the universe $\rho_{\rm tot}$, with $\rho_{\rm net} \simeq G\mu \rho_{\rm tot}$. The typical separation among long strings tracks the horizon length as the universe expands. 

Strictly speaking, the formula for the tension Eq.~(\ref{eq:tension}) is only valid for local strings, as these have a well localized spatial core profile (it is therefore also valid for fundamental strings from String Theory, which literally have zero width). Global strings, on the contrary, have a spatially extended core profile, with their tension receiving radial log corrections as $\propto \log[r/\eta^{-1}]$, due to the massless Goldstone field configuration. This correction extends up to the typical separation between infinite strings, i.e.~up to the Hubble scale $1/H$,  so it can be `sizeable'. 
Moreover, global string loops decay dominantly into the massless Goldstone modes~\cite{Saurabh:2020pqe,Baeza-Ballesteros:2023say}, which affects severely the loop number density. Both network evolution and GW and particle emission by global strings are under ongoing debate, and the final GWB spectrum can vary over six orders of magnitude for the same $\eta$ parameter, see {\it e.g.}~discussion in~\cite{Servant:2023mwt}. As our present work focuses on the precise reconstruction of GWB signals by LISA, we only consider GWB templates representing local strings. 

Local field theory strings have a finite core's width, corresponding to the inverse mass scale of the particles involved, and hence parametrically\footnote{The actual width of the string depends explicitly on the dimensionless coupling parameters, which are model dependent.} controlled by the microscopic length scale $\eta^{-1}$. As this width is much smaller than the typical separation among long strings (set by the cosmological horizon, or the inverse curvature length scale of the strings), the motion of either long strings or loops is expected to be well approximated by the dynamics of infinitely-thin strings, also known as Nambu-Goto (NG) strings. These are exactly one-dimensional objects, i.e.~with zero width, and hence their decay route can solely go through GW emission, since they are not made of any internal $dof$. All templates and results in this paper assume the well-studied Nambu-Goto strings, which by construction considers that particle emission (other than GWs) is absent. However, we note that recent lattice simulations clearly show that field-theory local strings formed after a phase transition can actually emit particles very efficiently~\cite{Baeza-Ballesteros:2024otj}, which will impact most directly in the density of loops along cosmic history. We postpone the calculation of the GWB spectrum from such field theory particle-emitter network for future work, and we limit ourselves here to the standard assumption of Nambu-Goto templates (representative of local strings, or fundamental strings from String Theory at most) with negligible particle emission.

Following, we recap generic formulae for calculating the GWB spectrum from a cosmic string network. We highlight that there are three major ingredients---loop number density, cosmic history, and loop properties---leading to three classes of templates that we discuss in later sections. For completeness, the derivation of the master formula Eq.~(\ref{eq:master_formula_GWB_strings}) is shown in appendix~\ref{app:GWB_derivation}.

\subsection{The GWB spectrum from a cosmic string network}

The GWB from a cosmic string network comes from the superposition of the GW signals emitted by the many loops that are produced along cosmic history. At each time $t$, there is a loop number density per unit length, $\texttt{n}(l,t)$,  so that $\#_{\rm loop}(t) = {\tt n}(l,t) dl$ indicates the number density of loops with length in range $[l,l+dl)$. As each loop oscillates and emits GWs at harmonic frequencies $f_e^{(j)} = 2j/l(t_e)$, $j = 1, 2, 3, ...$, with $l(t_e)$ the loop's length at the GW-emission time $t_e$, the redshifted frequency today of the $j$-th mode emitted at $t_e$, reads 
\begin{align}
    f = \left[\frac{2j}{l(t_e)}\right]\cdot\left[\frac{a_e}{a_0}\right],
    \label{eq:GW_frequency_loop_length}
\end{align}
with $a_e$ and $a_0$ the scale factor at $t_e$ and today, respectively. The master equation for the GWB spectrum today accounts for summing the GWs emitted by all loops produced along cosmic history till today, as 
\begin{align}
    \boxed{\Omega_{\rm GW}(f) = \frac{1}{3 H_0^2 m_{\rm Pl}^2}\sum_{j=1}^{\infty} \underbrace{\frac{2j}{f} ( G \mu^2 P_j )}_{\substack{{\rm GW ~ emission} \\ {\rm from ~ single ~ loops}}} \int_{a_{\rm ini}}^{a_0} da \, \underbrace{\frac{1}{H(a)}\left(\frac{a}{a_0}\right)^4}_{\rm cosmic ~ history} \, \underbrace{{\tt n}\left[\frac{2j}{f} \cdot \frac{a}{a_0},t(a)\right]}_{\rm loop ~ number ~ density}}~,
    \label{eq:master_formula_GWB_strings}
\end{align}
where $m_{\rm Pl}$ is the reduced Planck mass, $H_0$ is today's Hubble rate, and $a(t)$ is the scale factor. The lower integration bound $a_{\rm ini}$ is the scale factor when the cosmic-string network starts producing GWs. 
A derivation of this master equation can be found in appendix~\ref{app:GWB_derivation}.

From Eq.~\eqref{eq:master_formula_GWB_strings}, we note that the cosmic string GWB calculation requires the specification of three ingredients: {\bf a)} the loop number density ${\tt n}(l,t)$, {\bf b)} the cosmic history via the temporal dependence of the scale factor $a(t)$, and {\bf c)} the (harmonic) GW emission power from an individual loop, $P_j$, which we define later. Next, we discuss briefly the standard way to describe each of these ingredients in Sections~\ref{subsec:loopNumberDensity}, \ref{subsec:cosmicHistory}, \ref{subsec:singleLoopGWemission}. Following, we present in Section~\ref{sec:conventional} 
{\it conventional} templates of GWB spectra from cosmic string networks, based on conventional arguments for ${\tt n}(l,t), a(t)$ and $P_j$, whereas in Section~\ref{sec:beyondConventional} we present instead beyond-conventional templates, based on variations (individually or in combination) of the loop number density, the expansion history, and/or the loop properties, with respect to conventional expectations, as motivated by various BSM scenarios.

\subsection{Loop number density}
\label{subsec:loopNumberDensity}

The functional form of the loop number density ${\tt n}(l,t)$ can be obtained from analytical arguments,  numerical simulations, or from combination of the two. In many cases, it is the {\it loop production function} ${\tt f}(l_i,t_i)$, which is rather obtained from theoretical considerations and/or simulations. This is defined as the number density of loops per unit length and per unit time, at the loop-production time $t_i$. It is, in other words, the birth rate of loops, which are formed with initial length $l_i$, at the time $t_i$. The loop production function is related to ${\tt n}(l,t)$
at any time $t$ by 
    \begin{align}
        {\tt n}(l,t) \equiv \int_{t_{\rm min}}^{t} {\tt f}[l_i,t_i] \left[\frac{a(t_i)}{a(t)}\right]^3 dt_i\,,
        \label{eq:loop_number_density}
    \end{align}
where $l$ is a function of $\{l_i, t,t_i\}$ [see Eq.~\eqref{eq:loop_length_evo} below], and the redshift factor $[a(t_i)/a(t)]^3$ accounts for dilution of the loop number density from $t_i$ to $t$. The integral's lower boundary $t_{\rm min}$ is the earliest time that the network reached the scaling regime, which is assumed to be a time scale close enough to the formation time of the network. 

\subsection{Cosmic history}
\label{subsec:cosmicHistory}

According to the standard $\Lambda$CDM model of cosmology, the Hubble expansion rate $H \equiv d\log a/dt$ from the first Friedmann equation reads
\begin{align}
    H (t) = H_0 \sqrt{\Omega_{\rm rad}^{(0)} \mathcal{G}[T(t),T_0]\left[\frac{a_0}{a(t)}\right]^4 + \Omega_{\rm mat}^{(0)} \left[\frac{a_0}{a(t)}\right]^3 + \Omega_{\rm de}^{(0)}}\,,
    \label{eq:friedmann_LambdaCDM}
\end{align}
where $a(t)$ is the scale factor, the subscript/superscript ``0" denotes the time today, $\Omega_i^{(0)}$ is the current energy density fraction of the $i$-th component of the Universe, and the function $\mathcal{G}(T,T_0)$ describes the change of relativistic degrees of freedom ({\it dof}). In particular, 
\begin{align}
    \mathcal{G}(T,T_0) \equiv \frac{g_*(T)T^4}{g_*(T_0) T_0^4}\left[\frac{a(t)}{a_0}\right]^4= \frac{g_*(T)}{g_*(T_0)}\left[\frac{g_{*s}(T_0)}{g_{*s}(T)}\right]^{4/3},
\label{eq:bigG_expression}
\end{align}
where $g_*$ and $g_{*s}$ are the effective number of relativistic {\it dof} in energy density and entropy, respectively. 
We assume the standard $\Lambda$CDM cosmology for all templates, i.e.~$\Omega_{\rm rad}^{(0)} = 9.2 \times 10^{-5}$ for radiation, $\Omega_{\rm mat}^{(0)} = 0.308$ for non-relativistic matter, and $\Omega_{\rm de}^{(0)} = 0.692$ for dark energy, according to Planck 2018~\cite{Planck:2018vyg}. The only exception to this assumption is in section~\ref{sec:GWB_cosmic_history}, where we consider nonstandard early universe histories.

\begin{figure}[t]
\centering
    \includegraphics[width=10cm]{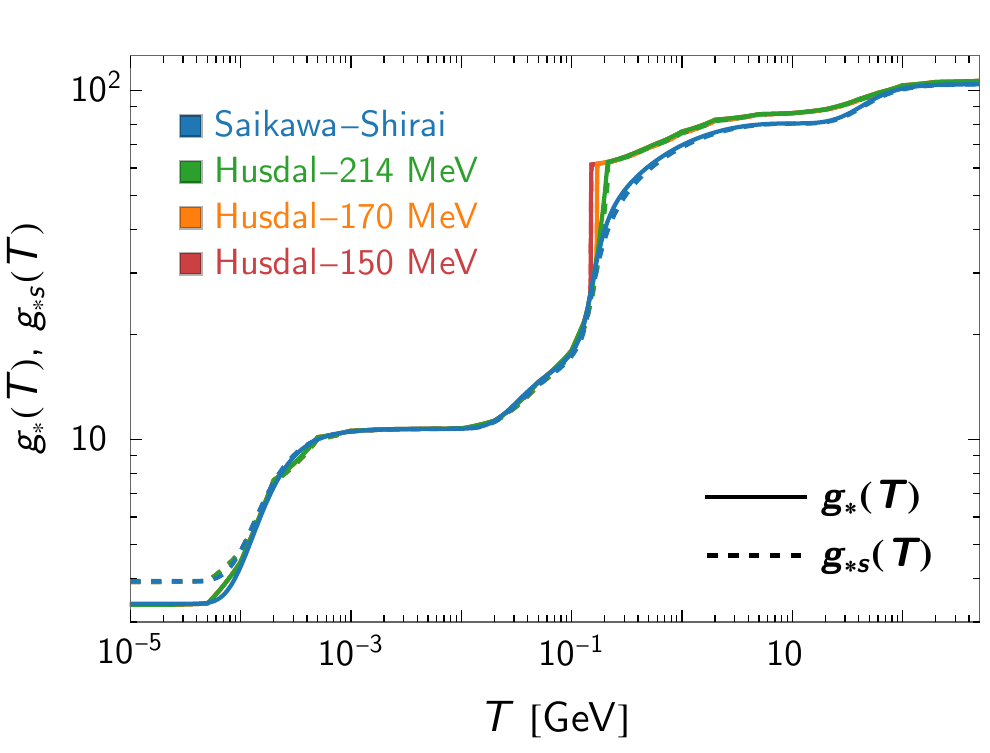}\\[-1em]
    \caption{Evolution of effective numbers of relativistic degrees of freedom in energy density (solid) and entropy density (dashed), obtained from Husdal \cite{Husdal:2016haj} and Saikawa-Shirai \cite{Saikawa:2018rcs}. In Husdal's result, several evolution profiles depend on the crossover temperature of the QCD phase transition.}
    \label{fig:g_gs}
\end{figure}

In this work we take the evolution of $g_*$ and $g_{*s}$ based on the SM {\it dof}, from Saikawa-Shirai \cite{Saikawa:2018rcs} and Husdal \cite{Husdal:2016haj} results, which are shown in Fig.~\ref{fig:g_gs}.
The latter also has several profiles depending on the crossover temperature for the QCD phase transition. Unless stated otherwise, our templates use by default the results of Saikawa-Shirai~\cite{Saikawa:2018rcs}, which improved previous calculations---using the ideal gas approximation for the Standard Model (SM) thermal bath---by including the effect of particle interactions. 
In section~\ref{sec:modeling_DOFs} we compare, in any case, the impact on the GWB spectrum from considering different prescriptions for the $g_*$ and $g_{*s}$ evolution, comparing the Husdal~\cite{Husdal:2016haj} and the Saikawa-Shirai~\cite{Saikawa:2018rcs} prescriptions. In  Section~\ref{sec:GWB_cosmic_history} we further modify the prescription for $g_*, g_{*s}$ at high energies, assuming the presence of extra relativistic {\it dof} motivated by BSM scenarios.

\subsection{GW emission from a single loop}
\label{subsec:singleLoopGWemission}

The GW emission power $dE_j/dt$ from the $j^{\rm th}$ harmonic mode and the total power $P_{\rm GW}$, can be parametrized as \cite{Vachaspati:1984gt}
\begin{align}
    \frac{dE_j}{dt} = G\mu^2 P_j ~~~~~~ \Rightarrow ~~~~~~ P_{\rm tot} = \sum_j {P_j} G\mu^2 \equiv \Gamma G \mu^2,
    \label{eq:GW_emission_power_single_loop}
\end{align}
with $P_j$ a dimensionless factor, independent of the loop size and loop age [such that it can be taken out of the integration in Eq.~\eqref{eq:master_formula_GWB_strings}], and where $\Gamma \equiv \sum_j P_j$ is a dimensionless constant. There are two standard ways to include the GW emission power: 
\begin{enumerate}[i)]
\item Using the high-frequency asymptotic expansion for high harmonics, $P_j \propto j^{-q}$, each mode is assumed to emit GWs with power  
\begin{align}
    P_j = \frac{\Gamma}{\zeta(q)} j^{-q}\,,
    \label{eq:grav_emission_power_loop_j}
\end{align}
where $\zeta(q) = \sum_k k^{-q}$ is the Riemann zeta function and $q$ depends on the features of the loop, with $q=4/3, 5/3$ and $2$, for loops with cusps, kinks, and colliding kinks, respectively. One typically uses $\Gamma \simeq 50$, which corresponds to the approximated value obtain in simulations~\cite{Blanco-Pillado:2017oxo}.

\item A more refined functional form of the GW emission power $P_j$ has been extracted from the numerical simulations in Ref.~\cite{Blanco-Pillado:2017oxo}, where it was found that $j^{4/3} P_j$  has a local maximum around $j \simeq 3$ and becomes constant for $j \gtrsim 100$, see figure~4 of \cite{Blanco-Pillado:2017oxo}. In this case we also use the precise value $\Gamma = \sum_j P_j \simeq 51.43$, as extracted from the simulations~\cite{Blanco-Pillado:2017oxo}.
Note that recent $P_j$ results \cite{Wachter:2024aos} including the backreaction from GW emission show that the functional form of $P_j$ evolves as loop shrinks. While we do not incorporate this new behavior in our template calculation, the loop-lifetime dependence of $P_j$ can lead to up to $20\%$ refinement in the final $\Omega_{\rm GW}$~\cite{Wachter:2024zly}.
\end{enumerate}

In most templates, we use Eq.~\eqref{eq:grav_emission_power_loop_j} for a more efficient calculations of the GW spectra, except in section \ref{sec:GW_simulated}, where we calculate the GWB template using the refined $P_j$ function.\footnote{Note that we do not have access to the data from \cite{Blanco-Pillado:2017oxo} and, instead, we have digitalized the functions from their figure 4.}
Moreover, we typically assume $q=4/3$ because a cusp can naturally appear on a loop after a few oscillations~\cite{Vilenkin:2000jqa}, except in section~\ref{subsec:loopProperties} where we explore other possibilities.\vspace*{-0.3cm}\\

{\bf Loop-length shrinkage.}---Upon emitting GWs, a loop of length $l$ shrinks in time as its energy $E = \mu l$ is lost in the form of gravitational radiation. 
As its tension is constant, we can write 
\begin{align}
  \frac{d E}{d t} = \mu\frac{d l}{dt} = - \Gamma G \mu^2 ~~~~~ \Rightarrow ~~~~~ l(t) = l_i - \Gamma G\mu(t-t_i),
\label{eq:loop_length_evo}
\end{align}
where the second step is the solution of the length evolution, with $t_i$ the initial time when the loop was created with length $l_i \equiv l(t_i)$.
Due to the smallness of the gravitational coupling, a loop emits GWs slowly and hence it lives for a long time $\Delta t \equiv t_f - t_i$, where the final decay time is defined by the condition $l(t_f) = 0$. It then follows that 
\begin{align}
    \Delta t = \frac{l_i}{\Gamma G\mu} = \frac{\alpha}{\Gamma G \mu} t_i\,,
\end{align}
where $\alpha \equiv l_i/t_i$ is the initial loop size in units of cosmic time. We use the value $\alpha \simeq 0.1$ ($\gg \Gamma G\mu$) as obtained in simulations~\cite{Blanco-Pillado:2011egf} for loops contributing to the GWB~\cite{Blanco-Pillado:2013qja,Auclair:2019wcv}, so typically $\Delta t / t_i \gg 1$.

We note that, given the strongest current constraint on the tension from pulsar timing arrays, $G\mu \lesssim 8\times 10^{-11}$~\cite{NANOGrav:2023hvm,EPTA:2023xxk,EPTA:2023hof,Figueroa:2023zhu,Ellis:2023oxs}, only loops produced during RD contribute {\it dominantly} to the GWB spectra amplitude today within the LISA frequency window $\sim\,10^{-5}$\,--\,$1$ Hz. As shown in Fig.~\ref{fig:BOS_each_contribution} in Appendix~\ref{app:pieces_CS_GWB}, the contribution to the GWB from loops produced after matter-radiation equality is typically peaked at much smaller frequencies, and is always negligible within the LISA frequency window, independently of the tension.

\section{Conventional templates}
\label{sec:conventional}

\begin{figure}[t!]
    \centering
    {\sffamily Conventional templates}\\[0.5em]
    \hspace*{-0.2cm}\includegraphics[width=7.5cm]{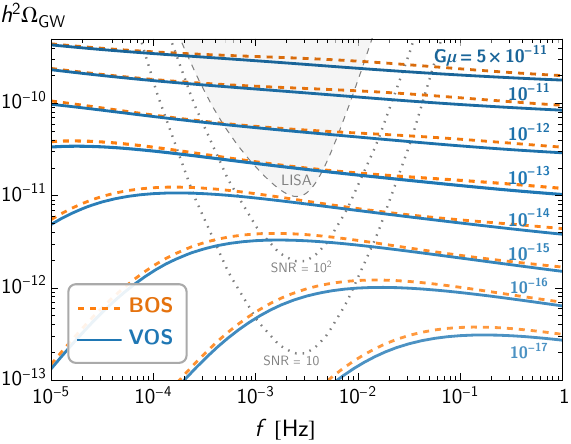}~~~ 
    \includegraphics[width=7.4cm]{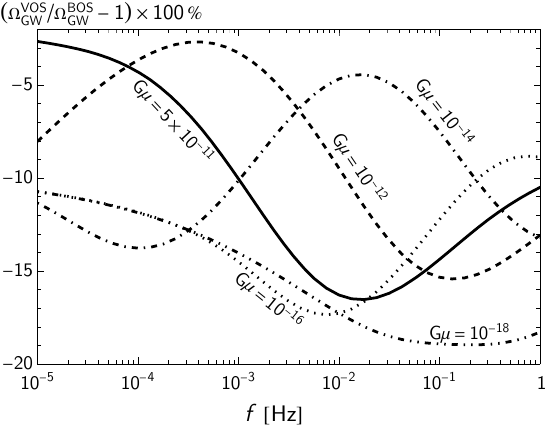}
    \\[-1em]
\caption{{\it Left Panel:} GWB spectra from a local cosmic string network for $G\mu = 5\times10^{-11}, 10^{-11},10^{-12},..., 10^{-18}$, obtained using the VOS model (Sect.~\ref{sec:GW_semi_analytic}, solid blue) and the BOS model (Sect.~\ref{sec:GW_simulated}, dashed orange). The $g_*,g_{*s}$ evolutions correspond to Saikawa-Shirai~\cite{Saikawa:2018rcs}. 
The top gray dotted line is the LISA (AA-channel) noise sensitivity, while the other two gray lines are the PLS curves with ${\rm SNR} = 100$ and $10$, respectively, and with $T_{\rm obs} = 75\% \times 4 = 3$ years (see Eq.~\eqref{eq:PLS_curve}  for LISA PLS curve definition). 
{\it Right Panel:} The \% relative difference of the predictions from the VOS and BOS modelings, for different tensions $G\mu$. 
Note that $\Omega_{\rm GW}^{\rm VOS}(f) < \Omega_{\rm GW}^{\rm BOS}(f)$ for all LISA frequencies. 
}
\label{fig:standard_spectra}
\end{figure}

The loop number density ${\tt n}(l,t)$, {\it cf.} Eq.~\eqref{eq:loop_number_density}, produced from a Nambu-Goto string network, can be obtained from either semi-analytic calculations or from fully numerical simulations.
This section focuses on two templates relying on each approach: Sect.~\ref{sec:GW_semi_analytic} considers the VOS template~\cite{Martins:1995tg,Martins:1996jp,Martins:2000cs}, based on analytical arguments and calibrated with numerical simulations, while Sect.~\ref{sec:GW_simulated} considers the BOS template~\cite{Blanco-Pillado:2013qja}, solely based on numerical simulations. 
For other dependencies discussed in the previous section, we use the cosmic history of the $\Lambda$CDM model [Eq.~\eqref{eq:friedmann_LambdaCDM}] and assume loops with cusps, i.e.~$q=4/3$. 
For the GW emission power of each mode, we use Eq.~\eqref{eq:grav_emission_power_loop_j} for VOS and the refined form obtained from the numerical simulations of Ref.~\cite{Blanco-Pillado:2017oxo} for BOS. We discuss the impact of different evolutions of the number of relativistic SM particle species in both the VOS and BOS templates in Sect.~\ref{sec:modeling_DOFs}. 
\begin{center}
    {\bf All templates in this section depend only on one parameter: $G\mu$}
\end{center}
The left panel of Fig.~\ref{fig:standard_spectra} shows the GWB spectra of the two models, VOS and BOS, which make very similar predictions within the LISA window. In the right panel of Fig.~\ref{fig:standard_spectra} we plot for the representative tensions $G\mu = 10^{-10}, 10^{-12}, 10^{-14}, 10^{-16}, 10^{-18}$, the relative difference in amplitude within the LISA frequency window, showing that is always less than $20\%$, independently of the tension. We note that $\Omega_{\rm GW}^{\rm BOS}(f) > \Omega_{\rm GW}^{\rm VOS}(f)$ for all LISA frequencies.  

\subsection{VOS: semi-analytic modeling}
\label{sec:GW_semi_analytic}
When considering cosmic strings as effectively one-dimensional Nambu-Goto strings, the dynamics of long strings can be described by the equations of their correlation length $L(t)\equiv \xi(t) t$ and mean velocity $\bar{v}(t)$ in the so-called \emph{Velocity-dependent One-Scale} (VOS) model~\cite{Martins:1995tg,Martins:1996jp,Martins:2000cs}. Details of said equations can be found in  Appendix~\ref{app:GWB_derivation}, or in the description of {\it Model I} in~\cite{Auclair:2019wcv}. By requiring that loops are created in such a way that they lead the network to evolve into the scaling regime, the loop number density---within time $(t_i,t_i+dt_i)$---can be written as~\cite{Martins:2000cs,Cui:2017ufi,Cui:2018rwi,Gouttenoire:2019kij}
\begin{align}
    \#_{\rm loop}^{\rm VOS}(t) = \mathcal{F}_\alpha \frac{C_{\rm eff}(t_i)}{\alpha_L \xi(t_i) t_i^4}\left[\frac{a(t_i)}{a(t)}\right]^3 dt_i = \underbrace{\mathcal{F}_\alpha \frac{C_{\rm eff}(t_i)}{\alpha_L \xi(t_i) t_i^4}\left[\frac{a(t_i)}{a(t)}\right]^3 \frac{dt_i}{dl}}_{= \mathtt{n}(l,t)} dl,
    \label{eq:vos_loop_number_density}
\end{align}
where the loop production coefficient,  defined by $C_{\rm eff}(t) = \tilde{c} \bar{v}(t)/[\gamma \xi^3(t)]$ with $\tilde{c} \simeq 0.23$ \cite{Martins:2000cs}, is the loop formation efficiency (taken from Nambu-Goto string simulations in an expanding Universe), while $\gamma \simeq \sqrt{2}$ is the typical Lorentz boost of loops.
As justified by the numerical simulations in Ref.~\cite{Blanco-Pillado:2013qja}, only a fraction $\mathcal{F_\alpha} \simeq 0.1$ of the total number of loops in the NG simulations, produced with initial size $l_i=\alpha t_i$ with $\alpha = \alpha_L \xi(t) t$ and $\alpha_L \simeq 0.37$, contributes significantly to GW production, while the other $\sim 90\%$ of loops are too small and their energy density simply redshifts away quickly, making negligible their contribution to the GWB.
We note that the simulation results contains uncertainties, which can be modeled phenomenologically as the fuzziness parameter \cite{Blanco-Pillado:2024aca}. This fuzziness parameter would modulate the overall amplitude of the GW spectrum and leads to some degeneracy with other model parameters. We simplify our analysis by not taking into account such degeneracy.

For loops produced during the scaling regime in the radiation era ($\xi_{r}\simeq 0.27$), we can simply use $\alpha = 0.1$~\cite{Blanco-Pillado:2017oxo}. 
We recall that for the tensions allowed by the current observational constraints, i.e.~$G\mu \lesssim 10^{-10}$, the GWB at the LISA window is generated by loops created during the radiation era, so it is reasonable to use $\alpha = 0.1$ in the calculation. 
Although the effect of relativistic DOFs around QCD and electroweak scales slightly changes $\xi(t)$ from its scaling regime,  
the relative difference in the GW spectra from using $\alpha(t) = \alpha_L \xi(t)$ and $\alpha = 0.1$ is less than 1\% (scanning over $G\mu$). 
We will therefore use $\alpha = 0.1$ by default, whenever using the VOS modeling, unless stated otherwise. 
Exceptions to this are some particular cases, where we will consider using $\alpha(t) = \alpha_L \xi(t)$, e.g.~when discussing current-carrying strings (Sect.~\ref{sec:current_carrying_string}), or the impact of cosmic history in the GWB signal (Sect.~\ref{sec:GWB_cosmic_history}), where $\xi$ deviates substantially from the scaling value in radiation.

Plugging Eq.~\eqref{eq:vos_loop_number_density} into Eq.~\eqref{eq:loop_number_density} and using also Eq.~\eqref{eq:loop_length_evo}, leads to a loop number density per unit length, at any given time, as
\begin{align}
    \boxed{{\tt n}_{\rm VOS}(l,t) =  \mathcal{F}_\alpha \frac{C_{\rm eff}(t_i)}{\alpha_L \xi(t_i) [\alpha_L \xi(t_i) + \alpha_L t_i \xi'(t_i) + \Gamma G\mu] t_i^4} \left[\frac{a(t_i)}{a(t)}\right]^3}\,~.
    \label{eq:loop_number_density_ModelI}
\end{align}
In the case of $\alpha = \alpha_L \xi = 0.1$, the second term in the denominator vanishes ($\xi'(t) = 0$).

We have generated templates for the GWB spectrum for the VOS model assuming the GW emission per mode as in Eq.~\eqref{eq:grav_emission_power_loop_j}, with $q=4/3$ and $\Gamma = 50$, and using Eq.~(\ref{eq:loop_number_density_ModelI}) with $\alpha = 0.1$, feeding $C_{\rm eff}(t_i)$ from solving the VOS equations in a $\Lambda$CDM Universe (see Appendix~\ref{app:generalized_VOS_current}). The resulting GWB spectra for representative values of the tension $G\mu$ are indicated by blue solid lines in the left panel of Fig.~\ref{fig:standard_spectra}.

In recent years, there have been developments on the analytical approximation for the expression of the GWB spectrum in Eq.~\eqref{eq:master_formula_GWB_strings} with the VOS input \eqref{eq:loop_number_density_ModelI} \cite{Sousa:2020sxs,Schmitz:2024gds}, reducing the computational resources for solving the system of VOS equations. The state-of-the-art result has  shown that the analytic approximation can reproduce the result where the VOS equations are solved fully numerically, with a relative error in $\Omega_{\rm GW}$ up to $\sim 20\%$ \cite{Schmitz:2024gds}.
As we shall show below, the uncertainty in the reconstruction of the GWB spectrum in LISA is however more precise, as good as $\sim 1 \% - 10$ \% level, depending on the tension. Hence, in our analysis we rather employ the full numerical solution of the VOS equations\footnote{We found that solving the VOS equations does not consume substantial computational time (a few seconds on a 4-core laptop), as the bottleneck in computation is the higher-harmonic summation, which we solve with a trick, as explained in App.\,\ref{app:mode_summation}.}.

\subsection{BOS: simulation-based modeling}
\label{sec:GW_simulated}
The loop number density can also be obtained directly from numerical simulations. In particular, using the simulations of a Nambu-Goto string network by Blanco-Pillado, Olum, and Shlaer (BOS)~\cite{Blanco-Pillado:2013qja}, the {\it loop production function} in the scaling regime (Eqs.~(17), (18), (30) in~\cite{Blanco-Pillado:2013qja}) during radiation and matter, reads
\begin{align}
    {\tt f}_{\rm BOS}[l_i,t_i] = \begin{cases}
    \frac{92.126}{d_H^5(t_i)} \,\delta\left[\frac{l_i}{d_H(t_i)}- 0.05\right] ~ ~ & {\rm (radiation~era)},\vspace{0.3cm}\\
    \frac{5.34}{d_H^5(t_i)[l_i/d_H(t_i)]^{1.69}} \,\Theta\left[0.06 - \frac{l_i}{d_H(t_i)} \right] \Theta\left[\frac{l_i}{d_H(t_i)} - \Gamma G \mu \right]  ~ ~ & {\rm (matter~era)},
    \end{cases}
    \label{eq:loop_production_function_bos}
\end{align}
where $d_H(t) = a(t)\int_{0}^t a^{-1}(t')dt'$ is the horizon scale. Using Eq.~\eqref{eq:loop_number_density}, the  number density of loops produced during radiation era reads
\begin{align}
    \boxed{{\tt n}_{r,\rm BOS}(l, t) =  (1842.52) \, \tilde{d}_H^{-4}  \left[\frac{a(t_i(\tilde{d}_H))}{a(t)}\right]^3 t_i'(\tilde{d}_H)}\,~~,
    \label{eq:BOS_loop_number_density}
\end{align}
where $\tilde{d}_H \equiv l_i/0.05$, and the last derivative is with respect to $d_H$. The time of production $t_i$ relates to other variables via Eq.~\eqref{eq:loop_length_evo}.
As discussed earlier, for signals within the LISA window for tensions $G\mu \leq 10^{-10}$ compatible with current PTA observations~\cite{Figueroa:2023zhu}, it is enough to consider only loops produced before the matter-radiation equality $t_{\rm eq}$,  as the loops produced after $t_{\rm eq}$ contribute only negligibly to the GWB spectrum around the mHz frequency window. 

Here a comment is in order. If one neglects the changes of the relativistic {\it dof}\, $g_*,g_{*s}$ during the radiation era, i.e.~if we use $d_H = 2t$ and $a\propto t^{1/2}$, we then obtain the well-known formula ${\tt n}_{r, \rm approx}(l, t) \simeq 0.18 t^{-3/2}(l + \Gamma G \mu t)^{-5/2}$~\cite{Blanco-Pillado:2013qja}. Although this simplified expression is sometimes used in literature, we want to emphasize that this formula misses the important effect on the loop number density's redshifting from the evolution of $g_*,g_{*s}$, and hence leads to a wrong GWB spectral amplitude, see Fig.~\ref{fig:diff_calculation} in Appendix~\ref{app:diff_model_calculation}. Specifically, the suppression of $h^2\Omega_{\rm GW}$ due to the {\it dof} evolution occurs at lower frequencies in the correct spectrum. As discussed in~\cite{Blanco-Pillado:2017oxo}, a way towards including a correct $g_*,g_{*s}$ evolution effect is by realizing that the Friedmann equation during radiation domination is $H^2(T) = H_0^2 \Omega_{r} \mathcal{G}(T) \left({a_0}/{a}\right)^4$, leading to ${a}/{a_0} = \left(2 H_0 \sqrt{\Omega_{\rm rad}^{(0)}}\right)^{1/2} \mathcal{G}^{1/4}(T) t^{1/2}$.
Using eq.~\eqref{eq:loop_number_density}, we obtain
\begin{align}
    {\tt n}_{r,\rm est}(l,t) \simeq \left[\frac{\mathcal{G}(t_i)}{\mathcal{G}(t)}\right]^{\frac{3}{4}} \int_{t_{\rm form}}^{t} {\tt f}(l_i,t_i) \left(\frac{t_i}{t}\right)^{\frac{3}{2}} dt_i \simeq \left[\frac{\mathcal{G}(t_i)}{\mathcal{G}(t)}\right]^{\frac{3}{4}} \left[\frac{0.18}{t^{\frac{3}{2}}(l + \Gamma G \mu t)^{\frac{5}{2}}}\right],
    \label{eq:BOS_loop_number_density_estimate}
\end{align}
where $\mathcal{G}(t)$ is assumed to vary slowly over time, reproducing Eq.~(32) of \cite{Blanco-Pillado:2017oxo}.
By using ${\tt n}_{r,\rm est}$ with the refined GW emission presented in~\cite{Blanco-Pillado:2017oxo}, we can reproduce the GWB spectra presented there. We highlight that, in any case, that our templates use Eq.~\eqref{eq:BOS_loop_number_density}, so that there is no approximation.

The number density of loops produced during the matter era (with the loop production function in the second line of Eq.~\eqref{eq:BOS_loop_number_density}) can be written as, ${\tt n}_{m,\rm BOS}(l,t) \simeq [0.27 - 0.45(l/t)^{0.31}]t^{-2}(l+\Gamma G\mu t)^{-2}$ \cite{Blanco-Pillado:2013qja}, where there is no correction from the DOF evolution in the scale factor during the matter era, which is simply taken as $a \propto t^{2/3}$. Their contribution to the GWB within the LISA window is at least $10^{-4}$ times smaller than that of loops produced during the radiation era for $G\mu \lesssim 10^{-10}$, see Fig.~\ref{fig:BOS_each_contribution} in Appendix\,\ref{app:pieces_CS_GWB}.

The GWB spectra from the BOS model with simulation-inferred $P_j$ and $\Gamma \simeq 51.43$ from the simulations of Ref.~\cite{Blanco-Pillado:2017oxo}, are shown as orange-dashed lines in the left panel of Fig.~\ref{fig:standard_spectra}.

We note that recent Nambu-Goto string simulations suggest that the GW emission can backreact on the loop shape by smoothing their small-scale-structure~\cite{Blanco-Pillado:2019nto,Wachter:2024aos}. Ref.\,\cite{Wachter:2024aos}, in particular, finds that loops typically start out with $\Gamma \gtrsim 50$ and the smoothing effect from backreaction reduces the GW emission power over time. Nonetheless, the emission rate is larger than the usual case for most of loops' lifetime, leading to faster energy loss and earlier GW emission.
As shown in~\cite{Wachter:2024zly}, the GWB spectrum with numerical backreaction reduces the amplitude of $\Omega_{\rm GW}$ by up to $\sim 20\%$, compared to the BOS method used in this work. This corresponds to changes up to $40\%$ in the inferred tension $G\mu$ from the LISA detectability analysis we present in Sect.~\ref{sec:recon_result_conven}. 

\subsection{Uncertainty on the evolution of the SM effective DOF}
\label{sec:modeling_DOFs}

Although the conventional templates from VOS and BOS modelings have their ingredients set by standard physics arguments, there is still some uncertainty in the GWB spectral amplitudes, coming from the uncertainty in the cosmic history, even when assuming only the SM of particle physics as the only particle content during radiation domination. In particular, we recall that there are several predictions for the evolution of $g_*$ and $g_{*s}$, which albeit quite similar among themselves, they still show differences, as depicted in Fig.~\ref{fig:g_gs} in Sect.~\ref{subsec:cosmicHistory}. 

\begin{figure}[t!]
    \centering
    \includegraphics[width=0.65\linewidth]{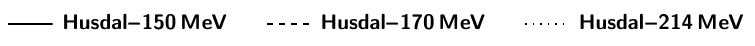}\\
    \includegraphics[width=7.4cm]{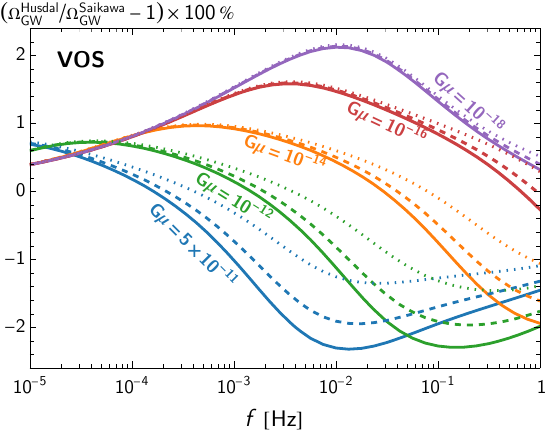}\hfill
    \includegraphics[width=7.4cm]{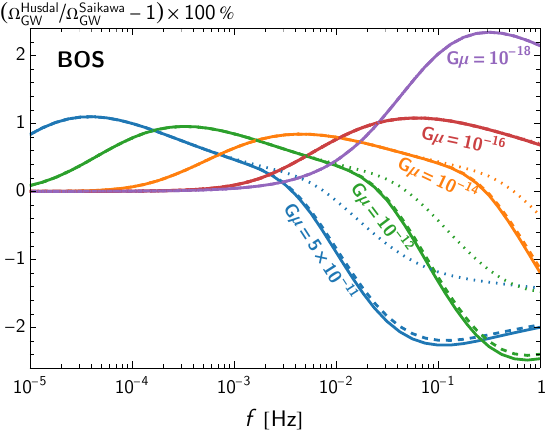}
    \\[-1em]
    \caption{\% relative difference of the GWB spectral amplitude $\Omega_{\rm GW}h^2$ in VOS and BOS models, assuming $g_*$ and $g_{*s}$ evolution from the Saikawa-Shirai and Husdal (of different QCD crossover temperature) results.}
    \label{fig:spectra_diff_g_gs}
\end{figure}

The resulting GWB spectra for each $g_*, g_{*s}$ modeling differ slightly. Fig.~\ref{fig:spectra_diff_g_gs} show the relative difference in amplitude between the GWB spectra from the Saikawa-Shirai  and the Husdal modelings (the latter with different QCD crossover temperatures). The differences $\lesssim 2\%$ for $G\mu < 10^{-10}$ in the LISA window. Therefore, a precise reconstruction of the SM thermal history requires the GWB reconstruction technique in LISA to an accuracy better than $2\%$. If we could distinguish observationally one DOF modeling from another, then we could probe the scale of the QCD quark-gluon transition in a completely independent manner than lattice QCD. In Sect.~\ref{subsec:ModelComparisonVOSvsBOSvsDOF} we address these interesting question and analyse the prospect capabilities of LISA to assess such refined distinctions. We anticipate here that even though LISA can reconstruct the GWB spectrum amplitude to a $\%$ accuracy level for $G\mu > 10^{-15}$ (in the absence of astrophysical foregrounds), a proper model comparison analysis does not allow however to make a clear distinction between one SM-DOF modeling or another, see Sect.~\ref{subsec:ModelComparisonVOSvsBOSvsDOF} and in particular Fig.~\ref{fig:bayes_bos_dofs}. 

\section{Beyond conventional templates}
\label{sec:beyondConventional}

As indicated in Sect.~\ref{sec:GWB_recap}, the computation of the GWB spectrum from a cosmic string network requires the specification of three main ingredients: the loop number density, the expansion rate, and the loop properties (namely birth size and GW emission power). In the previous Sect.~\ref{sec:conventional} we  presented {\it conventional} templates of GWB spectra based on standard physics arguments for such main ingredients. 
Since cosmic strings require BSM physics to begin with, it is perfectly possible that the BSM ingredients support more complex cosmic string 
scenarios than the cases discussed in Sect.~\ref{sec:conventional}. In this section, we introduce what we call {\it beyond-conventional} templates, based on variations of the main ingredients with respect to conventional expectations, typically motivated by BSM physics. These variations are grouped into three different categories, depending on whether we consider variations of {\bf a)} the loop number density,  {\bf b)} the expansion history, or {\bf c)} the loop properties. The different variations lead to different modified GWB's with certain spectral features that are smoking-gun signatures of the BSM scenarios behind the modifications. 

This section focuses first in the introduction of the non-conventional templates and their BSM origin. Later on we will also quantify the goodness of reconstruction of these BSM scenarios and see whether they can be distinguished from the conventional scenarios from Sect.~\ref{sec:conventional}. We postpone however such tasks for Sect.~\ref{sec:recons_beyond}, after the introduction of our reconstruction and model comparison techniques in Sect.~\ref{sec:reconstruction}.  

\subsection{Varying the loop number density}
\label{subsec:LoopNumberDensityBSM}

All templates in this category have the loop number density modified with respect to canonical scaling arguments, while for the other dependencies we assume a $\Lambda$CDM cosmic history, and consider cuspy loops with GW emission power $P_j \propto j^{-q}$ [{\it cf.}~Eq.~\eqref{eq:grav_emission_power_loop_j}] with $q=4/3$. 

\subsubsection{LRS model: smaller loop population}
\label{sec:LRS_model}

So far, the conventional models from the previous two sections considered the GWB from loops with an initial large length size at birth, $l \simeq 0.1 t$, i.e.~smaller but of the order of the horizon size, as inferred from the loop production function from the BOS simulations~\cite{Blanco-Pillado:2013qja}. Alternatively, the simulations by Lorenz-Ringeval-Sakellariadou (LRS)~\cite{Ringeval:2005kr,Lorenz:2010sm} obtain different results for the loop number density, via the direct extraction of the distribution of non-self-intersecting scaling loops (as apposed to extracting first the loop production function). 
In order to obtain the loop distribution Ref.~\cite{Lorenz:2010sm} solves a Boltzmann equation 
using a loop production function theoretically derived in Refs.~\cite{Polchinski:2006ee,Polchinski:2007rg,Dubath:2007mf}. For loop lengths smaller than the GW-emission scale $\Gamma G\mu t$, the loop production function then follows a power law, instead of a peak at the large initial loop's size. This modeling leads to produce smaller loops down to the gravitational backreaction scale $\gamma_c t$, where $\gamma_c \equiv \Upsilon (G\mu)^{1+2\chi}$, $\Upsilon \simeq 20$, and $\chi$ is  a numerical factor defined below.

The resulting loop number density per unit length consists then of \emph{three} loop populations, depending on the loop length scale~\cite{Lorenz:2010sm},
\begin{align}
    \boxed{{\tt n}_{\rm LRS}(l,t) = \begin{cases}
        \frac{C_0(1-\nu)^{1+2\chi}}{t^4(l/t + \Gamma G\mu)^{3-2\chi}} ~ ~ &{\rm for} ~ ~ \Gamma G\mu \ll l/t\\[0.5em]
        \frac{C_0(1-\nu)^{1+2\chi}(3\nu - 2\chi -1)}{(2-2\chi) \, \Gamma G\mu \, (l/t)^{2-2\chi}} ~ ~ &{\rm for} ~ ~ \gamma_c < l/t \ll \Gamma G\mu\\[0.5em]
        \frac{C_0(1-\nu)^{1+2\chi}(3\nu - 2\chi -1)}{(2-2\chi) \, \Gamma G\mu \, \gamma_c^{2-2\chi}} ~ ~ &{\rm for} ~ ~ l/t \ll \gamma_c \ll \Gamma G\mu
    \end{cases}
    }~\,,
    \label{eq:loop_number_density_LRS}
\end{align}
where a constant equation of state of the Universe was assumed, so that $a \propto t^\nu$, with $\nu = 1/2$ and $2/3$ for RD and MD, respectively. By calibrating with the Nambu-Goto simulations from Ref.~\cite{Ringeval:2005kr} on scales $l \gg \Gamma G\mu t$, 
the numerical factors in Eq.~\eqref{eq:loop_number_density_LRS} are
\begin{align}
    C_0 = \begin{cases}
        0.21^{+0.13}_{-0.12} ~ ~ &{\rm (radiation)}\\
        0.09^{+0.03}_{-0.03} ~ ~ &{\rm (matter)}
    \end{cases}, ~ ~ {\rm and } ~ ~ ~~ \chi = \begin{cases}
        0.200^{+0.075}_{-0.105} ~ ~ &{\rm (radiation)}\\
        0.295^{+0.035}_{-0.040} ~ ~ &{\rm (matter)}
    \end{cases}\,.
    \label{eq:LRS_details}
\end{align}
For large loops ($l \gg \Gamma G\mu t$), the loop number density is similar to that of the BOS model, though the BOS loops correspond to $\chi = 0.25$.
\begin{center}
    {\bf The templates in this model depend on: $G\mu$.}
\end{center}
While our template uses the central values of $\{C_0,\chi\}$ in Eq.~\eqref{eq:LRS_details}, other values of $\chi$ are also possible \cite{Auclair:2019zoz}. In particular, Ref.~\cite{Auclair:2020oww} discusses a model that depends on $\chi$'s value during radiation and matter eras, changing the shape of GW spectrum.

\begin{figure}[t!]
    \centering
    {\sffamily Including small loop population: LRS model}\\[0.25em]
    \includegraphics[width=7.3cm]{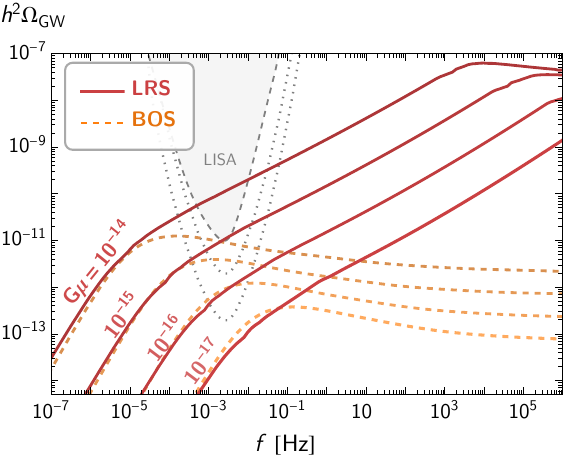}
    \hfill
    \includegraphics[width=7.4cm]{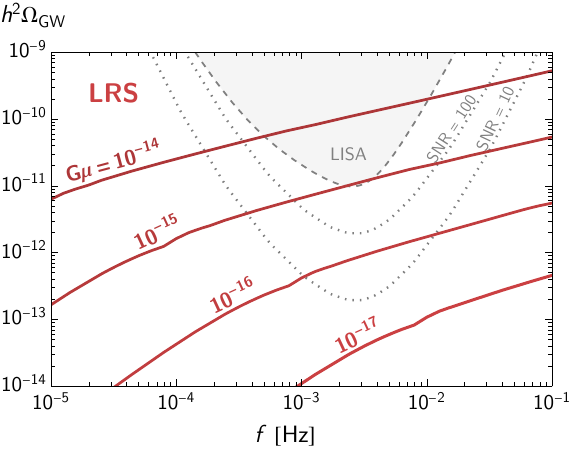}\\[-1em]
    \caption{GWB spectra from the LRS model (discussed in Section~\ref{sec:LRS_model}) of different $G\mu$ are shown in red, in comparison to the GWB spectra from the BOS model in dashed orange (which includes de running of $g_*, g_{*s}$, contrary to the LRS modeling). The LRS spectra have their high-frequency part enhanced by the existence of small-loop populations.}
    \label{fig:LRS_spec}
\end{figure}

The string network in the LRS model contains three populations of loops with different sizes at each time, and hence each population contributes to the GWB at different frequencies. Fig.~\ref{fig:LRS_spec} shows the GWB spectra from the LRS model.
The {low-frequency} part of the spectra resembles the BOS/VOS spectra, as this part corresponds to the GW emission from large loops. In the LRS model, the population of smaller loops contributes at higher frequencies and enhances the GWB amplitude. The strongest current constraint on the LRS cosmic strings comes therefore from the LVK bound,  and it reads $G\mu \lesssim 10^{-14}$~\cite{LIGOScientific:2021nrg}, which is much smaller than the $G\mu \lesssim 10^{-10}$ PTA bound~\cite{NANOGrav:2023hvm,Antoniadis:2023xlr,Figueroa:2023zhu} obtained for the VOS/BOS string network modelings. The validity of this stringent LVK bound rests however on the validity of the LRS modeling itself, which is under debate, see Refs.~\cite{Blanco-Pillado:2019vcs,Blanco-Pillado:2019tbi} and~\cite{Auclair:2021jud} for discussion. Ref.~\cite{Blanco-Pillado:2019tbi} in particular, finds consistency with the BOS modeling, and not LRS, even when the loop number density is extracted directly in their simulations.

We note that since the loop number density~\eqref{eq:loop_number_density_LRS} is derived by assuming a constant equation of the state, this prevents us from properly including the $g_*, g_{*s}$ evolution effect in the dilution of the loop number density, unlike in the VOS and BOS models.

\subsubsection{Small intercommutation probability (e.g.~super-strings)} 
\label{subsubsec:super_string}

Cosmic strings in general interact among themselves so that when they cross with each other, this typically leads to exchange their segments. This is referred as the \emph{intercommutation} of strings.
One can quantify the \emph{reconnection} probability $p$ to exchange partners when strings cross. The templates from conventional cosmic strings in Sect.\,\ref{sec:conventional} assumed $p=1$, because if NG strings are simply understood as the infinitely thin limit of field theory local strings, the latter reconnect with probability of order unity when they cross, due to the microphysical interaction properties of field theory~\cite{Vilenkin:2000jqa}. Nonetheless, in some setups, strings can have $p<1$ and can occasionally pass through each other without intercommuting. 
As a well-known example, cosmic super-strings in String Theory~\cite{Jackson:2004zg} can avoid their crossings in our (3+1)-dimensional Universe, by moving in the extra spatial dimension(s). Color flux tubes in pure ${\rm SU}(N)$ Yang-Mills theory~\cite{Yamada:2022aax, Yamada:2022imq} are also known to have a suppressed reconnection probability, which scales (in the large-$N$ limit) as $p \sim 1/N^{2}$ and $\exp(-N)$ for F- and D-strings, respectively.
For this work, we only consider cosmic strings with a single value of $p$, where in fact different types of superstrings can coexist in the same theory, leading to junction formation and adding more structures to the GW spectrum \cite{Copeland:2006if,Binetruy:2010bq,Avgoustidis:2011ax,Sousa:2016ggw,Avgoustidis:2025svu}. We acknowledge that our choice should not be interpreted as a realistic modeling of superstring networks, as more sophisticated and physically motivated treatments such as in Ref.~\cite{Avgoustidis:2025svu}, are needed for robust conclusions about superstrings in the LISA window.

With a smaller $p$, long strings possess more small-scale structures or wiggles, leading to an effective reconnection probability that controls the loop production of the network \cite{Avgoustidis:2004zt,Avgoustidis:2005nv},
\begin{align}
    p_{\rm eff} \simeq 1 - (1- p)^{10} \xrightarrow[]{p\ll 1} 10 p,
    \label{eq:effective_inter_com_prob}
\end{align}
where the factor 10 estimates the number of small-scale intersections per long-string collision.
Motivated by string theories~\cite{Jackson:2004zg}, one can have $p \sim \mathcal{O}(10^{-3})$, leading to $p_{\rm eff} \simeq \mathcal{O}(10^{-2})$.
Consider the VOS model in section~\ref{sec:GW_semi_analytic}, with ${\tt n}_{\rm loop} \propto C_{\rm eff}/(\alpha t_i^4) \propto \tilde{c}/L_i^4$, where $L_i$ is the long-string correlation length.
The less-frequent long-string collisions lead to less efficient loop production $\tilde{c} \to p_{\rm eff}\tilde{c}$, but the long strings becomes denser i.e., the correlation length scales as $L \propto \sqrt{p_{\rm eff}}$ \cite{Yamada:2022aax, Yamada:2022imq}.
Overall, \emph{the loop number density is boosted,} ${\tt n}_{\rm loop} \propto p_{\rm eff}^{-1}$. 
Similar studies on effects of $p<1$ on the string network can be found in  \cite{Sakellariadou:2004wq,Avgoustidis:2004zt,Avgoustidis:2005nv}.

We will assume that small values of $p$ affect only the loop number density and does not change the GW emission\footnote{This assumption is purely based on simplicity. We note that the initial loop size might be however affected~\cite{Sousa:2016ggw}. Moreover, a specific model of string compactification could also lead to the time-dependent string tension \cite{Revello:2024gwa,Ghoshal:2025tlk}.}.
The GWB from cosmic strings with $p < 1$ can be approximated as 
\begin{align}
    \boxed{\Omega^{p<1}_{\rm GW}(f,p_{\rm eff}) = p_{\rm eff}^{-1} \Omega_{\rm GW}^{p=1}(f)}~\,.
    \label{eq:GWB_superstring}
\end{align}
\begin{center}
    {\bf The templates in this model depend on: $G\mu$, $p_{\rm eff}$.}
\end{center}
The GWB spectrum in this case gets boosted for smaller $p_{\rm eff}$, while its frequency profile is the same independently of $p_{\rm eff}$. We note that because the energy density of long strings in a scaling network goes as $\rho_{\rm network} \propto L^{-2} \propto p_{\rm eff}^{-1} G\mu \rho_{\rm tot}$, it must hold that $p_{\rm eff} \gg  \sqrt{G\mu}$. Considering~\cite{Jackson:2004zg}, we adopt a range $10^{-3} \leq p < 1$ as a prior.

\subsubsection{Metastable cosmic strings (e.g.~string-monopole networks)}
\label{subsubsec:metastable_strings}

\begin{figure}[t!]
    \centering
    {\sffamily Metastable cosmic strings}\\
    \includegraphics[width=0.495\textwidth]{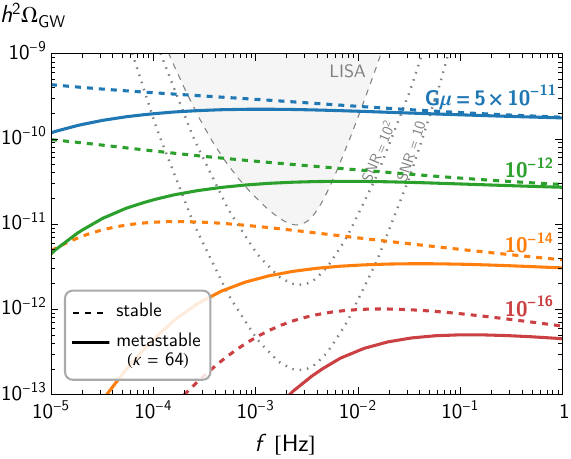}\hfill
    \includegraphics[width=0.495\textwidth]{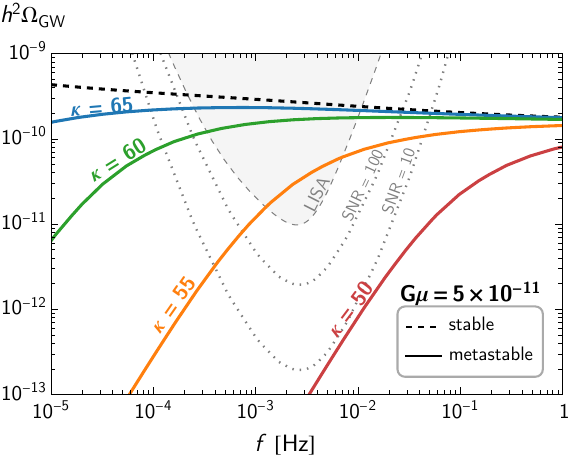}\\[-1em]
    \caption{GWB from the metastable cosmic-string network---decaying via monopole-pair creation---for fixed $\kappa = 64$ (left) and fixed $G\mu = 5\times10^{-11}$ (right). In both plots, the dashed curves show the case of stable cosmic strings or the metastable string network with $\kappa \gg 70$ (i.e., GWB spectrum looks like stable strings in LISA's window).
    }
    \label{fig:metastable_spectrum}
\end{figure}

In the conventional scenarios discussed in Sect.\,\ref{sec:conventional}, the string network has been existing from the formation at energy scale $\sim \eta$, producing loops and GWs all along cosmic history till today. However, there are well-motivated models (e.g.~in Grand Unified Theories) where the symmetry breaking pattern of the theory happens in multiple steps \cite{Lazarides:1981fv,Vilenkin:1982hm,Kibble:1982ae,Vilenkin:1984ib,Chitose:2025cmt}, rather than in just one breaking that leads to generate cosmic strings. At each step, other type of cosmic defects (e.g.~monopole or domain walls) can be formed and attached to cosmic strings. Such string-domain and string-monopole networks are unstable and can decay, stopping the creation of loops and hence the production of GWs at late times. As a result, the spectral amplitude of the GWB is suppressed at smaller frequencies.

Here we focus on an example of string-monopole network, see Refs.~\cite{Buchmuller:2019gfy,Buchmuller:2020lbh,Buchmuller:2021mbb,Buchmuller:2023aus,Chitose:2023dam}, where symmetry breaking happens as follows. A first symmetry breaking event generates monopoles at an energy scale $m_M$, but then an inflationary process takes place diluting away their density, preventing the monopole problem~\cite{Guth:1980zm}. Cosmic strings are then produced at a lower energy scale $\eta$, below which the process of producing loops (and hence GWs) commence. At much later times, however, monopole-antimonopole pairs start nucleating on cosmic strings, cutting strings into smaller segments\footnote{We do not include GW contributions from string segments \cite{Leblond:2009fq,Martin:1996cp,Buchmuller:2021mbb}, as its uncertainty is recently under debate \cite{Servant:2023tua,Chitose:2025qyt}.}, and shutting off loops and GW production.
The nucleation rate (per unit length) is given by
\begin{align}
    \Gamma_d = \frac{\mu}{2\pi}e^{-\pi \kappa} ~ ~ ,~{\rm with} ~ ~ \kappa \equiv \left(\frac{m_M}{\eta}\right)^2.
    \label{eq:metastable_kappa_def}
\end{align}
The breaking of strings of length $l$ mostly happens when $\Gamma_d l(t_{\rm brk}) \sim H(t_{\rm brk})$ or at the time $t_{\rm brk} \sim \Gamma_d^{-1/2}$ [using $l(t_{\rm brk}) \sim t_{\rm brk} \sim H^{-1}(t_{\rm brk})$]. The loop number density in this scenario can be written as~\cite{Buchmuller:2019gfy,Buchmuller:2020lbh,Buchmuller:2021mbb,Buchmuller:2023aus}
\begin{align}
    \boxed{{\tt n}_{\rm meta}(l,t) = {\tt n}_{\rm stable}(l,t) \, \Theta(t_{\rm brk} - t_i) \,  \mathcal{E}(l,t),
    }
    \label{eq:loop_number_density_metastable}
\end{align}
where the ${\tt n}_{\rm stable}(l,t)$ is the loop number density of the stable network (we use the VOS model for simplicity, but the BOS model is also equally applicable). Two suppressing factors in Eq.\,\eqref{eq:loop_number_density_metastable} are: \emph{i)} the step-function $\Theta(t_{\rm brk} - t_i)$, which models the termination of loop production, and \emph{ii)} the function $\mathcal{E}(l,t)$ which characterizes loops that get segmented by monopole pairs, and is given by $\mathcal{E}(l,t) = e^{-\Gamma_d [l(t)(t-t_{\rm brk})+\frac{1}{2}\Gamma G\mu(t-t_{\rm brk})^2]}$ ~\cite{Leblond:2009fq,Buchmuller:2021mbb}.

\begin{center}
    {\bf The templates in this model depend on: $G\mu$, $\kappa$.}
\end{center}
Using the loop number density \eqref{eq:loop_number_density_metastable} in the master formula \eqref{eq:master_formula_GWB_strings}, Fig.\,\ref{fig:metastable_spectrum} shows the GWB spectra of GWB from our metastable cosmic-string templates. With no GW emission at late times, the GWB spectrum exhibits a low-frequency (or \emph{infrared}) cutoff, which is found from a numerical fitting as~\cite{Servant:2023tua}
\begin{equation}
    f_{\rm meta} \simeq 64.7 ~ {\rm \mu Hz} \left({10^{-11}}/{G\mu}\right)^\frac{1}{2} e^{-\frac{\pi}{4}(\kappa-64)}.
    \label{eq:frequency_metastable_string}
\end{equation}
{This frequency corresponds to loops emitting GWs at the time $t_{\rm sup} \sim \sqrt{2/(\Gamma_d \Gamma G\mu)}$, when the loop number density is suppressed substantially by the function $\mathcal{E}(l,t)$, see~\cite{Leblond:2009fq,Buchmuller:2021mbb,Servant:2023tua}. Importantly, the segmentation of strings does not affect the GW spectrum, and the spectrum resembles that of the stable string network (e.g.~a conventional template) and has no metastable-string feature, if $t_{\rm sup} > t_0$, which holds for low enough tensions as
\begin{equation}
    G\mu \lesssim e^{\pi\kappa/2}/\sqrt{2\Gamma t_0^2 m_{\rm Pl}^2} \simeq 2.84 \times 10^{-18}\, e^{\pi(\kappa-64)/2}.
    \label{eq:metastable_stable_network}
\end{equation}
}

\subsubsection{Current-carrying cosmic strings}
\label{sec:current_carrying_string}

There are cosmic string models which can accommodate a current---of massless particles---propagating along the core of the strings~\cite{Witten:1984eb}. Here we focus on neutral current-carrying strings (e.g.~$B-L$ strings in $ SO(10)$ GUT theory~\cite{Davis:1996sp,Chavez:2002sm}) as considered in~\cite{Auclair:2022ylu}, as the charged carriers do not feel any long-range interaction, and cosmic strings only lose their energy through GW emission. We calculate the GWB of such current-carrying strings using a generalized of VOS equations that include a current~\cite{Martins:2020jbq}, see App.\,\ref{app:generalized_VOS_current}. The intensity of the current is indicated by the dimensionless strength defined as $Y \equiv (Q^2+J^2)/2$, with $Q^2$ and $J^2$ the total charge and current energy density in units of the string tension. We will assume for simplicity that the current only alter the dynamics of long strings, though it has been noted that the current can also change loops' dynamics, e.g., suppressing the GW emission power~\cite{Rybak:2022sbo,Rybak:2024our} and producing vortons~\cite{Auclair:2020wse}, see also \cite{Sousa:2024ytl}. We consider our templates in this section simply as a proof of concept for this shape of GWB spectrum.\footnote{A similar peak GWB shape can be generated by the effect of an intermediate kination era which arises in the rotating-axion model \cite{Gouttenoire:2021wzu,Gouttenoire:2021jhk,Co:2021lkc}. However, such a kination era could also induce the secondary peak at higher frequency \cite{Gouttenoire:2021jhk,Co:2021lkc}.}
This current-carrying string template is a simple model, where the effect of current is incorporated only in the dynamics of the string network and not in the GW emission. In many studies of \emph{superconducting} strings \cite{Rybak:2022sbo,Rybak:2024our}, which can arise in realistic UV completions, the current's effect can further suppress the GW emission, leading to different GWB predictions.

We consider an agnostic mechanism for the current generation and assume that the current is switched on at temperature $T_{\rm ini}$ with $Y_{\rm ini} = Y(T_{\rm ini})$ and quenched at $T_{\rm off}$, {\it i.e.}~the current operates for a finite duration characterized by $r = T_{\rm ini}/T_{\rm off}$. Before the current is switched on, the string network reaches its standard scaling regime. Once an initial current is established, if $Y_{\rm ini} > 0.46$ (for $\tilde{c} = 0.23$)~\cite{Auclair:2022ylu} then the system evolves to a new scaling regime with the current stabilizing at $Y(T) = 1$. For $Y_{\rm ini}$ smaller than this threshold, there is no attractor behavior of string network, which makes the result highly initial-condition dependent~\cite{Auclair:2022ylu}, so we don't consider such cases. The string network in the new scaling regime (when $Y_{\rm ini} > 0.46$) has $\xi(t) = L(t)/t \propto t^{\delta_\xi}$ and $\bar{v} \propto t^{\delta_{\bar{v}}}$ where $\delta_\xi \simeq \delta_{\bar{v}} \simeq -0.15$ for $\tilde{c} = 0.23$ (for other $\tilde{c}$ values, see~\cite{Auclair:2022ylu}). After the current is quenched at $T_{\rm off}$, the string network evolves back to the standard scaling regime.

\begin{figure}[t!]
    \centering
    {\sffamily Current-carrying local cosmic strings}\\
    \includegraphics[width=0.495\textwidth]{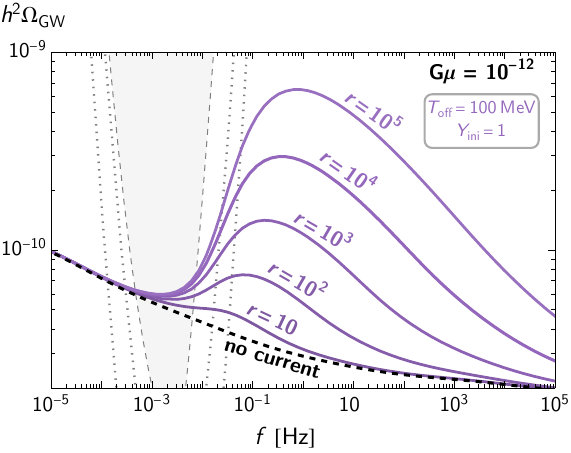}\hfill
    \includegraphics[width=0.495\textwidth]{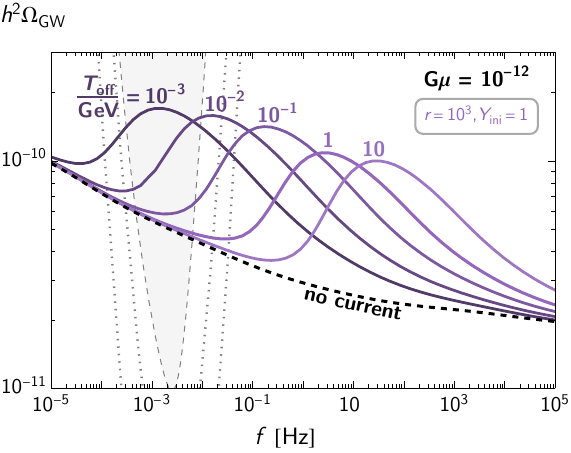}\\
    \includegraphics[width=0.495\textwidth]{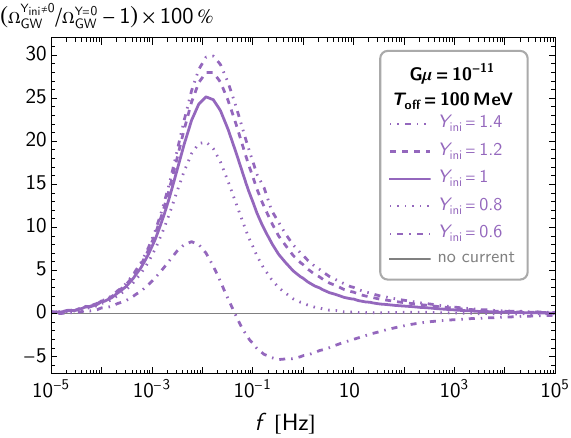}\\[-1em]
    \caption{
    GWB spectra from current-carrying cosmic strings depend on $G\mu$, $Y_{\rm ini}$, $T_{\rm off}$, and $r$.
    We fix $Y_{\rm ini} = 1$ in the top row and keep $G\mu = 10^{-11}$ and $T_{\rm off} = 100\,{\rm MeV}$ in the bottom row. The top-left panel shows the effect of duration $r$ when the current is active, and the top-right panel varies $T_{\rm off}$. The bottom row shows the relative error between the GW spectrum without current and those with different value of $Y_{\rm ini}$.}
    \label{fig:current_carrying}
\end{figure}

\begin{center}
    {\bf The templates in this model depend on: $G\mu$, $Y_{\rm ini}$, $T_{\rm off}$, $r$}.
\end{center}
We use the loop number density of the VOS model in Eq.\,\eqref{eq:vos_loop_number_density} which inputs into the master equation~\eqref{eq:master_formula_GWB_strings}. As generating templates varying the above four parameters is computationally expensive, we have produced only a subset of these, with either: \emph{i)} fixed $Y_{\rm ini} = 1$, or \emph{ii)} fixed $G\mu = 10^{-11}$ to illustrate the effect of $Y_{\rm ini}$. We show examples of these spectra in Fig.\,\ref{fig:current_carrying}, where the peak signature appears distinctively.
Although the variation of $Y_{\rm ini}$ changes the spectrum up to 30\%, this feature is less prominent than the variation of other parameters. In the signal reconstruction part of this work, we consider only the case where $\{G\mu,r,T_{\rm off}\}$ vary and the initial current is fixed to the attractor solution $Y_{\rm ini} = 1$. 
The peak position corresponds to the loop population formed around the time when the current is quenched. 
Using the frequency-temperature relation (see e.g.~Eq.\,(28) in~\cite{Gouttenoire:2019kij}), one can write the peak position as 
\begin{align}
    f_{\rm peak} \simeq (67  ~ {\rm mHz}) \,  r^{-\delta_\xi} \left(\frac{T_{\rm off}}{\rm GeV}\right) \left(\frac{50 \times 0.1 \times 10^{-11}}{\Gamma \alpha_L \xi_{\rm RD} G\mu}\right)^\frac{1}{2} \left(\frac{g_*(T_{\rm off})}{g_*(T_0)}\right)^{\frac{1}{4}}\,.
    \label{eq:peak_freq_current_string}
\end{align}

\subsection{Exploring cosmic histories}
\label{sec:GWB_cosmic_history}
Since the cosmic-string network produces GWs continuously from the time of its formation, the resulting GWB spectrum is highly sensitive to the cosmic history.
Both in Sect.\,\ref{sec:conventional} and Sect.\,\ref{subsec:LoopNumberDensityBSM}, we assumed that the Universe evolves according to $\Lambda$CDM, as described in Sect.\,\ref{subsec:cosmicHistory}.
However, the history of our Universe is constrained by observations only up to Big-Bang Nucleosynthesis (BBN) around the energy scale of  $\sim 1 ~ {\rm MeV}$. Deviations from the $\Lambda$CDM model may arise in many BSM scenarios at  higher energy scales, during the expansion history prior to BBN. Such deviations lead in general to spectral distortions of the GWB with distintive signatures, see e.g.~\cite{Allahverdi:2020bys,Simakachorn:2022yjy} for reviews.
In this work, we consider representative examples, like a background expansion rate that differs from RD during the reheating stage after inflation (Sect.~\ref{subsec:nonst_cosmo_after_inflation}), or the presence of extra relativistic degrees of freedom (DOF) in the particle spectrum, i.e.~BSM DOF, which can either thermalize with the SM species (Sect.~\ref{subsec:nonst_cosmo_thermalized_DOFs}), or be completely secluded (Sect.~\ref{subsec:nonst_cosmo_dark_DOFs}).

\subsubsection{Non-standard post-inflationary era}
\label{subsec:nonst_cosmo_after_inflation}

In many inflationary models, the field responsible for inflation -- the inflaton -- ends up oscillating around the minimum of its potential, during the reheating stage following after the inflationary period. For a homogeneous inflaton oscillating in a potential $V(\phi) \propto |\phi|^p$, its energy density averaged over oscillations behaves as $\rho_\phi \propto a^{-3(1+w)}$, with $w = (p-2)/(p+2)$ the effective equation of state (EoS) of the inflaton~\cite{Turner:1983he}. The expansion rate of the Universe during the oscillatory regime can be equivalent, in the simplest cases of $p = 2$ or $p = 4$, to matter- or radiation-domination, with $w=0$ or $w = 1/3$, respectively. Interestingly, for $p > 4$, the expansion rate corresponds to kination-domination, characterized by a {\it stiff} EoS $w > 1/3$ (though verifying $w \leq 1$). The durability of any oscillatory regime, and hence of a given period with EoS $w \neq 1/3$, depends on the inflaton's potential and its interactions with other species, see e.g.~Ref's~\cite{Lozanov:2016hid,Antusch:2020iyq,Antusch:2021aiw,Antusch:2022mqv,Figueroa:2024yja}. We also note that non-oscillatory reheating models~\cite{Figueroa:2016dsc,Opferkuch:2019zbd,Dimopoulos:2018wfg,Bettoni:2021zhq,Laverda:2023uqv,Figueroa:2024asq}, or rotating-axion scenarios~\cite{Co:2019wyp,Co:2020jtv,Co:2021lkc,Gouttenoire:2021wzu,Gouttenoire:2021jhk,Duval:2024jsg} can also lead to a phase of kination-domination after inflation. From the point of view of observability of the cosmic strings' GWB, a period of kination domination with $w > 1/3$ is actually the most interesting situation.  

We consider two modelings of a nonstandard era, based on: \emph{i)} an instantaneous transition to RD, as often used in literature, or more realistically, \emph{ii)} a smooth transition towards RD.

\begin{figure}[t!]
    \centering
    {\sffamily Nonstandard era after inflation}\\
    \includegraphics[width=0.495\textwidth]{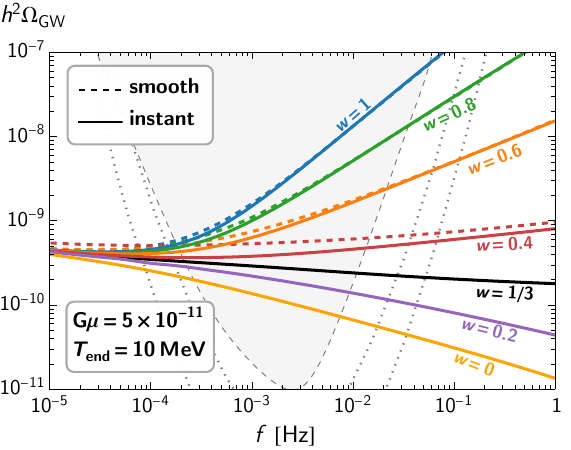}
    \hfill
    \includegraphics[width=0.495\textwidth]{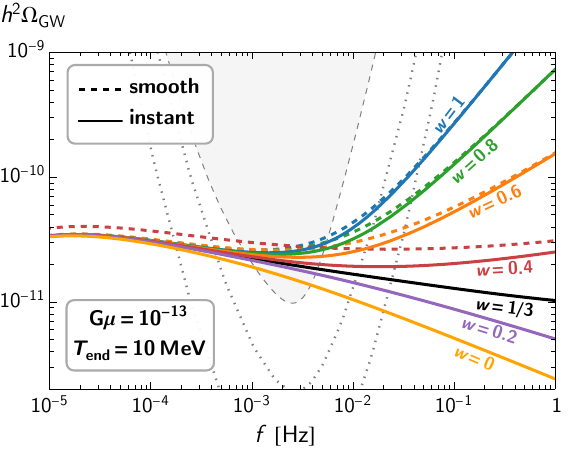}\\[-1em]
    \caption{GWB spectra from cosmic strings of $G\mu = 5 \times 10^{-11}$ and $10^{-13}$, assuming the non-standard era with equation of state $w$ at temperature $T > T_{\rm end} = 10\,{\rm MeV}$. The solid and dashed lines correspond to the instantaneous~\eqref{eq:non-standard_instant} and smooth~\eqref{eq:non-standard_smooth} transitions, respectively. For the smooth transition case, $w\geq 0.4$ to ensure that the $w$ is stiff enough not to dominate the Universe after BBN.}
    \label{fig:non-standard-era}
\end{figure}

{\underline{\it Instantaneous transition}}. This assumes that a nonstandard era transits into RD in a given moment, at some temperature $T_{\rm end}$. The total energy density hence changes its scaling `instantaneously' from $a^{-3(1+w)}$ to $a^{-4}$ in that moment, as reflected in the following expression
\begin{align}
    \boxed{\rho_{\rm tot}(a) = \begin{cases}
        \rho_{\rm \Lambda CDM}(a(T_{\rm end}))\left[\frac{a(T_{\rm end})}{a}\right]^{3(w+1)} &~ ~ {\rm for} ~ a < a(T_{\rm end}),\\
        \rho_{\rm \Lambda CDM}(a) &~ ~ {\rm for} ~ a\geq a(T_{\rm end}),
    \end{cases}
    }
    \label{eq:non-standard_instant}
\end{align}
where the energy density of the $\Lambda$CDM components is $\rho_{\rm \Lambda CDM} = 3 m_{\rm Pl}^2 H^2$ with $H$ given by Eq.\,\eqref{eq:friedmann_LambdaCDM}.
{Note that the GWB-signal reconstruction study for this setup has been considered also in~\cite{Blanco-Pillado:2024aca}.}

{\underline{\it Smooth transition}}. In more realistic reheating scenarios, see for example~\cite{Lozanov:2016hid,Antusch:2020iyq,Antusch:2021aiw,Antusch:2022mqv,Figueroa:2024yja} or ~\cite{Figueroa:2016dsc,Opferkuch:2019zbd,Dimopoulos:2018wfg,Bettoni:2021zhq,Laverda:2023uqv,Figueroa:2024asq}, the energy budget of the universe transits gradually from being dominated by the inflaton to being dominated by an ensemble of relativistic particles. The end of reheating depends on the energy transfer mechanism from the inflaton into relativistic species. 
We can parametrically define when a nonstandard era ends when the energy densities of the inflaton and the relativistic ensemble become equal (with the inflaton's energy becoming subdominant subsequently). For simplicity one assumes that the relativistic species form a thermal bath, as it is often the case that the interaction rate among the relativistic species is higher than the expansion rate. The total energy density can then be written as
\begin{align}
    \boxed{\rho_{\rm tot}(a) =  \rho_{\rm \Lambda CDM}(a) + \rho_{\rm NS}(a(T_{\rm end}))\left[\frac{a(T_{\rm end})}{a}\right]^{3(w+1)},}
    \label{eq:non-standard_smooth}
\end{align}
where $\rho_{\rm NS}(a(T_{\rm end})) = \rho_{\rm \Lambda CDM}(a(T_{\rm end})) = \frac{\pi^2}{30}g_*(T_{\rm end})T_{\rm end}^4$. The last equality is justified as the nonstandard era is assumed to end in RD above the BBN scale $\sim 1$ MeV. This assumption means that the non-standard energy density co-exists with the $\Lambda$CDM components. 

\begin{center}
    {\bf The templates in this model depend on: $G\mu$, $w$, and $T_{\rm end}$.}
\end{center}

We calculate the GWB templates for both instant and smooth transitions, using the VOS model and the GW emission from a loop given by Eq.\,\eqref{eq:grav_emission_power_loop_j}, with $q=4/3$. The reason for applying the VOS modeling is that other models of loop number density in Sect.~\ref{sec:conventional} only calibrate their results in radiation or matter eras, and do not work for an arbitrary equation of state of the Universe. Fig.\,\ref{fig:non-standard-era} shows the cosmic-string GWB spectra from our templates.

\begin{figure}[p!]
\centering
{\sffamily Thermalized extra relativistic DOF}\\[0.25em]
\includegraphics[width=0.495\textwidth]{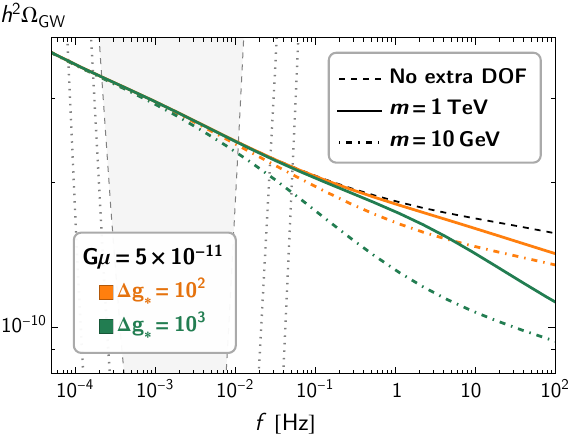}\hfill
\includegraphics[width=0.245\textwidth]{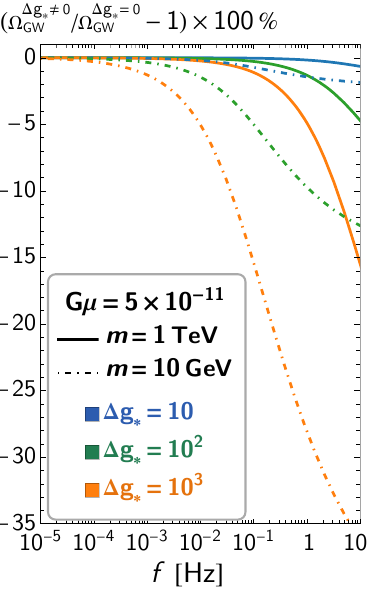}
\includegraphics[width=0.245\textwidth]{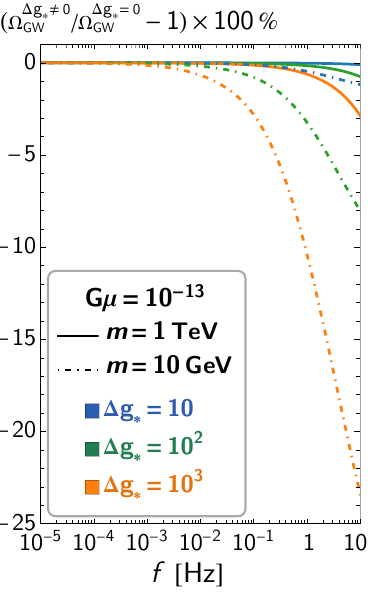}\\[-1em]
\caption{\emph{Left Panel:} Spectra of GWB from cosmic strings, assuming the presence of the extra thermalized DOFs with mass $m$ and the effective number of DOFs $\Delta g_*$. Our calculation includes the cosmic history with the realistic DOF evolution \eqref{eq:realistic_DOF_evo}. Although the extra DOFs of mass $m<{\rm TeV}$ lead to observable spectral-suppression feature in LISA, they can interact with SM particles and would be restricted by collider constraints $m\gtrsim 1\,{\rm TeV}$. \emph{Right Panels:} For $m\gtrsim 1\,{\rm TeV}$, the extra thermalized-DOF effect leads to the the relative difference in GWB amplitudes, compared to the standard prediction, by $\lesssim 1\%$ for $G\mu \lesssim 5 \times 10^{-11}$ and $\Delta g_* \lesssim 10^3$ within LISA window ($f \leq 0.1\,{\rm Hz}$). These plots are based on the work in progress \cite{servantSpectro}.}
\label{fig:CS_dof}
\vspace*{1cm}
{\sffamily Completely secluded sector}\\[0.25em]
\includegraphics[width=0.5\textwidth]{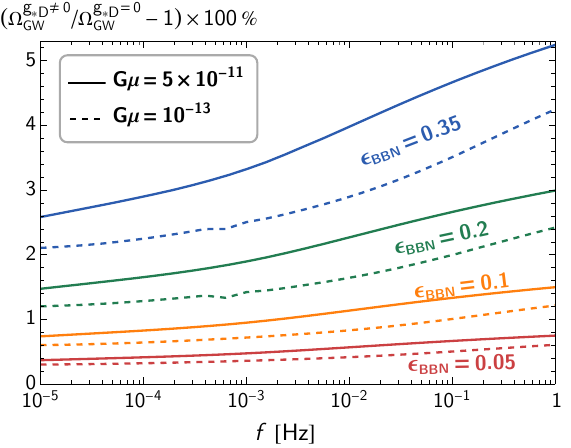}\\[-1em]
\caption{Each colored line shows the relative difference between GWB ampitudes of the standard prediction and the one with extra secluded DOFs.
The completely secluded DOFs with its abundance is quantified by $\epsilon_{\rm BBN}$ enhances the GWB spectrum slightly. As constrained by $\Delta N_{\rm eff}$ bound, $\epsilon_{\rm BBN} \leq 0.35$, leading to enhancement $\lesssim 5\%$ within LISA window.  This plot is based on the work in progress \cite{servantSpectro}.}
 \label{fig:CS_dark_dof}
\end{figure}

\subsubsection{Thermalized extra relativistic DOF}
\label{subsec:nonst_cosmo_thermalized_DOFs}

The cosmic history can also change even when the energy density of the background behaves as radiation. A well-motivated case is when there are massive particles which thermalize efficiently with the SM particles. When they become non-relativistic, i.e.~when the temperature drops below their mass, $T < m$, their abundance becomes Boltzmann suppressed, and they decay into lighter SM species. Because they are efficiently coupled to the SM, their masses must be $m \gg \mathcal{O}({\rm TeV})$, so that collider constraints~\cite{ATLAS:2024fdw,CMS:2024zqs} are avoided.

Being agnostic about these BSM particles, their effective number of relativistic DOF in energy and entropy densities,  $\delta g_*, \delta g_{*s}$, can be 
calculated by integrating over their phase space distribution~\cite{Kolb:1990vq}. Assuming all particles have the same mass $m$\footnote{As discussed in \cite{servantSpectro}, fermions and bosons give $\lesssim 1\%$ relative difference in $\Omega_{\rm GW}$ for cosmic-string GWB due to the slight difference in $g_*$ and $g_{*s}$ evolution around $T\sim m$. Our analysis considers for simplicity that all particles are bosons.}, the extra species modify the effective number of DOFs in Eq.\,\eqref{eq:bigG_expression} by
\begin{align}
    \boxed{g_*(T) = g_*^{\rm SM}(T) + \delta g_*(T,m) ~ ~ ; ~ ~ g_{*s}(T) = g_{*s}^{\rm SM}(T) + \delta g_{*s}(T,m)}\,,
    \label{eq:realistic_DOF_evo}
\end{align}
where $ g_*^{\rm SM}$ and $g_{*s}^{\rm SM}$ are the effective numbers for the SM particles (for which we assume Saikawa-Shirai results \cite{Saikawa:2018rcs}). We obtain $\delta g_*(T,m), \delta g_{*,s}(T,m)$ by performing the aforementioned phase-space integrals~\cite{Kolb:1990vq}, with $\delta g_*(T \gg m) = \delta g_{*s}(T \gg m) = \Delta g_*$ = {\it constant}. In summary we parametrize the extra DOFs by their mass $m$ and their effective number $\Delta g_*$ in the high-temperature limit.

\begin{center}
    {\bf The templates in this model depend on: $G\mu$, $m$, and $\Delta g_*$.}
\end{center}

We produce our templates by using the VOS model, the GW emission from a loop as in Eq.\,\eqref{eq:grav_emission_power_loop_j} with $q=4/3$, and the cosmic history as in Eq.\,\eqref{eq:friedmann_LambdaCDM} with the above $g_*(T), g_{*s}(T)$. The GWB spectra have a distinct step-suppression feature, as shown in Fig.~\ref{fig:CS_dof}. The reason for this is that the additional DOFs make the Universe expand faster for a longer time\footnote{Consider the times when $T_i\gg m$ and $T_f\ll m$. Using entropy conservation, we have $a_f/a_i = T_i/T_f \times [(g_{*s}^{\rm SM}+\Delta g_*)/g_{*s}^{\rm SM}]^{1/3}$. Therefore, $\Delta g_*$ prolongs the cosmic expansion.}. This dilutes the number density of loops and the GW energy density that existed prior to the decay of the extra DOFs. The step suppression takes place at frequencies $f > f_{\rm tp}$, above a turning point frequency given by~\cite{Gouttenoire:2019kij,servantSpectro},
\begin{align}
    f_{\rm tp} \simeq 0.2 ~ {\rm Hz} \left(\frac{m}{\rm TeV}\right)\left(\frac{10^{-11}}{G\mu}\right)^{\frac{1}{2}}\left[\frac{g_*(T_{\rm dec})}{g_*(T_0)}\right]^{\frac{1}{4}}\,,
    \label{eq:thermalized_dof_frequency}
\end{align}
with $T_{\rm dec}$ the temperature where $\delta g_*(T_{\rm dec}) \to 0$. Numerical results show  that $\delta g_*(T_{\rm dec})\lesssim 0.01$ for $T_{\rm dec} \simeq 0.1m$, when considering  $\Delta g_*$ up to $10^4$~\cite{servantSpectro}.
A particle of mass $m \gtrsim 10$ TeV---which can thermalize with the SM thermal bath and evades collider constraints---imprints a signature on the GWB at a frequency $\gtrsim 1$ Hz for $G\mu < 10^{-11}$. If one considers the metastable strings that can evade the PTA window, $G\mu$ can be as high as $10^{-9}$, in order not to be excluded by the LVK constraints. Still, the signature of particles heavier than 10 TeV resides outside the LISA window~\cite{servantSpectro}.

\subsubsection{Completely secluded sector}
\label{subsec:nonst_cosmo_dark_DOFs}

The extra DOFs can also decouple from the SM thermal bath while being relativistic. If they have very small mass or are massless, they contribute to the dark radiation of the Universe, and leave the same effect as if there is a completely-decoupled sector.

A completely secluded (relativistic) sector which has its own thermal equilibrium at a temperature $T_D$ (which can be different from the temperature $T$ of the SM sector) contributes to the total energy density of the Universe and change the Friedmann equation as follows 
\begin{align}
    \boxed{H^2 (t) = H_0^2 \text{\huge$[$}\underbrace{\Omega_{\rm rad}^{(0)} \mathcal{G}[T(t),T_0]\left(\frac{a_0}{a(t)}\right)^4 + \Omega_{\rm mat}^{(0)} \left(\frac{a_0}{a(t)}\right)^3 + \Omega_{\rm de}^{(0)}}_{\rm \Lambda CDM} + ~ \Omega_{D}^{(0)}\mathcal{G}_D[T(t),T_0]\left(\frac{a_0}{a(t)}\right)^4\text{\huge$]$}}\,,
    \label{eq:friedmann_LambdaCDM_dark}
\end{align}
where $\Omega_D^{(0)} = \rho_D(T_0)/\rho_{\rm tot,0}$ with 
$\rho_D = \pi^2g_*^D(T_D)T_D^4/30$ being the energy density of the dark sector. The function  $\mathcal{G}_D(T,T_0)$ parametrizes the thermal history of the completely secluded sector and is defined analogously as to Eq.~\eqref{eq:bigG_expression}, but with the effective number of species corresponding to the decoupled sector, $g_{*}^{D}, g_{*s}^{D}$.

The extra-radiation component raises the total energy density of the Universe, compared to the $\Lambda$CDM model. 
The maximally allowed energy density is bounded by the dark-radiation constraint which is typically written in terms of an effective number of extra neutrinos $\Delta N_{\rm eff}$, defined by $\rho_D = 7\pi^2\Delta N_{\rm eff} T_\nu^4/120$. At the BBN scale $T \simeq {\rm MeV}$, $\Delta N_{\rm eff} \lesssim 0.2$ \cite{Mangano:2011ar,Peimbert:2016bdg,Planck:2018vyg}, leading to $g_*^D(T_D/T)^4\lesssim 0.35$ just at the onset of BBN. The energy density of the string network in the scaling regime attains a constant fraction of the total energy density of the Universe. Thus, the network produces more GWs and leads to an enhancement of the GWB amplitude, in contrast to the SM-thermalized DOFs.
As shown in \cite{servantSpectro}, this enhancement can be up to $\sim 7\%$ as maximally allowed by the $\Delta N_{\rm eff}$ bound.
For simplicity, in this work we assume constant $g_{*}^{D}, g_{*s}^{D}$, as this maximizes the spectral signature\footnote{As shown in~\cite{servantSpectro}, the evolution of $g_{*}^{D},g_{*s}^{D}$ can also lead to extra small suppression features.}. We parametrize the effect of the dark sector by considering its energy density as $\rho_D = \rho_{D,\rm BBN}(a_{\rm BBN}/a)^4$ with $\rho_{D,\rm BBN} = \pi^2 \epsilon_{\rm BBN} T_{\rm BBN}^4/30$ and $\epsilon_{\rm BBN} = \left.g_*^D(T_D/T)^4\right|_{\rm BBN}$.

\begin{center}
    {\bf The templates in this model depend on: $G\mu$ and $\epsilon_{\rm BBN}$}.
\end{center}

We calculate the template assuming the GW emission as in Eq.\,\eqref{eq:grav_emission_power_loop_j} with $q=4/3$, and the VOS model as obtained from solving the VOS equations in an expanding background with Hubble rate given by Eq.\,\eqref{eq:friedmann_LambdaCDM_dark}.
Fig.~\ref{fig:CS_dark_dof} shows the relative enhancement due to secluded dofs, with respect to the $\Lambda$CDM prediction.

\subsection{Changing the loop properties}
\label{subsec:loopProperties}

So far, our cosmic-string GWB spectra assumed loops with an initial loop size $\alpha = 0.1$, and with cusps, i.e.~with GW emission power index $q=4/3$. These assumptions are based on the fact that a cusp is typically expected to be formed in a Nambu-Goto loop after a few oscillations~\cite{Vilenkin:2000jqa}, as supported by the numerical simulations of Nambu-Goto string networks from Ref.~\cite{Blanco-Pillado:2013qja}. Furhtermore, the value $q = 4/3$ is also singled out as the relevant effective value to use for loops formed in Nambu-Goto simulations\footnote{Technically, it is found that $j^{4/3} P_j$ is not constant but rather has a local maximum around $j \simeq 3$ and decays asymptotically down to a constant for $j \gtrsim 100$, see figure~4 of \cite{Blanco-Pillado:2017oxo}. The amplitude of $j^{4/3} P_j$ at the maximum is however only a factor $\sim 1.7$ larger than the value the constant asymptotic value.}, see Ref.~\cite{Blanco-Pillado:2017oxo}. Among the variety of cosmic-string models and early-Universe scenarios, however, small-scale structures could potentially also arise on a loop, hence changing the GW emission per loop oscillation. Furthermore, in the VOS modeling, the initial loop size $\alpha$ can apriori be treated as a free parameter. 

In the following we consider scenarios where we vary $q$ and $\alpha$ with respect to their canonical values $\alpha = 0.1$ and $q = 4/3$, either individually or simultaneously. 

\subsubsection{Loop's GW emission power}
\label{subsubsec:loop_emission_power}

We first consider templates with various values of $q$, which we recall it represents the exponent of the power-law in Eq.~(\ref{eq:GW_emission_power_single_loop}) characterizing the GW power emission of the $j$-th  harmonic, {\it cf.}~$P_j \propto j^{-q}$. In particular the value of $q$ can change to e.g.~$q = 5/3,$ if there are kinks in the loop, or to $q = 2$ if kink-kink collisions take place~\cite{Vilenkin:2000jqa}. Here we consider such canonical values $q = 4/3, 5/3$ and $2$, and explore also the  agnostic possibility that effectively $q$ could take values around and in-between the canonical NG ones. For practical reasons we sample in intervals of $\Delta q = 0.1$, and exclude $q = 1$ as $\sum_{j = 1}^\infty j^{-1}$ diverges. %Here we consider such canonical values $q = 4/3, 5/3$ and $2$, and explore also the possibility that effectively $q$ could take any value between (some value above) $1$ and $2$, which for practical reasons we sample in intervals of $\Delta q = 0.1$. We exclude $q = 1$ as $\sum_{j = 1}^\infty j^{-1}$ diverges, so we start at $q = 1.1$.

\begin{figure}[t!]
    \centering
    {\sffamily Varying the GW emission index $q$}\\
    \includegraphics[width=0.495\textwidth]{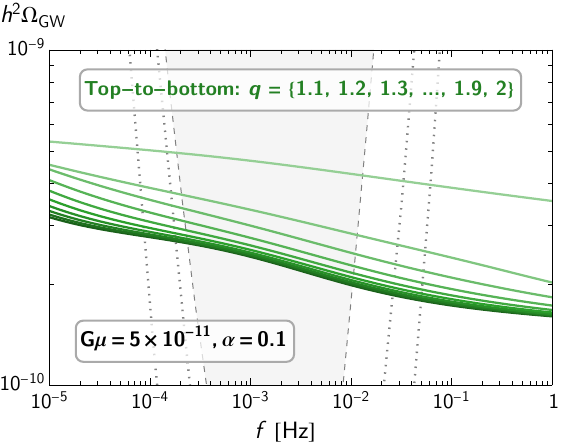}\hfill
    \includegraphics[width=0.495\textwidth]{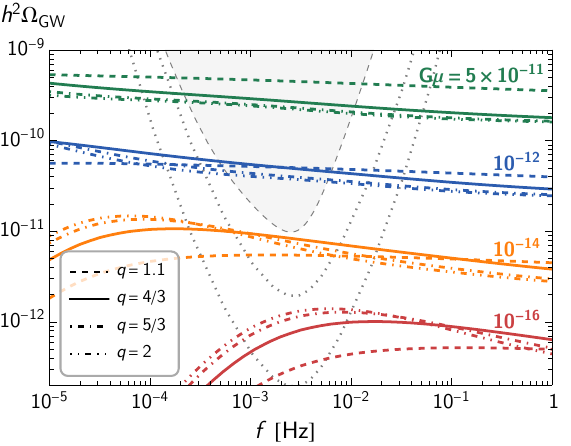}\\[-1em]
    \caption{{\it Left panel:} GWB from cosmic-string network of $G\mu = 5 \times 10^{-11}$ and $\alpha = 0.1$ for various values of $q$ in Eq.\,\eqref{eq:grav_emission_power_loop_j}, calculated from VOS model (Sect.~\ref{sec:GW_semi_analytic}).
    {\it Right panel:} Similar to the left panel, but with different $G\mu$ values. The green spectra  are the same as those in left panel.}
    \label{fig:spec_q_plot}
\end{figure}

\begin{center}
    {\bf The templates in this model depend on: $G\mu$, $q$}.
\end{center}
We use the loop number density of the VOS model in section~\ref{sec:GW_semi_analytic} with $\alpha = 0.1$ and $\Lambda$CDM cosmic history. In the left-panel of Fig.~\ref{fig:spec_q_plot} we show  the effect of considering non-conventional values $q = 1.1, 1.2, ....,$ $1.9, 2.0$ on the GWB spectrum, for $G\mu = 10^{-10}$ around the LISA's frequencies. The spectral amplitudes converge for $q > 1.5$. In the right-panel we also plot the GWB spectra for $q = 1.1, 4/3, 5/3$ and $2$, for the various tensions $G\mu = 10^{-16}, 10^{-14}, 10^{-12}, 10^{-10}$, again around the LISA frequency window. While the spectra for $q = 5/3$ and $2$ are very similar for all tensions, differing at most by $\sim 20\,\%$, the spectrum for $q = 4/3$ is clearly distinguishable from the $q = 5/3, 2$ cases. 

\begin{figure}[t!]
    \centering
    {\sffamily Varying initial size $\alpha$ of loops}\\
    \includegraphics[width=0.495\textwidth]{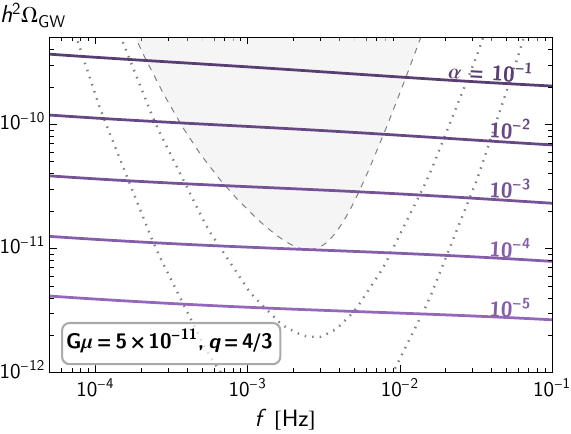}\hfill
    \includegraphics[width=0.495\textwidth]{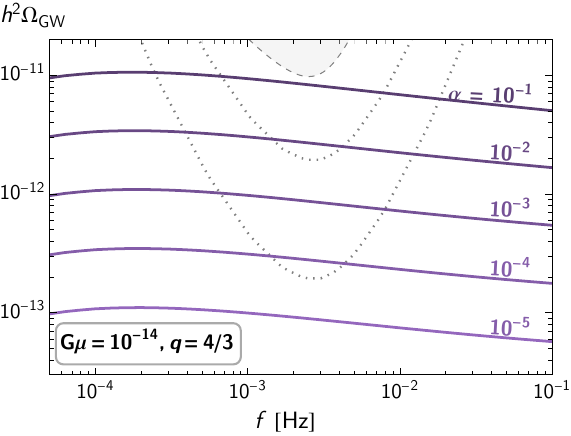}\\[-1em]
    \caption{GWB spectra from cosmic strings of $G\mu = 5\times 10^{-11}$ and $10^{-14}$ with a variation of the initial size $\alpha$ of loops. The spectral amplitudes follow the scaling of $\Omega_{\rm GW} \propto \sqrt{\alpha}$ for $\alpha \gg \Gamma G\mu$.
    }
    \label{fig:spec_alpha_plot}
\end{figure}

\subsubsection{Initial loop size}
\label{sec:atypical_loop_size}

Here we consider  templates with various values of the $\alpha\equiv l_i/t_i$, which represents the initial loop size in units of cosmic time, when a loop is first `born' out of the network. In the VOS modeling, $\alpha$ can be treated as a free parameter, so we consider a variation from $\alpha = 0.1$ as determined by Nambu-Goto simulations~\cite{Blanco-Pillado:2013qja}, down to $\alpha = 5 \times 10^{-6}$, which is sufficient for the reconstruction analysis (i.e., the signal has SNR $<10$ for $G\mu < 10^{-11}$ ).\footnote{In principle, the value of $\alpha$ can be as small as the GW emission scale $\Gamma G\mu$.}
We use the loop number density of the VOS model in Sect.~\ref{sec:GW_semi_analytic} (with constant $\alpha$) and the $\Lambda$CDM cosmic history (we cannot consider the BOS modeling here, as the loop number density and the initial loop size are fixed consistently from numerical simulations). 

\begin{center}
    {\bf The templates in this model depend on:} $G\mu$, $\alpha$ (and $q$)
\end{center}

Fig.\,\ref{fig:spec_alpha_plot} shows the effect of changing $\alpha$ in the GWB spectra, for $q = 4/3$. The impact on the signal is quite noticeable, as for {GW produced during RD era} and $\alpha \gg \Gamma G\mu$, the spectral amplitude scaling as $\Omega_{\rm GW} \propto \sqrt{\alpha}$ holds  independently of the frequency~\cite{Auclair:2019wcv,Gouttenoire:2019kij}. A degeneracy between $\alpha$ and $G\mu$ emerges therefore if $\alpha$ is allowed to vary. For example, while signals with standard $\alpha = 0.1$ for tensions $G\mu = 10^{-13}$ and $G\mu = 10^{-15}$ can be detected and reconstructed  within LISA to better than $\sim 4\%$ and $\sim 20\%$ of precision in $G\mu$ respectively (see Sect.~\ref{sec:recon_result_conven}), the degeneracy between $\alpha$ and $G\mu$ degrades substantially the simultaneous reconstruction of these two parameters when both are allowed to vary (see Sect.~\ref{subsec:recons_loopProperties}).

Furthermore, from an agnostic point of view, one can even speculate with considering the variation of $\alpha$ together with the previous variation of $q$. In our detailed analysis in Sect.~\ref{subsec:recons_loopProperties} we thus also consider templates depending simultaneously on $\alpha$ and $q$, so that we quantify the degeneracy in these two parameters, when reconstructing an arbitrary signal under a VOS template.

\subsection{Field theory local string network}
\label{subsec:FieldTheoryNetworks}

We note that in this work we have computed all template signals under the assumption that cosmic strings are correctly described by Nambu-Goto string description. With the sole exception of super-strings in String Theory, cosmic strings are however field-theory objects, and hence we would like to draw a word of caution on this respect. We note that recent field-theory simulations~\cite{Baeza-Ballesteros:2024otj} of  the GW emission of local-string loop configurations in an Abelian-Higgs field theory, show indeed that square-shape loops\footnote{Created from the crossing between pairs of straight boosted infinite strings, and hence containing four well localized kink-like features.}, emit GWs with a total power of the same order as (although slightly above) the Nambu-Goto prediction $P_{\rm tot} = \Gamma G\mu^2$, {\it cf.}~Eq.~(\ref{eq:GW_emission_power_single_loop}), evaluated for $\mathit{\Gamma} = 50$ and $\mu=\pi v^2$, with $v$ the vacuum expectation value of the Higgs-like field in the simulations. The emission of GWs by such squared loops supports therefore the GW emission power computed {\it \`a la} Nambu-Goto (at least in what concerns the total power emitted). However, Ref.~\cite{Baeza-Ballesteros:2024otj} has also shown that randomly shaped loops created naturally in a simulated phase transition (hence expected to be more realistic than squared-loops), emit GWs with power $\sim 2-3$ times larger than the same Nambu-Goto prediction. The different results between square-shape and random-shape loops suggest then that the Nambu-Goto description is not guaranteed
to be a precise characterization of the GW emission of field theory loops. From this point of view, the sampling of values $q = 1.1, 1.2, ....,$ $1.9, 2.0$ in Sect.~\ref{subsec:loopProperties} could be loosely interpreted as a way to accommodate different GW emission powers, whilst still enforcing the functional form of the Nambu-Goto $j$-th harmonic GW emission, $P_j \propto j^{-q}$, so that $P_{\rm tot} = \sum_j P_j \equiv \Gamma G\mu^2$. The GW emission of field theory loops could be accommodated therefore into the variation of the loop's GW emission power category of beyond-conventional templates as considered in Sect.~\ref{subsec:loopProperties}.  Strictly speaking, one should perform, of course, a re-analysis of the signal considering the power spectrum of the exact GW emission of the field theory loops, as measured in the simulations of~\cite{Baeza-Ballesteros:2024otj}, and include this as one more case in Sect.~\ref{subsubsec:loop_emission_power} .

The above consideration ignores however a major distinctive aspect that distinguishes field theory loops from those of Nambu-Goto. Namely, local field theory string loops emit particles of the scalar and gauge fields they are actually made of. This has been shown repeatedly by 
large scale numerical simulation of Abelian-Higgs field theory networks~\cite{Vincent:1997cx,Moore:2001px,Bevis:2006mj,Bevis:2010gj,Daverio:2015nva,Correia:2019bdl,Correia:2020gkj,Correia:2020yqg}, and more recently by isolating string loop configurations~\cite{Matsunami:2019fss,Hindmarsh:2021mnl,Baeza-Ballesteros:2024otj}. The latter works show that particle emission actually leads to the decay of the loops on time scales $\Delta t_{\rm dec} \propto L^p$, where $L$ is the loop length. It is found that $p \simeq 2$ for artificial ({\it i.e.}~square-shape) loops~\cite{Matsunami:2019fss,Hindmarsh:2021mnl,Baeza-Ballesteros:2024otj}, whilst $p \simeq 1$ for network loops~\cite{Hindmarsh:2021mnl,Baeza-Ballesteros:2024otj}. This implies that below a critical length, artificial loops decay primarily through particle production, whilst for larger loops GW emission dominates~\cite{Matsunami:2019fss,Baeza-Ballesteros:2024otj}\footnote{This implies a breakdown in the GWB spectrum, which is then suppressed at high frequencies, though way above the LISA frequency window~\cite{Auclair:2019jip,Auclair:2021jud,Gouttenoire:2019kij}. A similar unobservable effect is also found in Abelian-Higgs strings with a cusp, though with different critical scale and breakdown frequency~\cite{Olum:1998ag,Auclair:2019jip}.}. However, for network loops, which presumably represent more realistic configurations, particle emission always dominates, as supported by the state-of-the-art data from Ref.~\cite{Baeza-Ballesteros:2024otj}, which sustained a separation of scales between the length of the loops $L$ and their core-radius $r_c$, up to ratios as large as $L/r_c \lesssim 6000$. In other words, for network loops particle production is observed even when the core's width represents only $\sim 0.01\%$ of the loop's length, in a situation where Nambu-Goto is clearly expected to be very good description of the motion of the loop. This seems to imply that just because the motion of the loop's core is well described by Nambu-Goto, it does not mean that particle emission is absent, as it is implicitly assumed when describing a Nambu-Goto string. Dominance of particle emission over GW emission, implies that the GWB from a local string network should be greatly suppressed compared to estimations that ignore particle emission\footnote{Analogous studies of the decay of global string loops can be also found in Refs.~\cite{Saurabh:2020pqe,Baeza-Ballesteros:2023say}, suggesting as well a suppression of the overall GWB from a global string network. Contrary to local loops, however, the rate of particle production always dominates over the rate of GW emission for global loops, independently of the shape of the loop, for at least length-to-width ratios that reach up to $L/r_c \lesssim 1700$~\cite{Baeza-Ballesteros:2023say}.}. For example, fitting PTA data to Nambu-Goto cosmic strings that only emit GWs leads to the tight constraint $G\mu \lesssim 10^{-10}$~\cite{NANOGrav:2023hvm,Antoniadis:2023xlr,Figueroa:2023zhu}, whilst fitting to an Abelian-Higgs field theory network that allows for particle production,  
loosens the constraint to 
$G \mu \lesssim 10^{-7}$~\cite{Kume:2024adn}. 

A proper quantification of the spectral shape of the GWB emitted by a local field theory string network requires, however, not only the determination of the particle and GW emission rates of the loops, but also an understanding of the underlying loop production function of the network. One needs to incorporate all together  the results on particle and GW emission from the individual loops into a calculation pipeline of the spectrum of the GWB, making a number of assumptions about the loop production function. This will affect the loop number density along cosmic history, and in this respect, the GWB from a field theory cosmic string network could be seen as another case of beyond-conventional templates of Sect.~\ref{subsec:LoopNumberDensityBSM}, where variations of the loop number density are considered. As no definite answer about the spectral shape of the GWB from a local field theory network is presently available in the literature, we exclude from this paper 
the study of LISA's ability to probe such a case. We simply highlight here that the only thing we can truly assess at the present about the GWB from a field theory local string network, is that the particle and GW emission rates as measured in the lattice for network loops~\cite{Hindmarsh:2021mnl,Baeza-Ballesteros:2024otj} will greatly reduce the expected amplitude of the GWB at all frequencies. We postpone any quantitative assessment of the reconstruction of this signal by LISA for separate work, once the spectral form of the GWB becomes available.  

\newpage
\section{Reconstruction of conventional templates 
at LISA}
\label{sec:reconstruction}

In this section we address the question of how well cosmic string GWB signals can be reconstructed by the LISA detector, using SBI techniques. We first focus on the simplest cases of cosmic string GWB's, given by the conventional templates discussed in Sect.~\ref{sec:conventional}: the semi-analytic {\bf VOS model} (Sect.~\ref{sec:GW_semi_analytic}) and the  Nambu-Goto simulated {\bf BOS modeling} (Sect.~\ref{sec:GW_simulated}), both of which depend solely on one parameter, the string tension $G\mu$.
We chose the prior range for both templates to be 
\begin{center}
    \fbox{\bf ~ String tension $G\mu$: log--uniform\,$[10^{-18},10^{-9}]$}
\end{center}
The upper boundary is chosen to be somewhat larger than the observational bound from PTA's, $G\mu \leq 7.9 \times 10^{-11}$~\cite{NANOGrav:2023hvm}.
We choose the lower bound where we expect a signal with low detectability, so that the reconstruction capability is  substantially degraded. This happens when the signal-to-noise ratio (SNR, defined below) becomes smaller than unity. We will justify this choice later, in Sect.\,\ref{sec:recon_result_conven}, by showing that a signal with ${\rm SNR} < 1$ leads to a reconstruction error $> 100\%$. 

We present our results on signal reconstruction in Sect.~\ref{sec:recon_result_conven}, where we discuss in detail LISA's ability to reconstruct conventional templates as a function of the tension, and define various metrics to characterize the quality of signal reconstruction (Sect.~\ref{subsec:quality_recon_conven}). We also compare our results to other detection/reconstruction methods, such as the use of the power-law integrated sensitivity curve based on the signal-to-noise ratio (Sect.~\ref{subsec:sbi_vs_snr}), or Markov-Chain Monte Carlo techniques (Sect.~\ref{subsec:sbi_vs_MCMC}). We further discuss how model comparison between templates is done via SBI in 
Sect.~\ref{subsec:ModelComparisonVOSvsBOSvsDOF}. There we quantify LISA's ability to discriminate among conventional templates, {\it i.e.}~VOS vs BOS, given a collection of (simulated) datasets. 

\begin{figure}
    \centering
    \includegraphics[width=0.9\linewidth]{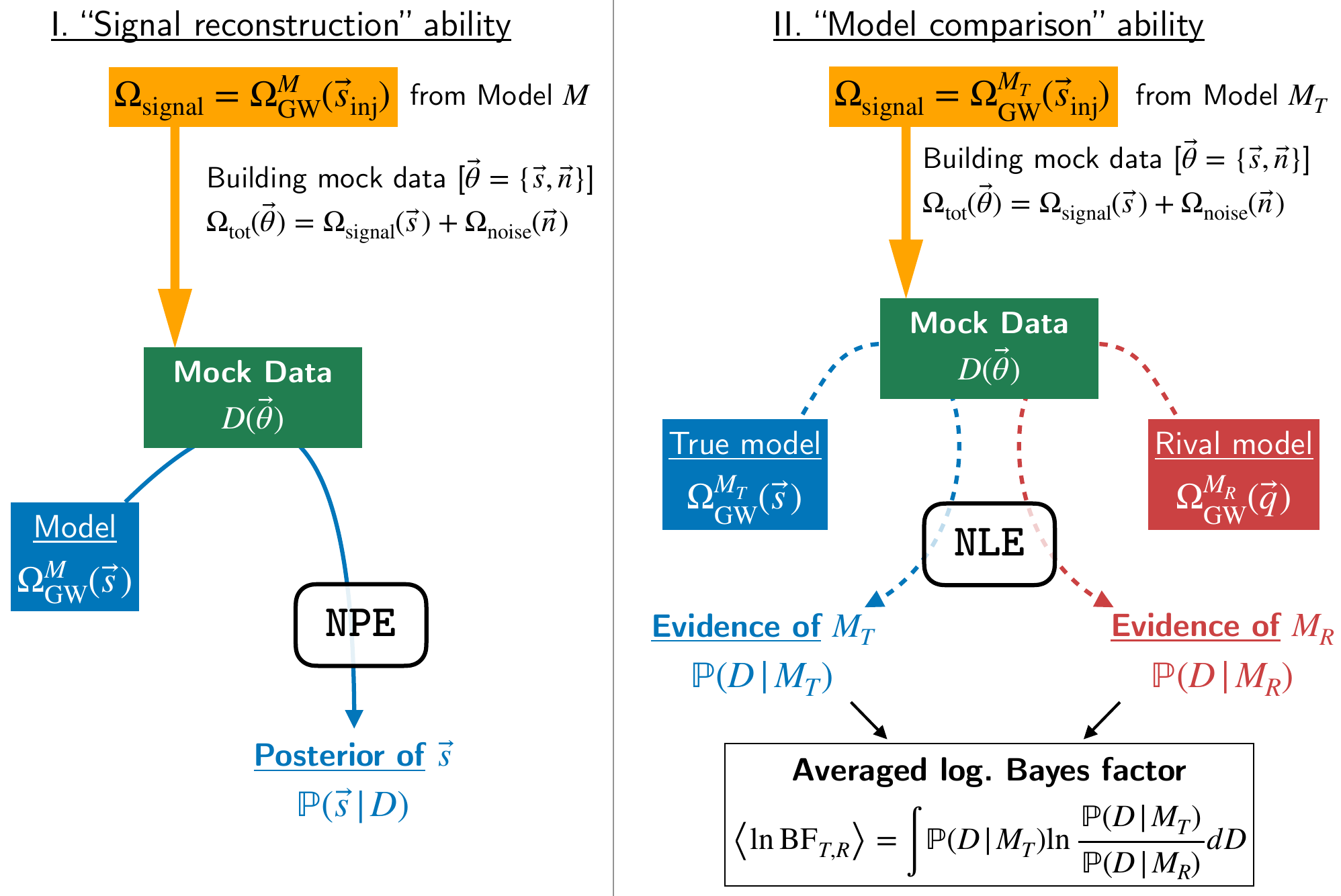}\\[-0.75em]
    \caption{Flow charts for determining the signal-reconstruction and model-comparison ability of LISA, discussed in Sects.~\ref{sec:recon_result_conven} and \ref{subsec:ModelComparisonVOSvsBOSvsDOF}, respectively. For the former, the posterior probability distributions on the  parameters $\{\theta_i\}$ of model $M$ are obtained via Bayesian inference on the mock data generated by the template $M$ with a set of injected parameter $\{\theta_i\}_{\rm inj}$. For the latter, we use Bayes factor (BF) to describe how well a rival model $M_{R}$ can explain the data generated by a true template $M_{T}$ with a set of injected paramters $\{\theta_i\}_{\rm inj}$. We explore the model parameter space by scanning over $\{\theta_i\}_{\rm inj}$.}
    \label{fig:flowchart}
\end{figure}

Before we dive into our reconstruction/comparison results, we point the reader to Appendix \ref{app:LISA_noise}, where, for completeness, we have recapped briefly details on the expected noise of LISA, as well as on  
our procedure to generate mock data. A summary of our SBI reconstruction technique, which graphically is shown in Fig.\ref{fig:flowchart}, is also confined to Appendix \ref{app:SBI_summary}. For a fully detailed explanation of our procedure, we refer  the reader in any case to Ref.~\cite{Dimitriou:2023knw}.

\begin{figure}[t!]
    \centering
    {\sffamily Reconstructing the BOS model (with Saikawa-Shirai DOF's)}\\
    \includegraphics[width=0.5\textwidth]{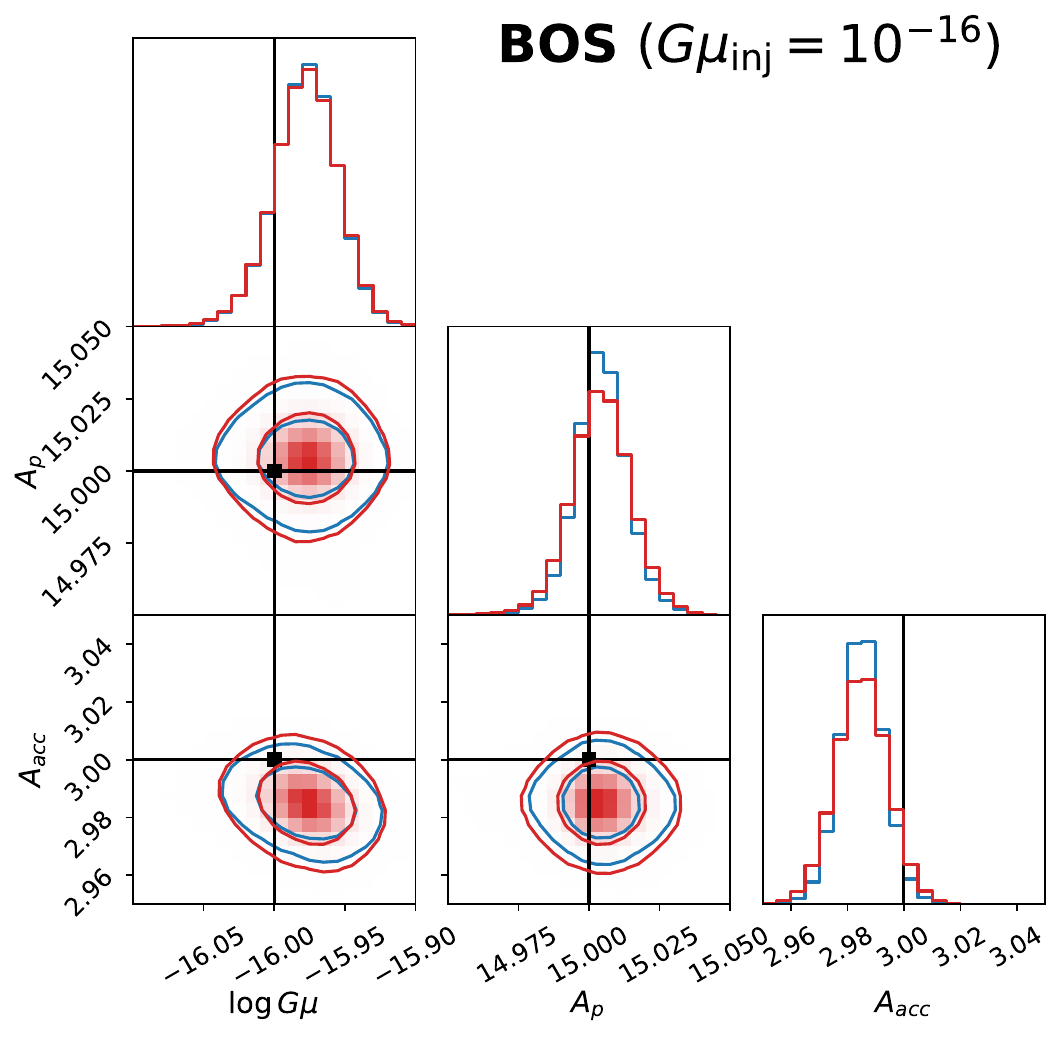}\hfill
    \includegraphics[width=0.5\textwidth]{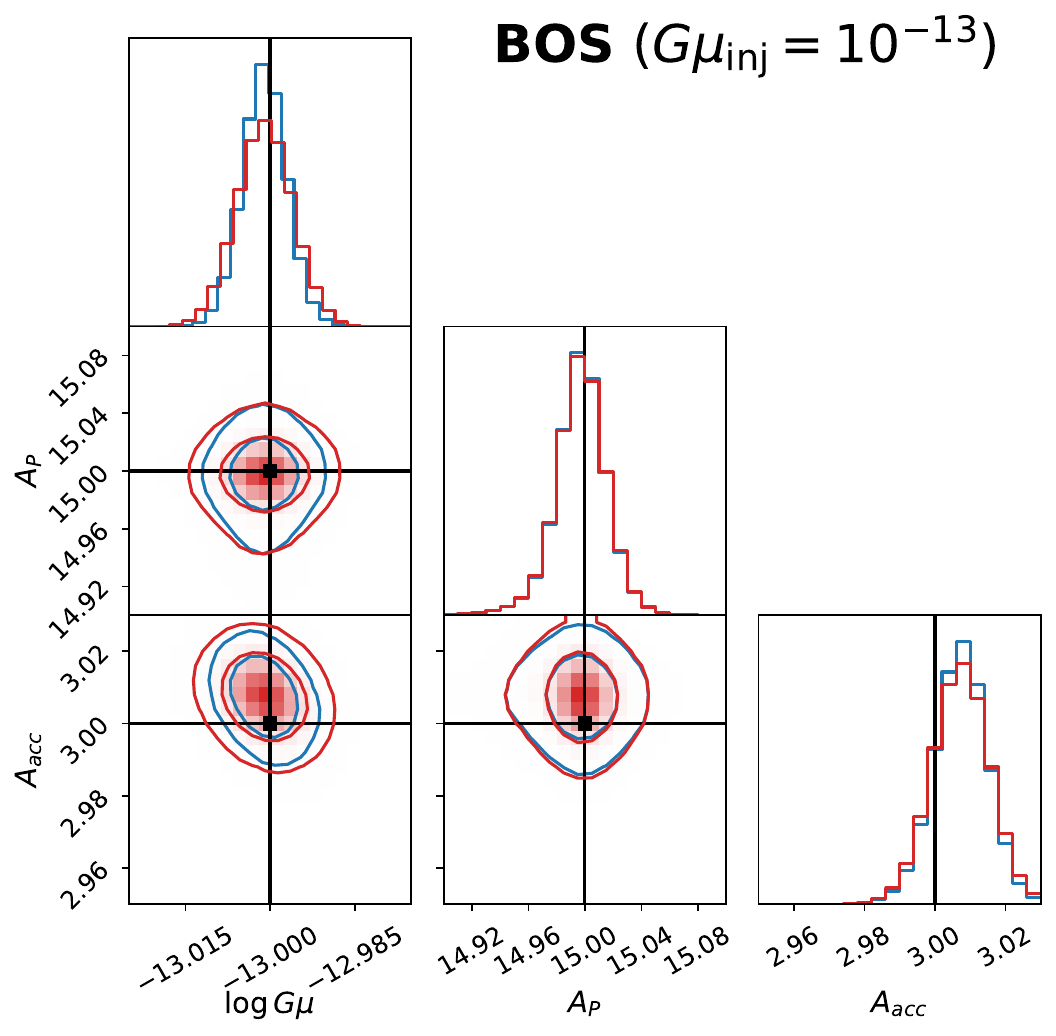}\\[-1em]
    \caption{Reconstruction of cosmic-string GWB (BOS model) at LISA, shown as 1D and 2D posteriors of the network tension $\{G\mu\}$ and the noise parameters $\{A_P,A_{\rm acc}\}$. The injected values of parameters are shown by black lines and dots. 
    Inner and outer red contours show the 68\% and 95\% reconstruction uncertainty regions using SBI, while in comparison the blue contours show the results from a MCMC technique (see Sect.~\ref{subsec:sbi_vs_MCMC} for details). The colored regions represent the probability density in the inner part of the posterior.}
    \label{fig:recon_bos_potato}
\end{figure}

\subsection{Reconstruction with SBI} 
\label{sec:recon_result_conven}

\begin{figure}[t!]
\centering
\hspace{-1.5cm}
\includegraphics[width=1.1\textwidth]{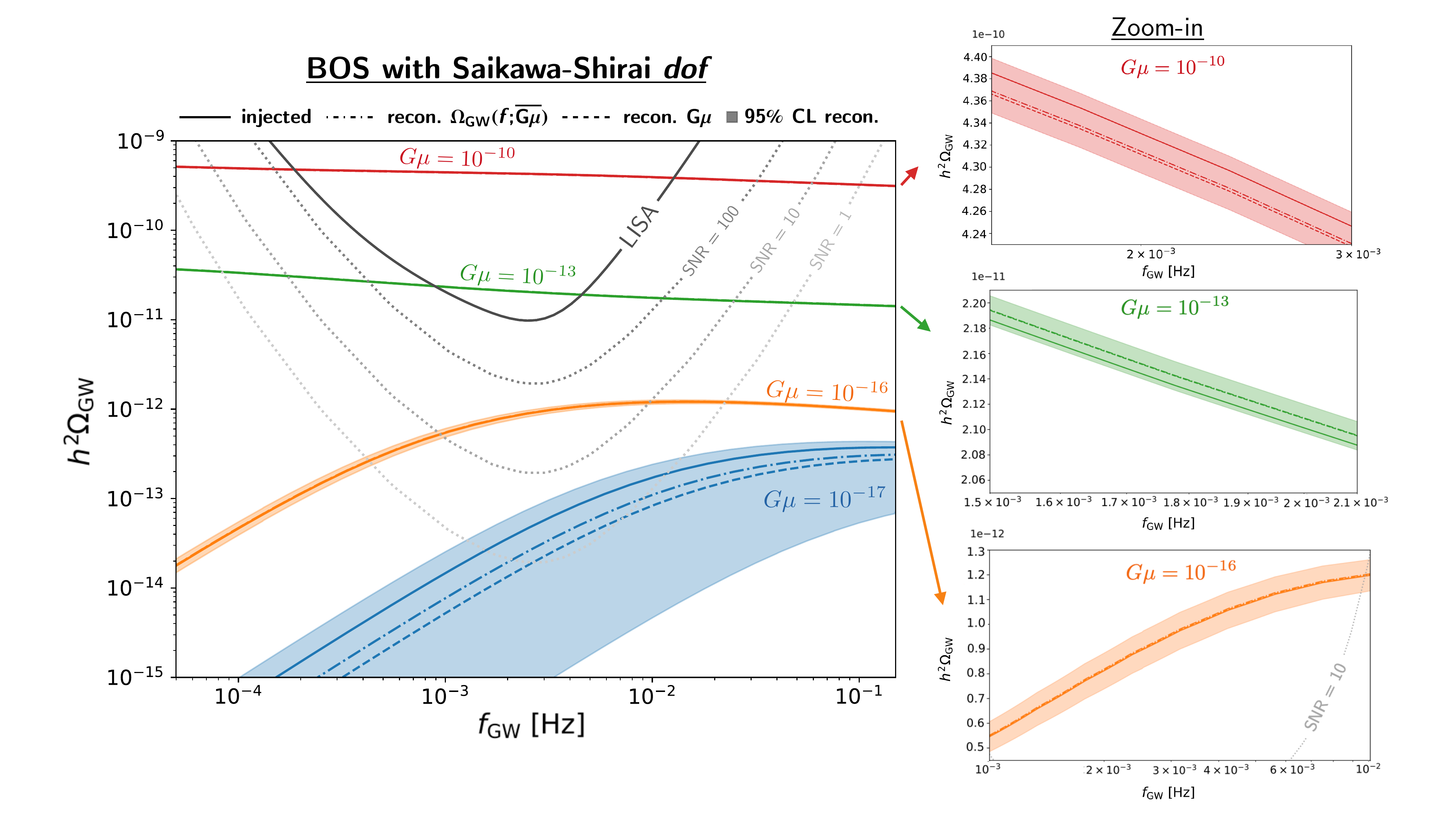}
\\[-1em]
    \caption{{\it Left Panel:} The reconstruction of cosmic-string spectra for different $G\mu$ using the BOS templates (with Saikawa-Shirai dof). The solid colored lines are the injected signal. Colored dashed lines are the signal templates evaluated at the mode of $G\mu$ from the marginalized $G\mu$-posterior. Colored dot-dashed lines correspond to the signal templates evaluated at the mean value of $G\mu$ from the marginalized $G\mu$-posterior. The colored bands illustrate the regions where the signal is evaluated within the 95\% CL of the posterior of $G\mu$.
 The black solid line is the LISA noise in the AA channel, while the dotted gray lines are the PLS of different SNR. {\it Right Panel:} The zoom-in spectra of the left panel, where the vertical axis is shown in linear scale.}
\label{fig:recon_conven_spec}
\end{figure}

In Fig.\,\ref{fig:recon_bos_potato} we show the posteriors on signal and noise parameters obtained using our SBI reconstruction for injected signals (BOS with Saikawa-Shirai DOF) with $G\mu_{\rm inj} = 10^{-16}$ (left) and $10^{-13}$ (right). The diagonal plots show the 1D marginalized posterior for each parameter, and the off-diagonal the 2D posteriors in terms of 68\% (inner) and 95\% (outer) credible intervals. The black lines represent the true values of the parameters.
We see that we are able to reconstruct the true values of both signal and noise parameters within the 68\% confidence regions, and with a high precision (more on this below). Note that the case with  $G\mu_{\rm inj} = 10^{-13}$ has a smaller $G\mu$-posterior but wider ($A_P$, $A_{\rm acc}$)-posteriors than those from the $G\mu_{\rm inj} = 10^{-16}$ case. This is expected as the signal with $G\mu = 10^{-13}$ dominates over LISA's noise sensitivity, while the opposite happens for the signal with $G\mu = 10^{-16}$, see the signals' spectra in Fig.~\ref{fig:recon_conven_spec}.

Fig.\,\ref{fig:recon_conven_spec} shows reconstructed signals, where solid colored lines represent GWB spectra of the injected signals (BOS model with Saikawa-Shirai DOF) for the tensions $G\mu = 10^{-17}, 10^{-16}$, $10^{-13}$, and $10^{-10}$. Colored dashed lines are the signal templates evaluated at the mode of $G\mu$ from the marginalized $G\mu$-posterior. Colored dot-dashed lines, on the other hand, correspond to the signal templates evaluated at the mean value of $G\mu$ from the marginalized $G\mu$-posterior, and hence represent more correctly the reconstructed signal. The colored bands illustrate the regions where the signal is evaluated within the 95\% CL of the posterior of $G\mu$.

\subsubsection{Quality of reconstruction}
\label{subsec:quality_recon_conven}

When reconstructing a signal, we can define a measurement of the `reconstruction quality' or `error', using the posterior probability distribution of a given 
{parameter}, see {\it e.g.}~Fig.~\ref{fig:recon_bos_potato}. In particular, we define the {\bf precision} of a parameter $X$ by the range of confidence in its posterior, {\it i.e.}~by the difference with respect to the true/injected value, defined by 
\begin{align}
   \Delta X \equiv \frac{X_{\rm upper} - X_{\rm lower}}{|X_{\rm inj}|}~,
   \label{eq:precision_recon}
\end{align}
where $X_{\rm upper}$ and $X_{\rm lower}$ are the upper and lower values of $X$, as extracted from its posterior distribution, for fixed confidence level. From now on, we will report the precision for 95\% confidence levels only. We note that we chose the absolute value $|X|$ in the denominator to enforce $\Delta X$ to be positive. 

If $X$ is positive, in analogy to Eq.~(\ref{eq:precision_recon})  we can also define the precision of $\log(X)$, as $\Delta\log X \equiv [(\log X)_{\rm upper}-(\log X)_{\rm lower}]/|\log X_{\rm inj}|$. It is however worth mentioning that the precision in $\log(X)$ can differ largely from the precision in $X$ itself. With an easy algebra and using Eq.~\eqref{eq:precision_recon}, the relation between $\Delta X$ and $\Delta \log X$ reads,
\begin{align}
    \Delta X = \frac{X_{\rm lower}}{X_{\rm inj}}\left(\left(10^{|\log X_{\rm inj}|}\right)^{\Delta{\log X}} -1 \right)\,.
    \label{eq:logX_X_precision_relation}
\end{align}
For example, for $G\mu_{\rm inj} = 10^{-13}$, $G\mu_{\rm lower} = 10^{-13.05}$ and $G\mu_{\rm upper} = 10^{-12.95}$, so we have $100\times \Delta {\log G\mu} \simeq 0.77\%$, whereas for the tension directly we have $100\times\Delta {G\mu} \simeq 23\%$.

Alternatively, one could think also to use as an error of reconstruction the {\bf accuracy} of an observable $X$. This describes how close the most probable value of the observable, i.e.~the mode\footnote{Similarly, one could define the accuracy with respect to the mean value, instead of the mode.} $X_{\rm mode}$ in the posterior distribution, approaches the correct injected value $X_{\rm inj}$. Using an absolute value to ensure its positiveness, we define the accuracy of a variable $X$ as  
\begin{align}
        {\rm Acc}(X) \equiv \left|\frac{X_{\rm mode} - X_{\rm inj}}{X_{\rm inj}}\right|\,.
\label{eq:accuracy_recon}
\end{align}
By construction, the accuracy is smaller than the precision if the injected value is contained within the 95\% interval. If this is not the case, the accuracy can be arbitrarily large, reflecting a strong bias in the inference procedure, leading to a wrong assessment of the model parameter reconstruction error. 

As we will see later, especially in the case of multi-parameter templates in Sect.~\ref{sec:recons_beyond}, the parameter reconstruction suffers some degradations due to degeneracy, {\it i.e.}~many parameter sets can lead to similar GW signals within the LISA window. The posterior probability distribution recovered from such cases gets populated across the degenerate set of parameters, with the worst case being when the posterior flattens over the prior range(s). In these cases, the precision defined in~\eqref{eq:precision_recon} cannot capture the true reconstruction ability, as it will be prior dependent. To quantify the quality of the recovered posterior in those more challenging cases, we define a third quantity which acts as a {\bf quality indicator},
\begin{equation}
 {\cal Q}(X) \equiv
 \frac{X_{\rm upper} - X_{\rm lower}}{\delta X_{\rm prior}}~,
 \label{eq:quality_indicator}
\end{equation}
where the 95\% interval of the posterior (in the numerator) is compared to the full prior range $\delta X_{\rm prior}$. This quantity is positive definite and, by construction, ranges within
the interval $Q(X)\in[0,0.95]$, such that a flat posterior corresponds to
$Q(X)=0.95$.
For parameters of a template where the reconstruction is not satisfactory ({\it e.g.}~precision larger than 100$\%$), we will report the quality indicator. In these cases the signal is weaker than the reconstruction capabilities of our procedure, and the
corresponding outcome is a posterior which roughly resembles the prior distribution. In this situation, a quality indicator as we defined, is more meaningful, since it reports how close the
posterior is to the prior. But the quality indicator does not contain information about the
true (injected) value, which is what the precision metric captures. That’s why the two
metrics are complementary and worth presenting, whenever convenient. 
We also note, in any case, that $Q$ has no particular utility for the reconstruction of conventional templates here in Sect.~\ref{sec:reconstruction}, and it will become only useful in Sect.~\ref{sec:recons_beyond}, where some of the signals have multi-parameter space dependencies. Finally, all reconstruction quantities reported in this section are obtained by averaging over 20 independent realizations of the mock data, each analyzed using the full reconstruction procedure.

\begin{figure}[t!]
    \centering
\includegraphics[width=0.495\textwidth]{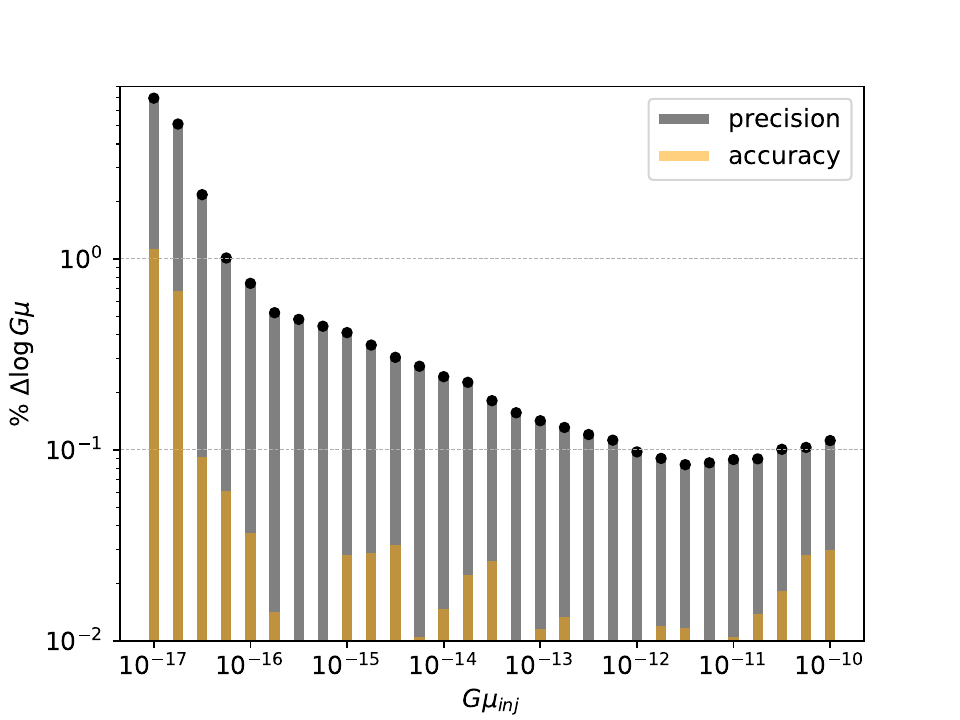}
\includegraphics[width=0.495\textwidth]{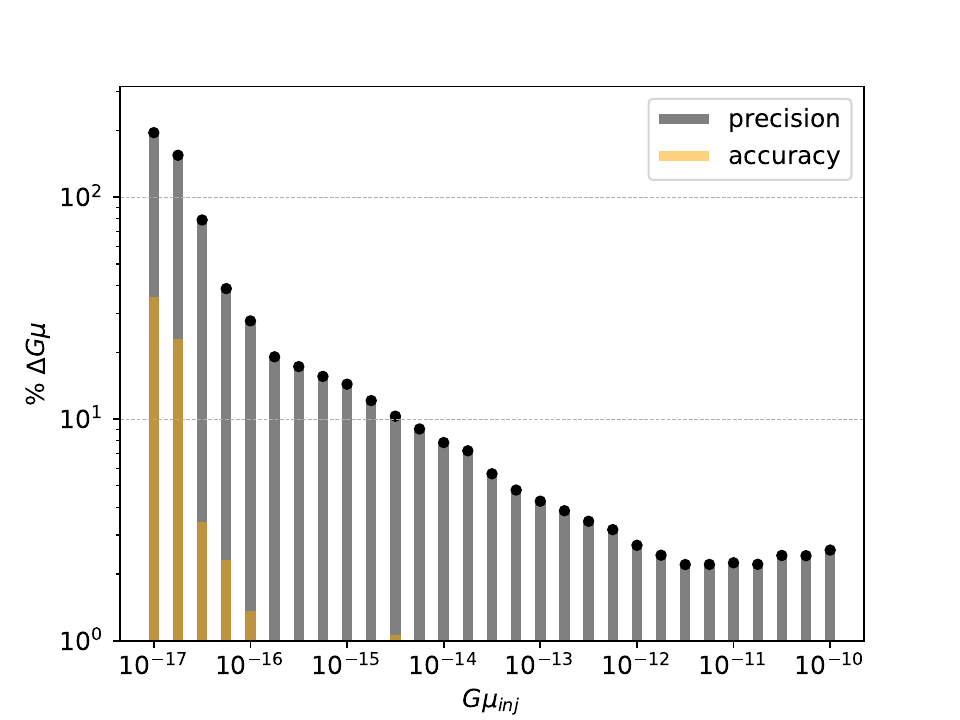}
\\[-0.25em]
\includegraphics[width=0.495\textwidth]{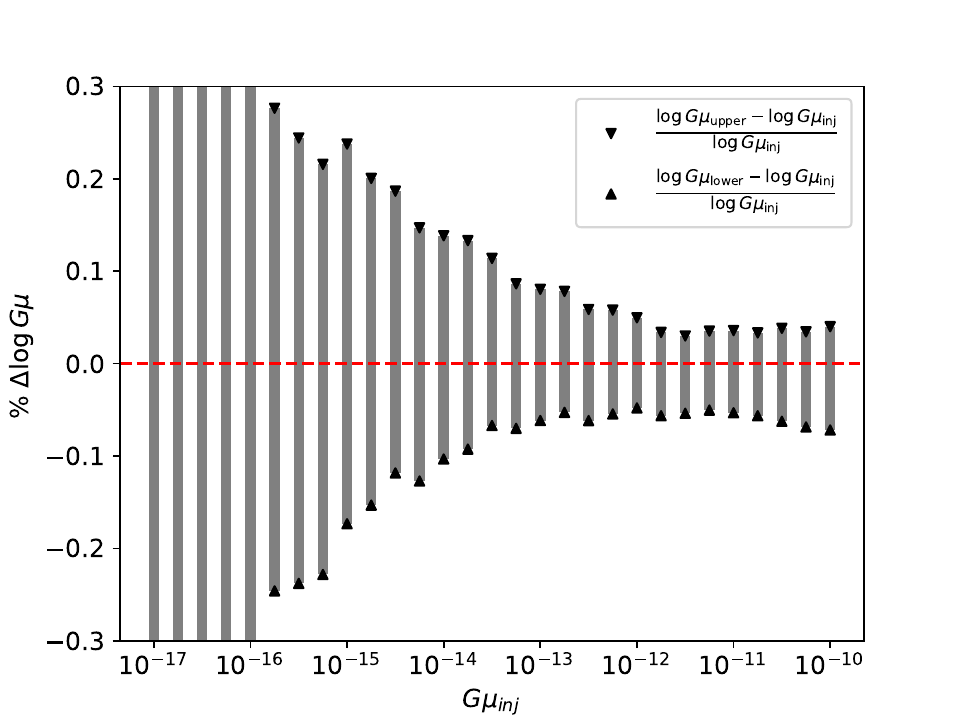}\hfill
\includegraphics[width=0.495\textwidth]{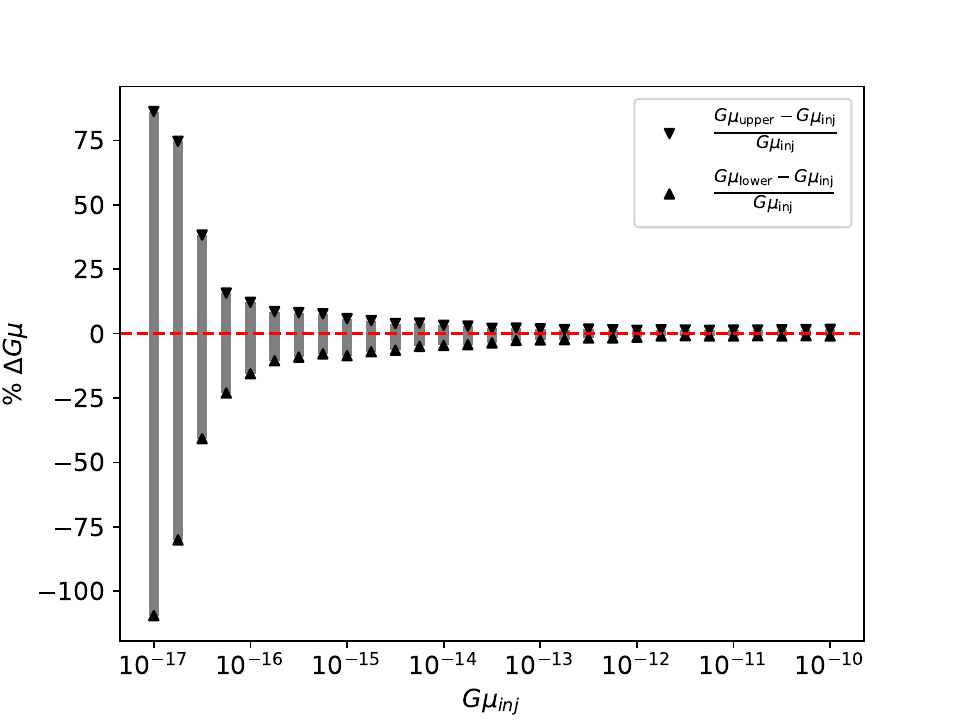}
\\[-1em]
    \caption{
    \emph{Top panel:} For each BOS signal with Saikawa-Shirai dof and  $G\mu_{\rm inj}$, the reconstruction precisions [defined in Eq.~\eqref{eq:precision_recon}] in $\log G\mu$ (left) and $G\mu$ (right) for each true signal of $G\mu_{\rm inj}$ are shown by the gray bars.
    The gold bars represent the  accuracy [Eq.~\eqref{eq:accuracy_recon}].
    \emph{Bottom panel:}  The \% relative errors in reconstructing $\log G\mu$ (left) and $G\mu$ (right) as gray bars whose minimum and maximum correspond to the  difference between each of the 95\% boundaries of the posterior and the injected $G\mu$ relative to the latter.
    }
\label{fig:keynote_accuracy_precision_reconstruction}
\end{figure}

{\bf Results.}-- Fig.~\ref{fig:keynote_accuracy_precision_reconstruction} shows the reconstruction precision and accuracy by LISA of the BOS template with Saikawa-Shirai DOF. In the top-left panel, we show the precision in  \% of 
$\log G\mu$. 
For very weak signals (lowest values of $G\mu_{\rm inj}$) both precision and accuracy improve when the signal becomes stronger. We observe that while the accuracy indicator improves monotonically up to $G\mu_{\rm inj} \simeq 10^{-16}$, for larger values it stops improving and develops a random pattern reflecting the statistical fluctuations of the mock data. For this reason we do not track the accuracy further in our study, as it fluctuates with each realization. We note however that Ref.~\cite{Blanco-Pillado:2024aca} reported the reconstruction quality using a similar notion of accuracy, where the mean of the posterior is employed instead of mode in Eq.~\eqref{eq:accuracy_recon} (we have checked of course that the accuracy forecast using the mean value is equally prone to statistical fluctuations). The precision, on the other hand, improves monotonically until it reaches a minimum around 
$G\mu_{\rm inj} \gtrsim 10^{-12}$, and then settles down and even grows slightly (more on this below). We will use therefore, from now on, the precision as the metric to quantify the quality of reconstruction of a signal. If the reconstruction is however bad, say with precision larger or way larger than $100\%$, then we will use instead $Q$ as the indicator of the quality of reconstruction (this will be the case in several scenarios presented in Sect.~\ref{sec:recons_beyond}).

\begin{figure}[t!]
    \centering
\includegraphics[width=0.5\textwidth]{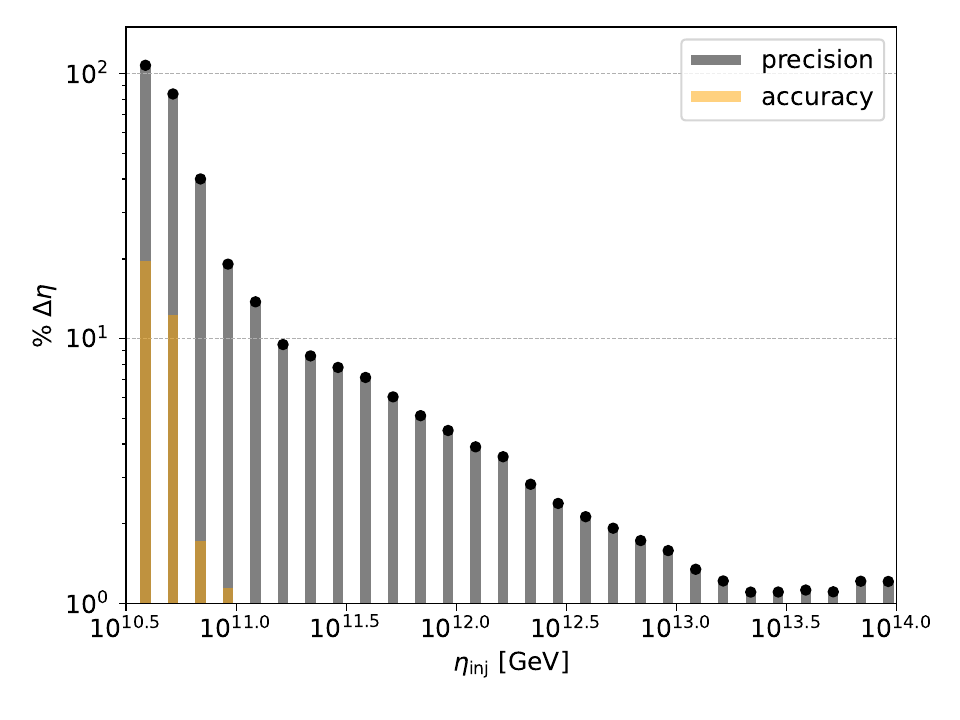}
\\[-1em]
    \caption{For BOS template with Saikawa-Shirai dof, the reconstruction precision in the energy scale of the symmetry breaking $\eta$ for each true signal with $\eta_{\rm inj}$ is obtained from Fig.~\ref{fig:keynote_accuracy_precision_reconstruction}-top-right and Eq.~\eqref{eq:tension}.  The gold bars show the reconstruction accuracy.
    }
\label{fig:keynote_accuracy_precision_eta}
\end{figure}

We find that the adopted SBI technique can reconstruct the cosmic-string GW signal with precision in $\Delta\log G\mu$ as $\lesssim 5\%$ for $G\mu_{\rm inj} \gtrsim 10^{-17}$, $\lesssim 1\%$ for $G\mu_{\rm inj} \gtrsim 10^{-16}$, and $\lesssim 0.1\%$ for $G\mu_{\rm inj} \gtrsim 10^{-13}$.
While this precision in $\log G\mu$ looks rather good, we want to stress that it is somewhat misleading to think that such small numbers reflect directly the quality of reconstruction of the signal, like {\it e.g.}~the uncertainty in $h^2\Omega_{\rm GW}$ shown in Fig.~\ref{fig:recon_conven_spec}. The precision in the amplitude of the GWB spectrum relates more closely to the precision in $G\mu$, as the background amplitude depends parametrically on some soft-power of $G\mu$, {\it e.g.}~$\Omega_{\rm GW} \propto (G\mu)^{1/2}$ for loops produced and emitted during RD. As the relation between $\Delta X$ and $\Delta \log X$ is not linear, {\it cf.}~Eq.~(\ref{eq:logX_X_precision_relation}), it seems more appropriate to really assess the quality of reconstruction of the signal by means of the precision of $G\mu$, and not of $\log G\mu$. This circumstance is clearly appreciated in Fig.~\ref{fig:keynote_accuracy_precision_reconstruction}, where the top-right panel shows that the reconstruction precision in $G\mu$ is in reality worse than suggested by the small numbers characterizing the precision of $\log G\mu$ in the top-left panel. For instance, $\Delta G\mu \lesssim 200\%$ for $G\mu_{\rm inj} = 10^{-17}$, $\lesssim 30\%$ for $G\mu_{\rm inj} = 10^{-16}$, and $\lesssim 5\%$ for $G\mu_{\rm inj} = 10^{-13}$, are much larger in comparison to the corresponding precisions quoted before for $\log G\mu$ for the same injected tensions, $\lesssim 5\%$, $\lesssim 1\%$, and $\lesssim 0.1\%$, respectively. Similarly as in the case of $\log G\mu$, we note again that the accuracy of $G\mu$ [{\it c.f.}~Eq.~(\ref{eq:accuracy_recon})] improves monotonically up to $G\mu_{\rm inj} \simeq 10^{-16}$, but becomes a random variable for larger tensions, depending on the realization of the mock data. The random pattern of the accuracy of $G\mu$ for large tensions is however not shown in  Fig.~\ref{fig:keynote_accuracy_precision_reconstruction}-top-right, simply because the fluctuations are all smaller than $1\%$, and hence they are cut in the plot (recall however the randomness shown for $\log G\mu$ in Fig.~\ref{fig:keynote_accuracy_precision_reconstruction}-top-left). Hence, we reinforce again the notion that accuracy is not a useful reconstruction indicator.

Furthermore, we also note that the uncertainty in $G\mu$ can be directly translated into that of the symmetry breaking scale $\eta$, $\Delta G\mu = 2 \Delta \eta$, using Eq.~\eqref{eq:tension}. For example, LISA can reconstruct a signal with  $\eta_{\rm inj} \gtrsim 2\times 10^{11} \, {\rm GeV}$ with a precision $\Delta \eta \lesssim 10\%$. The reconstruction precision for energy scales (tensions) injected between $\eta_{\rm inj} \sim 3\times 10^{10}$ GeV and $\eta_{\rm inj} \sim 10^{14}$ GeV, can be appreciated in Fig.~\ref{fig:keynote_accuracy_precision_eta}, reaching $\Delta \eta \lesssim 2\%$ for $\eta_{\rm inj} \gtrsim 5\times 10^{12} \, {\rm GeV}$.

The lower panel of Fig.~\ref{fig:keynote_accuracy_precision_reconstruction} also provides the distribution of the relative reconstruction uncertainty with respect to each $G\mu_{\rm inj}$.
The upper makers are calculated from the upper boundaries of each 95\% CL posterior, while the lower bound comes from the lower limit.
The sizes of these gray bars are essentially the values of the precision amplitudes shown in the top panels. The distribution of the uncertainty shows that the reconstruction posteriors are not exactly symmetric around the injected values, though the level of asymmetry depends on the realization.  

Another visible feature in Fig.~\ref{fig:keynote_accuracy_precision_reconstruction} is that the precision reaches a minimum around $G\mu \simeq 5\cdot 10^{-12}$ before degrading ({\it i.e.}~increasing) for larger values. As we argue next, we believe this arises from the data generation process. In the context of Bayesian inference, if we consider a given dataset and a fitting function of some parameters $\vec{\theta}$, the corresponding posterior width is partially controlled by the data variance, {\it i.e.}~a larger variance leads to wider posteriors. For $G\mu_{\rm inj} \gtrsim 10^{-12}$, the GW signal starts dominating over the LISA noise, see Fig.~\ref{fig:recon_conven_spec}. The data variance is thus determined by the signal amplitude $\Omega_{\rm GW}$, see Eq.~\eqref{eq:data} in Appendix~\ref{app:LISA_noise}. 
For a larger $G\mu_{\rm inj}$, the signal becomes larger, and the data variance increases correspondingly. As it turns out in our case, for this signal-dominated regime the posterior width grows mildly with $G\mu_{\rm inj}$, resulting in the observed degradation of the precision for tensions above $G\mu \simeq 5\cdot 10^{-12}$ (numerically the growth actually scales close to linearly with $\log G\mu_{\rm inj}$). This feature is therefore also expected to be present when reconstructing the tension parameter for beyond-conventional templates with amplitude of the GWB signal larger than the conventional template amplitudes for the same tension. This is the case e.g.~of the LRS modeling ({\it cf.}~Fig.~\ref{fig:LRS_spec}), which overtakes the LISA noise at much lower tensions, thus exhibiting this degradation effect of the reconstruction precision more prominently, as it starts at lower tensions (this is clearly shown in Fig.~\ref{fig:recon_precision_LRS}).

In summary, we see that LISA is  capable of reconstructing the conventional templates with great precision.
If we consider a signal with $G\mu \gtrsim 10^{-15}$, LISA can reconstruct it with a precision of $\Delta G\mu \lesssim 20\%$, which is translated {\it e.g.}~for radiation-loops to $\Delta \Omega_{\rm GW} \sim \Delta G\mu/2 \lesssim 10\%$. This range of precision falls below the difference between the spectra predicted by VOS and BOS models, as shown in the right panel of Fig.~\ref{fig:standard_spectra}.
However, as we shall show in Sect.~\ref{subsec:ModelComparisonVOSvsBOSvsDOF}, the decisive distinction by LISA between the two modelings requires yet a more precise reconstruction, of the order $\Delta G\mu \sim 3\%$, which becomes only realized  for signals with tensions as large as $G\mu \gtrsim 10^{-13}$. We compare and contrast the BOS and VOS models more quantitatively and systematically in Sect.~\ref{subsec:ModelComparisonVOSvsBOSvsDOF}.

Before moving on, we discuss briefly the third quantity of reconstruction, the quality indicator $\mathcal{Q}$ defined in Eq.~\eqref{eq:quality_indicator}.
The reconstruction results for conventional templates in Fig.~\ref{fig:keynote_accuracy_precision_reconstruction} leads to $100 \times \mathcal{Q}(\log G\mu) \approx 13\%$ for $G\mu_{\rm inj} = 10^{-17}$ and $100 \times \mathcal{Q}(\log G\mu) \lesssim 1\%$ for $G\mu_{\rm inj} \gtrsim 10^{-16}$. As discussed above, the signal with $G\mu_{\rm inj} = 10^{-17}$ cannot be reconstructed well ({\it i.e.}~only the order of magnitude of $G\mu$ is reconstructible). The quality indicator will be useful in Sect.~\ref{sec:recons_beyond}, where the degeneracies between model parameters are manifest, preventing the precision from being a good measure of the reconstruction result. As we shall see, such cases will have systematically $100 \times \mathcal{Q} \gtrsim 10\%$.

Continuing with our study of conventional signals, we compare next the SBI result to common detection techniques in the literature, including comparison against signal-to-noise ratio arguments (Sect.~\ref{subsec:sbi_vs_snr}) and the mandatory consistency check with Markov-Chain Monte Carlo method (Sect.~\ref{subsec:sbi_vs_MCMC}), highlighting whenever appropriate the advantage(s) of the SBI technique. 

\begin{figure}[t!]
    \centering
    \includegraphics[width=0.495\textwidth]{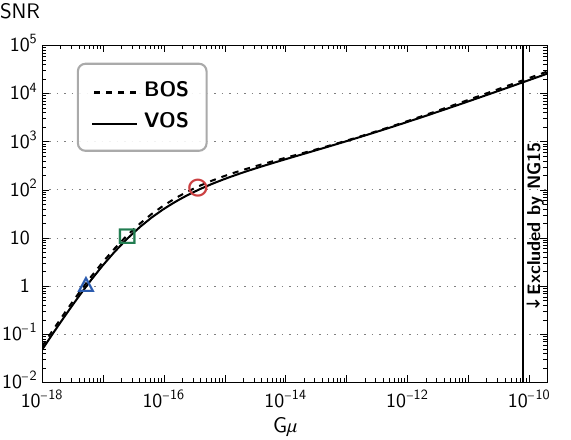}\hfill
    \includegraphics[width=0.495\textwidth]{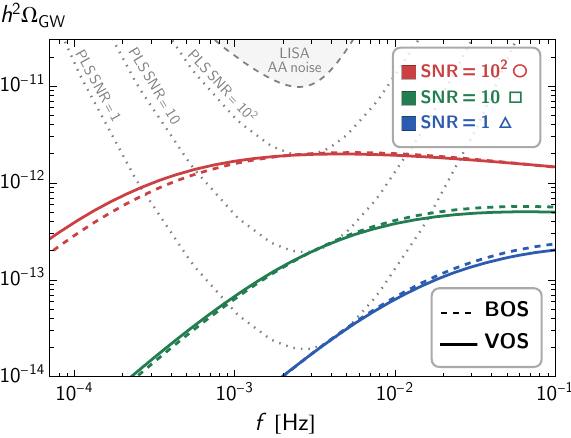}\\[-1em]
    \caption{\emph{Left Panel:} SNR at LISA based on Eq.\,\eqref{eq:SNR_definition}, for the GWB signals from the two conventional models, VOS (solid) and BOS (dashed). The region on the right of the black solid vertical line is excluded by NANOGrav 15-year data \cite{NANOGrav:2023hvm}. The benchmark points $\triangle$,  $\square$,  \raisebox{0.2ex} {$\ocircle$} correspond to the GWB spectra with SNR $\simeq 1,\,10,\, 100$. \emph{Right Panel:} The GWB spectra correspond to the benchmark points in the left panel. The gray dashed curve shows the LISA noise curve of the AA channel, and the dotted gray lines are the corresponding PLS curves.}
    \label{fig:standard_spectra_SNR}
\end{figure}

\subsubsection{SBI versus Signal-to-Noise Ratio (SNR)}
\label{subsec:sbi_vs_snr}

A convenient method in the literature for claiming detectability of a signal is to use the signal-to-noise ratio (SNR) criterion. For a GW signal with amplitude $\Omega_{\rm GW}$ observed at some detector with noise  $\Omega_{\rm noise}$, and observation duration $T_{\rm obs}$, SNR is defined by
\begin{align}
    {\rm SNR} = \sqrt{T_{\rm obs}\int_{f_{\rm min}}^{f_{\rm max}} df \left[\frac{\Omega_{\rm GW}(f)}{\Omega_{\rm noise}(f)}\right]^2},
    \label{eq:SNR_definition}
\end{align}
where $T_{\rm obs} = 3$ years for LISA, we take the AA channel for $\Omega_{\rm noise}(f)$, and $f_{\rm min}$ ($f_{\rm max}$) is the minimum (maximum) of the detector's frequency range (for LISA, $f_{\rm min} = 30 ~ \mu{\rm Hz}$ and $f_{\rm max} = 0.5 ~ {\rm Hz}$, respectively). We note that while formula~\eqref{eq:SNR_definition} is often used in literature, it is strictly only valid in the weak-signal regime ($\Omega_{\rm GW} < \Omega_{\rm noise}$) \cite{Romano:2016dpx}. We will use it to calculate the SNR as we aim to compare with other works, but we stress that for a generic SNR computation, the denominator receives an additional contribution from the variance of the signal itself~\cite{Romano:2016dpx, Brzeminski:2022haa}.

Fig.\,\ref{fig:standard_spectra_SNR}--left shows the SNR of VOS and BOS templates. A signal with $G\mu \gtrsim 3 \times 10^{-17}$ has ${\rm SNR}\,>10$, which is often the threshold used in the literature to claim detectability in LISA. However, as shown by the realistic reconstruction in Fig.~\ref{fig:keynote_accuracy_precision_reconstruction}-top-right, the reconstruction error ({\it i.e.}~the precision) in the tension is $\Delta G\mu \sim 100\%$. That is, only the order of magnitude of the signal/tension can be determined, not the precise amplitude/value. Signal reconstructions with better precision, say $\lesssim 10\%$ and $\lesssim 5\%$ in $G\mu$, are obtained for backgrounds with $G\mu_{\rm inj} \gtrsim 10^{-14}$ and $\gtrsim 10^{-13}$, respectively. These correspond to ${\rm SNR} \gtrsim 5\cdot 10^2$ and $\gtrsim 10^3$, respectively. Unfortunately, there is no simple correspondence between $\Delta G\mu$ and SNR, other than observing than increasing SNR implies typically a smaller value of $\Delta G\mu$, {\it i.e.}~a better reconstruction. From Fig.~\ref{fig:keynote_accuracy_precision_reconstruction}--top-right and Fig.~\ref{fig:standard_spectra_SNR}-left, we can infer from fitting our results that $\Delta G\mu \simeq 6/{\rm SNR}$ for $5\cdot 10^{-18} \lesssim G\mu \lesssim 5\cdot 10^{-16}$, whereas $\Delta G\mu \simeq 22/{\rm SNR}$ for $5\cdot 10^{-16} \lesssim G\mu \lesssim 5\cdot 10^{-12}$. For $G\mu \gtrsim 5\cdot10^{-12}$ the trend $\Delta G\mu \propto 1/{\rm SNR}$ does not even hold. This is likely due to the degradation of the reconstructed precision for very high tensions in the signal-dominated regime -- see explanation in Sect.~\ref{subsec:quality_recon_conven} --, and also possibly due to the fact that the SNR calculation given by Eq.~(\ref{eq:SNR_definition}) becomes invalid in the said regime.

The usage of SNR as a detection criterion does not describe therefore the capability of signal or parameter reconstruction, 
as {\it e.g.} the precision and confidence levels we  discussed in Sect.\,\ref{sec:recon_result_conven}. Furthermore, the SNR cannot be used for distinguishing between signals from different templates, because Eq.~\eqref{eq:SNR_definition} integrates over the GWB spectrum and hence loses  information about the spectral different details among them. 

{\bf Power-Law Integrated Sensitivity (PLS) curve.}
From the SNR definition \eqref{eq:SNR_definition},
one can still introduce a graphical tool which has been commonly used in the literature. This is the so-called power-law integrated sensitivity (PLS) curve~\cite{Thrane:2013oya}, which represents a (spectral) threshold above which a signal with a power-law shape, {\it i.e.}~$\Omega_{\rm GW}(f) = \Omega_\beta\left({f}/{f_{\rm ref}}\right)^\beta$ with $\Omega_\beta$ the amplitude at the reference frequency $f_{\rm ref}$, and $\beta$ a spectral index, can be detected with given SNR (after some time of observation $T_{\rm obs}$).
To be precise, the PLS curve is defined as 
\begin{align}
\Omega_{\rm PLS}(f) \equiv&\, {\rm max} \left[f^\beta \frac{{\rm SNR}}{\sqrt{T_{\rm obs}}}\left(\int_{f_{\rm min}}^{f_{\rm max}} df \left[ \frac{f^\beta}{\Omega_{\rm noise}(f)} \right]^2 \right)^{-1/2}\right]\,.\label{eq:PLS_curve}
\end{align}
For the LISA PLS curves shown in this paper, we use $\beta \in [-5,5]$, $T_{\rm obs} = 3$ years, and the noise curve $\Omega_{\rm noise}(f)$ from the AA channel [Eq.\,\eqref{eq:Omeganoise}]. The minimum and maximum frequencies are $f_{\rm min} = 3 \cdot 10^{-5}$ Hz and $f_{\rm max} = 0.5$ Hz, respectively. Fig.\,\ref{fig:standard_spectra_SNR}--right shows LISA PLS curves and the signals from VOS and BOS templates that sustain SNR $= 1, 10, 100$, matching the selected signals from Fig.\,\ref{fig:standard_spectra_SNR}--left. 
One can observe clearly that conventional cosmic string spectra `touching' the PLS of given SNR threshold, have precisely such value of SNR, without having to re-evaluate the SNR integral~\eqref{eq:SNR_definition}. Unlike our SBI technique, this method has however no further information about the real reconstruction capabilities, such as the confidence levels of detection, or quantification of whether different signals (say with similar SNR) can be distinguished and/or preferred given a data set. We stress therefore that while PLS curves have been useful in the past to obtain graphically a rapid assessment of the detectability of signals in LISA, at the end of the day they only provide an estimation\footnote{The estimation could be even somewhat misleading for signals with frequency profiles very different from a power-law within the LISA sensitivity, though this is not the case for conventional cosmic string GWBs, as these are ``close enough'' to a power law within the LISA sensitivity range.} of the SNR, but not a quality reconstruction assessment of the signal itself, nor of the parameters that determine it.

\subsubsection{SBI versus Markov chain Monte Carlo (MCMC)}
\label{subsec:sbi_vs_MCMC}

We compare now the results from our SBI technique against more traditional MCMC-based reconstruction methods. This comparison enables us to test the goodness and efficiency of our procedure. However, as we noted in~\cite{Dimitriou:2023knw}, it does not constitute a proper consistency check of our method, since MCMC requires an explicit evaluation of the likelihood, which in general is implicit in the simulator. See appendix \ref{app:mcmc} for more details on this. 

Adopting the public MCMC library {\tt emcee}\footnote{with 6 random walkers, $nburn = 500$, $nsteps = 1000$,
$rconv = 10^{-3}$, and a maximum of 50 iterations which results in generating $\mathcal{O}(10^{4})$ samples} \cite{Foreman-Mackey_2013}, Fig.~\ref{fig:recon_bos_potato} shows the posterior of the BOS-template reconstruction from MCMC technique (blue curves).
We see that they agree well with our results using SBI method (red curves), while the latter tends to give marginally wider posteriors, as can be seen in the 2D contours. Nonetheless, we would like to emphasize that MCMC requires a fresh set of simulations to make parameter inference given a newly observed dataset, while a pre-trained, amortized SBI algorithm, does not. This is a crucial advantage motivating the use of SBI in our analysis.

\subsection{Model Comparison: VOS vs BOS vs SM-dof}
\label{subsec:ModelComparisonVOSvsBOSvsDOF}

As we have shown, LISA can reconstruct conventional  cosmic-string GW signals with precisions than range from $\Delta G\mu \sim 100\%$ for small tensions $G\mu \sim 10^{-17}$, down to $\Delta G\mu \simeq 2-3 \%$ for large tensions $G\mu \gtrsim 10^{-12}$. Two questions are now in order: 

{\it I)} \emph{If there was a GWB data set generated from the detection of a signal by LISA, say corresponding to a concrete cosmic-string model $M_{T}$, 
could we discriminate between this true model $M_{T}$ and another rival cosmic-string model $M_{R}$?} 

{\it II)} \emph{What if the data comes from a true -unknown- model $M_{T}$, different from the two models we are comparing against it?}

In our analysis we adopt the Bayesian model comparison framework, for which the evidence of the models needs to be computed somehow. A likelihood-based approach such as MCMC will require evaluating the likelihood (and thus performing a simulation) for every step of the numerical integration. On the other hand, our model comparison strategy with SBI leads to an optimized approximation of the true likelihood, which does not require further simulations to be evaluated. In that sense, the calculation of the evidence is computationally more efficient in SBI than in MCMC. 

We refer the reader to appendix \ref{app:model_comparison} for the the -standard- mathematical formalism corresponding to the Bayesian model comparison, adressing both questions \#1  and \#2 above. Note, however, that the latter is relatively less common in the literature, even though arguably much more useful, since typically the true data-generating model will be different from the models we want to compare with each other.

The ability for the models to be optimally discriminated depends on two properties: \emph{1) the difference between their spectra}, and \emph{2) their reconstruction uncertainty (quantified by either the precision or the quality indicator)}. 
Whenever the precision level is much larger than the spectral signal differences, it becomes impossible to draw any conclusion on the model comparison, as the information about the signal spectral differences cannot be retrieved due to the large uncertainty.

As an interesting example, we compare first the conventional templates VOS and BOS, and quantify LISA's ability to discriminate between them for a given tension. Following, we also explore whether LISA can distinguish the uncertainty in the conventional GWB spectra due to the different modeling of DOF evolution, as discussed in Sect.~\ref{sec:modeling_DOFs}.

\begin{figure}[t!]
    \centering
    {\sffamily Model comparisons between conventional VOS and BOS templates.}
\\[0.25em]   

\includegraphics[width=0.65\textwidth]{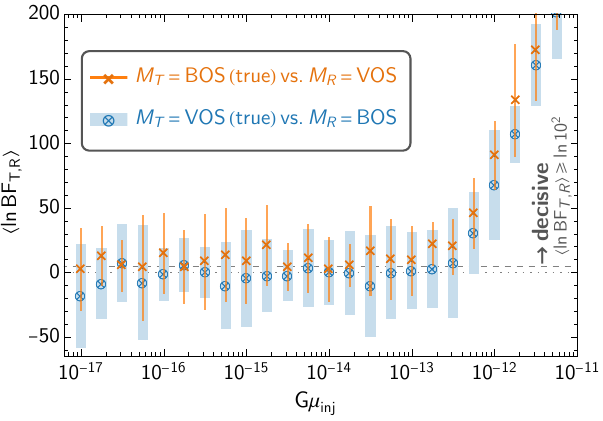}\\[-1em]
\caption{LISA's ability in discriminating different conventional templates of cosmic-string GWB is shown by the averaged Bayes factors $\langle \ln {\rm BF}_{T,R}\rangle$, calculated from Eq.~\eqref{eq:average_BF} with the method in Fig.~\ref{fig:flowchart}-right. Assuming the signal of the true model $M_T$ is from the BOS template with $G\mu_{\rm inj}$, each orange cross is the mean value of the averaged BF (sampled from 20 datasets) for the BOS model versus  the VOS template as its rival $M_R$, and the orange bar indicates the standard deviation of the averaged BF values. When the true model is instead the VOS template, each blue cross-circle and bar correspond to the mean and standard deviation of the averaged BF for VOS versus BOS templates, respectively. The horizontal gray dotted line corresponds to $\langle \ln {\rm BF}_{T,R} \rangle = 0$, which means both true and rival models are preferred equally by the data. Above the dashed gray line, $\langle \ln {\rm BF}_{T,R} \rangle \geq \ln 10^{2}$ means LISA can decisively discriminate the true model from the rival model.}
\label{fig:bayes_vos_bos}
\end{figure}

\vspace*{0.3cm}
\paragraph{Which conventional modeling, VOS or BOS ?} Following the method described in the flowchart~\ref{fig:flowchart}--right and discussed in Appendix~\ref{app:model_comparison}, we create simulated data where the true signal from model $M_T$ is in turn either the VOS template from Sect.~\ref{sec:GW_semi_analytic}, or the BOS template from Sect.~\ref{sec:GW_simulated}, for a given $G\mu_{\rm inj}$. Then we test the data against both models, and calculate the Bayes factor ${\rm BF}_{T,R}$ given in Eq.~\eqref{eq:bayes_factor_definition}.
We repeat the process for 20 sampled datasets and calculate the averaged logarithmic Bayes factor $\langle \ln {\rm BF} \rangle$ using Eq.~\eqref{eq:average_BF}.

Fig.~\ref{fig:bayes_vos_bos} shows LISA's ability in differentiating these two conventional templates.
The orange data points correspond to the case when the true signal is built from the BOS template, while the blue points correspond to having the VOS template as the true model.
We see that, in both cases, LISA can decisively distinguish between BOS and VOS models, {\it i.e.}~$\langle \ln {\rm BF}_{T,R}\rangle \gtrsim \ln 10^{2}$, when the true signal has $G\mu_{\rm inj} \gtrsim 5\times 10^{-13}$.
From Fig.~\ref{fig:keynote_accuracy_precision_reconstruction}-top-right, this value of $G\mu_{\rm inj}$ corresponds to the reconstruction precision of $\Delta G\mu \lesssim 3\%$.
By propagating the error by simply assuming $\Omega_{\rm GW} \propto \sqrt{G\mu}$ (for the spectrum from loop produced and emitting GW in radiation era), LISA can decisively discriminate between BOS and VOS models if the GWB spectra can be reconstructed with a precision of $\Delta \Omega_{\rm GW} \sim \Delta G\mu/2 \lesssim 1-2\%$.
We show in Fig.~\ref{fig:standard_spectra}--right that the relative difference between VOS and BOS spectra inside the LISA frequency window of the order of 10\% for $G\mu \gtrsim 10^{-18}$. This current example suggests that the decisive model-comparison ability of LISA requires the reconstruction precision roughly a factor $\sim 5-10$ smaller than the difference between the spectra of the two templates.

\begin{figure}[t!]
    \centering
    {\sffamily Model comparisons between the BOS templates with different SM dof evolutions.}
\\[0.25em]   
    \includegraphics[width=0.65\textwidth]{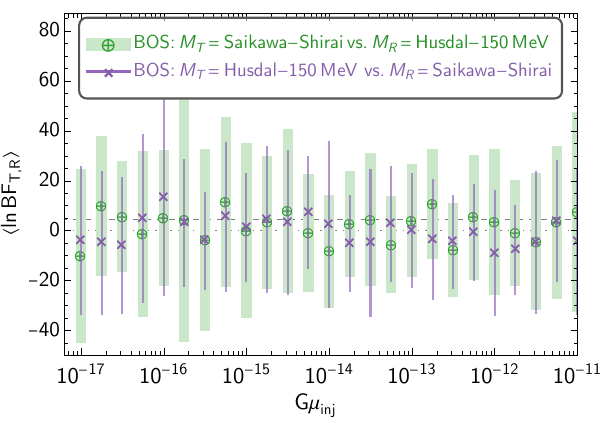}\\[-1em]
    \caption{LISA's ability in distinguishing  different SM dof evolutions (see Figs.~\ref{fig:g_gs} and \ref{fig:spectra_diff_g_gs}) is shown by the averaged BF $\langle \ln {\rm BF}_{T,R}\rangle$ in Eq.~\eqref{eq:average_BF}. Assuming the BOS template as the true model $M_T$ with $G\mu_{\rm inj}$, each green plus-circle is the averaged BF (sampled from 20 datasets) of the Saikawa-Shirai dofs versus the Husdal dofs with the QCD phase transition at 150 MeV, where the former is the true model. The green data points correspond to the opposite case where the BOS template with Husdal dofs is the true model. The error bars correspond to the standard deviations of the averaged BF. We see that both cases have their averaged BF fluctuate around zero; thus, it is impossible to distinguish the two SM dof evolutions from cosmic-string GWB with LISA.
    }
    \label{fig:bayes_bos_dofs}
\end{figure}

\vspace*{0.3cm}
\paragraph{Which DOF evolution?} In  Fig.~\ref{fig:spectra_diff_g_gs} of Sect.~\ref{sec:modeling_DOFs}, we showed how different modelings of the SM-DOF evolution lead to slightly different GWB spectra with $\mathcal{O}(1\%)$ relative differences in $\Omega_{\rm GW}$ among them. Fig.~\ref{fig:bayes_bos_dofs} shows that LISA cannot distinguish these modelings of SM-DOF evolutions, as the averaged Bayes factors are fluctuating around zero for all tensions. We only show the comparison between the DOF evolutions of the Saikawa-Shirai modeling and the Husdal modeling with QCD phase transition at 150 MeV, as this is the case where the GW spectra differ by the largest amount compared to other Husdal modeling, recall Fig.~\ref{fig:spectra_diff_g_gs}. Therefore, we expect that this case has the largest BF value for each $G\mu_{\rm inj}$, and yet we clearly observe that LISA cannot distinguish the two different DOF schemes.

The main reason for failing to discriminate between DOF modelings is that the signal reconstruction is not precise enough to distinguish the $\mathcal{O}(1\%)$ difference between the theoretical spectra. We just saw before that when comparing VOS and BOS modelings, a decisive differentiation between them required roughly $\sim 5-10$ times smaller precision than the difference between spectral signal amplitudes.
As the best reconstruction precision in $G\mu$ is $2-3\%$, this is not enough for differentiating among DOF modelings. We would need an improvement of precision by close to an order of magnitude, but this is not possible due to the degradation -- explained in Sect.~\ref{subsec:quality_recon_conven} --  of the reconstruction-precision for very high tensions in the signal-dominated regime.

\section{Reconstruction of beyond-conventional templates at LISA}
\label{sec:recons_beyond}

\begin{figure}[t!]
    \centering
    {\sffamily Templates with varying loop number densities}\\[0.25em]
    \includegraphics[width=0.485\linewidth]{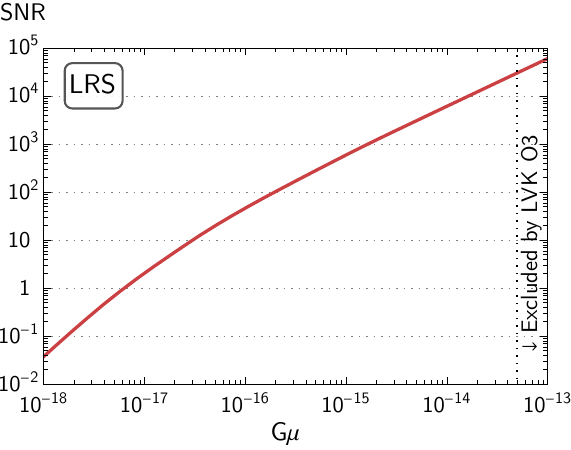}\hfill
    \includegraphics[width=0.47\linewidth]{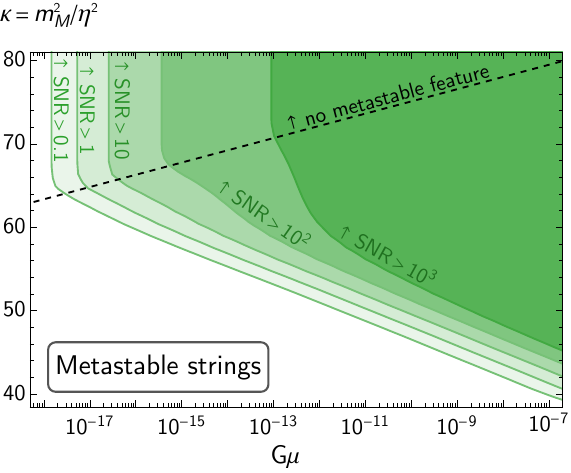}\\[-0.25em]
    \includegraphics[width=0.485\linewidth]{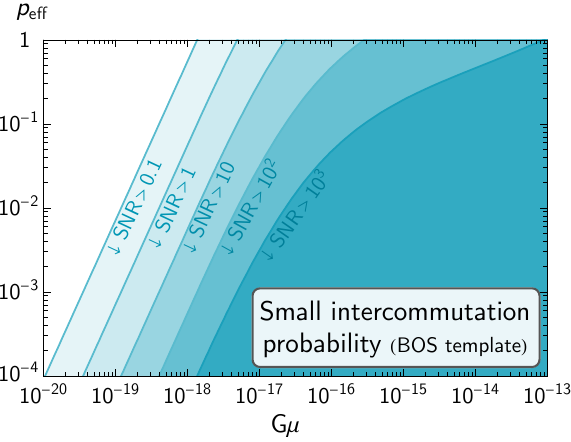}\hfill
    \includegraphics[width=0.48\linewidth]{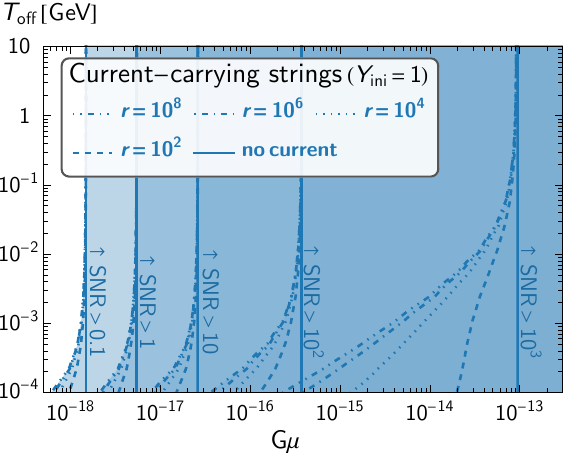}\\[-1em]
    \caption{SNR forecast [using Eq.~\eqref{eq:SNR_definition}] of GW signal observed at LISA in the beyond-conventional cosmic-string templates, where the loop number density gets modified; cf. Sect.~\ref{subsec:LoopNumberDensityBSM}.}
    \label{fig:snr_beyond_1}
\end{figure}

As demonstrated in the previous section for cosmic-string conventional signals, the SBI technique enables us to quantify precisely the ability of LISA to reconstruct cosmic-string GWB templates and to perform a comparison among models. Now we explore these two aspects for the beyond-conventional templates introduced in Sect.~\ref{sec:beyondConventional}, which are more complicated due to their multi-parameter dependencies (an exception of this is the LRS model, which depends only on $G\mu$). Table~\ref{tab:summary_templates_beyond} summarizes the parameters of each template and the ranges of priors used for our analysis.
In Sects.~\ref{subsec:recon_loop_number}, \ref{subsec:recon_explore_cosmic_hist}, and \ref{subsec:recons_loopProperties}, we present the reconstruction results for each of the three classes of beyond-conventional templates, depending on whether we consider, respectively, modifications of {\bf a)} the loop number density (LRS, super,  metastable, current-carrying strings), {\bf b)} the expansion history (non-standard cosmologies, extra degrees of freedom, either thermal or secluded), 
or {\bf c)} the loop properties (birth length, power emission), {\it c.f.}~Fig.~\ref{fig:template_summary}. While our major aim in this section is to provide a precise parameter-reconstruction analysis, as a starting point and for comparison purposes we also provide the SNR forecasts of each template, see Figs.~\ref{fig:snr_beyond_1}, \ref{fig:snr_beyond_2}, and \ref{fig:snr_beyond_3}. Lastly, in Sect.~\ref{sec:model_comparison_beyond_conventional}, we discuss LISA's ability to discriminate between conventional and beyond-conventional templates.

\paragraph{Signal-reconstruction ability for multi-parameter templates.}
The quantification of the reconstruction ability for  beyond-conventional templates can be done similarly to that for the conventional templates discussed in   Sect.~\ref{sec:recon_result_conven}. The reconstruction precision of each model parameter is obtained by using Eq.~\eqref{eq:precision_recon} with the posterior probability distribution of that parameter, marginalized over all other model and noise parameters. The marginalization is needed as we do not have \textit{a priori} knowledge about the true parameter values of an observed signal\footnote{One could of course also consider particular physics cases where some of the template parameters are fixed to  fiducial values. The analysis for such cases would be simpler than what we present in this work, as there would be less free parameters. While the methodologies would be very similar, we expect the results to be however quite different, as fixing some of the parameters would help to break degeneracies, allowing for a better reconstruction precision on the remaining uncertain parameters.}.  

For the single-parameter templates considered in Sect.~\ref{sec:recon_result_conven}, a signal with larger SNR acquires generally a better reconstruction precision\footnote{This is not true however in the signal dominated region, recall discussion in  Sect.~\ref{subsec:quality_recon_conven}. .}. This is however not guaranteed for multi-parameter templates, as their parameter reconstruction can suffer two difficulties: 
\begin{enumerate}[\it i)]
    \item \underline{Degeneracy between parameters.} There can be various sets of model parameters or a direction in the model's parameter space that lead to similar GW signals in a detector's sensitivity window; this also means that they have similar SNR. The simplest example is the case of power-law signal: $\Omega_{\rm GW}(f) = A(f/f_{\rm ref})^\beta$. For a fixed $\beta$, the direction in the parameter space where $\Omega_{\rm GW}(f)$ remains the same is $A \propto f_{\rm ref}^\beta$. We will refer to this kind of relation between parameters as a \emph{degenerate direction}.\\
    {\it Effects on posterior and precision}: 
    Whenever a degenerate direction is present, the reconstruction posterior will be extended from the true-signal's (injected) parameter coordinates in the parameter space and elongated along such direction. Although the posterior could be confined within this direction, the uncertainty of each parameter---which is defined from the marginalized posterior---can become large, degrading the reconstruction precision. Furthermore, if such a degenerate direction extends too far, the posterior could be truncated by the prior range, making the resulting precision prior-dependent. 
    
    \item \underline{No observable feature.} When there is a region in parameter space leading to no observable feature in the spectrum for the LISA's frequency range. This effect can be viewed as the extreme case of having degeneracy across this whole region of parameter space.
    \\
    {\it Effects on posterior and precision}: Since the GWB spectrum is independent of such parameters within the LISA window, their posterior will spread widely over the prior volume. 
    The reconstruction precision in this case becomes prior-dependent and does not reflect the true capability of the detector. 
\end{enumerate}

When reconstructing a given parameter $X$, we will report whenever possible its precision $\Delta X$ in linear scale [{\it c.f.}~Eq.~\eqref{eq:precision_recon}], since this typically relates directly to the uncertainty of the underlying physics parameters.
In some cases where the degeneracy appears and the precision in linear scale becomes large, we might still opt to report the precision of such parameter in logarithmic scale. The precision can indeed suffer severely from the two aforementioned reconstruction degradations, when the posterior spreads widely over the prior volume. In such cases, we present our result using the quality indicator $\mathcal{Q}(X)$ [Eq.~\eqref{eq:quality_indicator}], describing the size of the recovered posterior relative to the prior range. Note that the quality indicator could be used as well to distinguish between the two types of degradations.
When no feature falls within the LISA window [point $ii)$ above], the posterior flattens over the prior range leading to $\mathcal{Q} \approx 0.95$, which is the maximal value allowed by definition. On the other hand, the presence of the degenerate direction [point $i)$ above] leads typically to a smaller $\mathcal{Q}$, as the  posterior only extends along such direction, but not necessarily covering the whole prior range.

\begin{table}[p!]
\centering
{\underline{\bf Beyond Conventional templates} \\[0.75em]
\begin{tabular}{lll}
\hline
Parameter & Description & Prior\\ \hline \\[-0.5em]
\multicolumn{3}{l}{\underline{\bf \emph{i)} Varying loop number density}} \\[0.5em]
\multicolumn{3}{l}{{\bf Including small loop population or LRS model}
 (section~\ref{sec:LRS_model})} \\[0.25em]
$G\mu$ & String tension & log--uniform$[10^{-18},10^{-9}]$ \\[0.5em]
\multicolumn{3}{l}{{\bf Smaller intercommutation probability} (section~\ref{subsubsec:super_string})} \\[0.25em]
$G\mu$ & String tension & log--uniform$[10^{-18},10^{-9}]$ \\
$p_{\rm eff}$ & Effective intercommutation probability & log--uniform$[10^{-3},1]$ \\
& ($\Omega_{\rm GW}^{p_{\rm eff}<1} = p_{\rm eff}^{-1} \Omega_{\rm GW}^{p_{\rm eff}=1}$) & \\[0.5em]
\multicolumn{3}{l}{{\bf Metastable cosmic strings} (section~\ref{subsubsec:metastable_strings})} \\[0.25em]
$G\mu$ & String tension & log--uniform$[10^{-18},10^{-9}]$ \\
$\kappa$ & Ratio of monopole-to-string scales & uniform$[40, 80]$ \\[0.5em]
\multicolumn{3}{l}{{\bf Current-carrying cosmic strings} (section~\ref{sec:current_carrying_string})} \\[0.25em]
$G\mu$ & String tension & log--uniform$[10^{-10} ~ 10^{-18}]$ \\
$Y_{\rm ini}$ & Initial current strength & $Y_{\rm ini} = 1$ for varying $G\mu$ \\
$T_{\rm off}$ [GeV] & Temperature when the current switches off & log--uniform$[10^{-4},10^{3}]$ \\
$r$ & Temperature ratio of when & log--uniform$[10^{-0.3},10^{10.7}]$ \\
 & the current switches on and off &  \\[0.5em]
\multicolumn{3}{l}{\underline{\bf \emph{ii)} Exploring cosmic histories}} \\[0.5em]
 \multicolumn{3}{l}{{\bf Nonstandard era after inflation (instantaneous transition)} (section~\ref{subsec:nonst_cosmo_after_inflation})} \\[0.25em]
$G\mu$ & String tension & log--uniform$[10^{-18},10^{-9}]$ \\
$T_{\rm end}$ [GeV] & Temperature when the nonstandard era ends & log--uniform$[10^{-3},10^{5}]$\\
$w$ & Equation of state parameter & uniform$[0,1]$  \\[0.5em]

 \multicolumn{3}{l}{{\bf Thermalized extra relativistic DOFs} (section~\ref{subsec:nonst_cosmo_thermalized_DOFs})} \\[0.25em]
$G\mu$ & String tension & log--uniform$[10^{-17},10^{-5}]$ \\
$\Delta g_*$ & Effective number of extra DOFs & log--uniform$[10^{-2},10^{6}]$ \\
$m$ [GeV] & Mass scale of the extra DOFs & log--uniform$[10,10^{10}]$ \\[0.5em]
 \multicolumn{3}{l}{{\bf Completely-secluded extra relativistic DOFs} (section~\ref{subsec:nonst_cosmo_dark_DOFs})} \\[0.25em]
$G\mu$ & String tension & log--uniform$[10^{-18},10^{-9}]$ \\
$\epsilon_{\rm BBN}$ & $\epsilon_{\rm BBN} = \left.g_{*D}(T_D)(T_D/T)^4\right|_{\rm BBN}$ & log--uniform$[10^{-4},10^{-0.1}]$ \\[0.5em]
\multicolumn{3}{l}{\underline{\bf \emph{iii)} Changing loop's properties}} \\[0.5em]
\multicolumn{3}{l}{{\bf Loop's GW emission power} (section~\ref{subsubsec:loop_emission_power})}\\[0.25em]
$G\mu$ & String tension & log--uniform$[10^{-18},10^{-9}]$ \\
$q$ & GW emission power (i.e., $P_{\rm GW}^j \propto j^{-q}$) & uniform$[1.1,2]$ \\[0.5em]
\multicolumn{3}{l}{{\bf Initial loop size} (section~\ref{sec:atypical_loop_size})}\\[0.25em]
$G\mu$ & String tension & log--uniform$[10^{-18},10^{-9}]$ \\
$\alpha$ & Initial loop size (in unit of time) & log--uniform$[5 \times 10^{-6},10^{-1}]$\\[0.25em]
\hline \\[-2em]
\end{tabular}
}
\caption{Table of parameters and priors of beyond-conventional cosmic-string GWB templates, discussed in Sect.~\ref{sec:beyondConventional}.
}
\label{tab:summary_templates_beyond}
\end{table}

Finally, when discussing 
in Sect.~\ref{sec:model_comparison_beyond_conventional} the ability to discriminate between models, we will chart regions of parameter spaces where LISA can distinguish beyond-conventional GWB spectra from conventional signals. 

\subsection{Varying the loop number density}
\label{subsec:recon_loop_number}

Fig.~\ref{fig:snr_beyond_1} shows the SNR forecast for the GW signals in the templates with varying loop number density, namely the LRS modeling, superstrings, metastable strings, and current-carrying strings, see Sect.~\ref{subsec:LoopNumberDensityBSM}. The prediction for the LRS model differs from the conventional case for all $G\mu$ values, because the additional small-loop population adds extra GW contributions, {\it c.f.}~Fig.~\ref{fig:LRS_spec}.
Other scenarios of this class, can however lead to GWB spectra that approach the prediction of conventional templates inside LISA's window, hence becoming featureless within such frequencies. This is the case of {\it e.g.}~metastable strings with large values of $\kappa$ above the black dashed line in Fig.~\ref{fig:snr_beyond_1}-top-right [determined by the equality in   Eq.~\eqref{eq:metastable_stable_network}], strings with small intercommutation probability evaluated when $p_{\rm eff} \to 1$, as shown in Fig.~\ref{fig:snr_beyond_1}-bottom-left, and current carrying strings with $r \to 1$ or $T_{\rm off} > 1$ GeV, shown in Fig.~\ref{fig:snr_beyond_1}-bottom-right. In the conventional-template limit, their SNR predictions do not depend on any parameter except $G\mu$, and thus the SNR contour lines become vertical in the right panels of Fig.~\ref{fig:snr_beyond_1}.

In the following, we present the reconstruction precision for each template parameter where we scan all possibilities of the injected true signal within the model's parameter prior ranges. 
As we shall see, some parameters cannot be reconstructed well in some parts of the parameter space due to the degeneracy and/or lack of observable feature within LISA, as discussed before.

\begin{figure}
    \centering
    {\sffamily LISA's reconstruction precision for LRS templates}\\
    \includegraphics[width=0.7\linewidth]{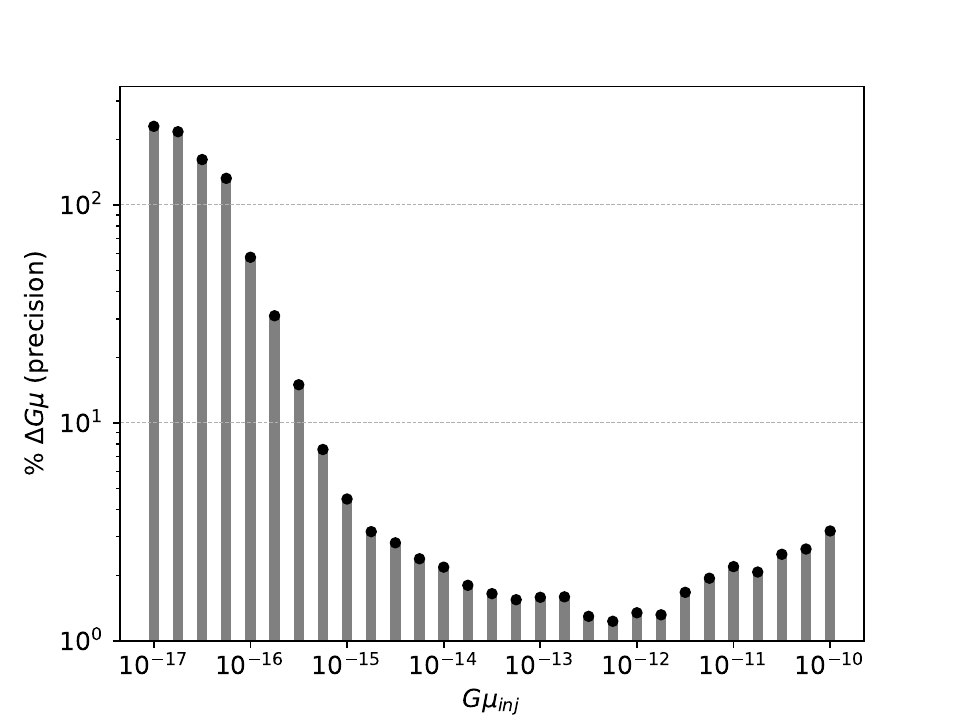}\\[-1em]
    \caption{The reconstruction precision [cf Eq.~\eqref{eq:precision_recon}] for $G\mu$ corresponding to the LRS template when the mock data has the true signal with $G\mu_{\rm inj}$, defined from the 95\% CL  interval of the reconstruction posterior.}
    \label{fig:recon_precision_LRS}
\end{figure}

\paragraph{Small-loop population(s): LRS model.}
The reconstruction precision of $G\mu$ for the LRS template (Sect.~\ref{sec:LRS_model}), is shown in Fig.~\ref{fig:recon_precision_LRS}.
The small-loop populations enhance the GWB signal, increasing the reconstruction quality at smaller $G\mu_{\rm inj}$ values than in the conventional templates, {\it c.f.}~Fig.~\ref{fig:keynote_accuracy_precision_reconstruction}.
We obtain a reconstruction precision of $G\mu$ $\lesssim 10\%$ for $G\mu_{\rm inj} \gtrsim 5\times 10^{-16}$ and $\lesssim 3\%$ for $G\mu_{\rm inj} \gtrsim 3 \times 10^{-15}$, corresponding to signals with ${\rm SNR} \gtrsim 500$ and $10^3$, respectively.
Interestingly, the precision degrades gradually for $G\mu_{\rm inj}\gtrsim 10^{-14}$, similarly to what happened in the conventional case shown in 
Fig.~\ref{fig:keynote_accuracy_precision_reconstruction}, but now earlier and more prominently. This happens once the GW signal starts dominating the LISA noise (c.f. Fig.~\ref{fig:LRS_spec}), consequently dominating the data variance (recall once again the discussion presented in Sect.~\ref{subsec:quality_recon_conven}).

\begin{figure}[t!]
    \centering
    {\sffamily Examples of reconstruction posteriors for superstring templates} \\[0.25em]
    \includegraphics[width=0.485\linewidth]{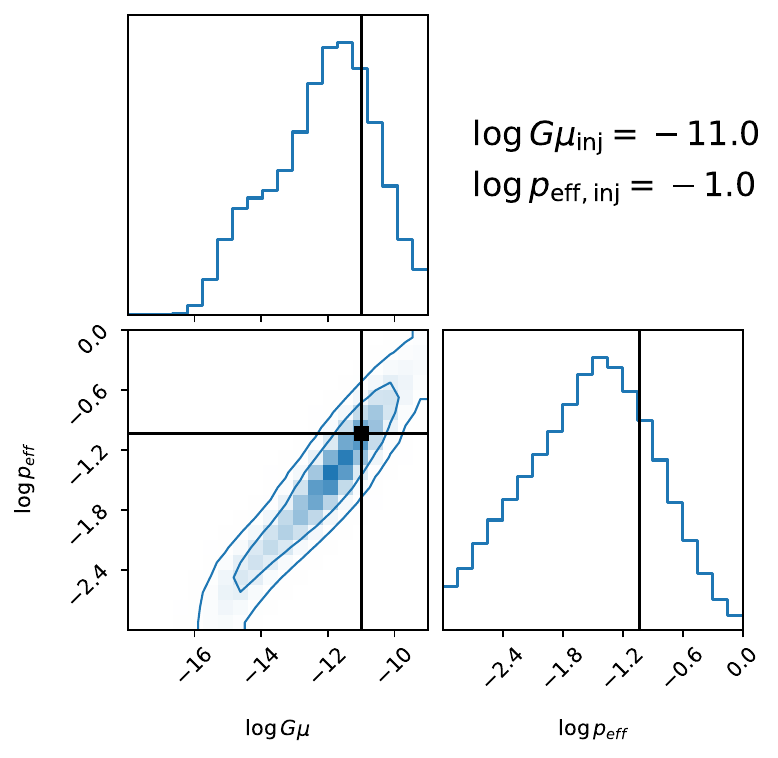}\hfill
    \includegraphics[width=0.485\linewidth]{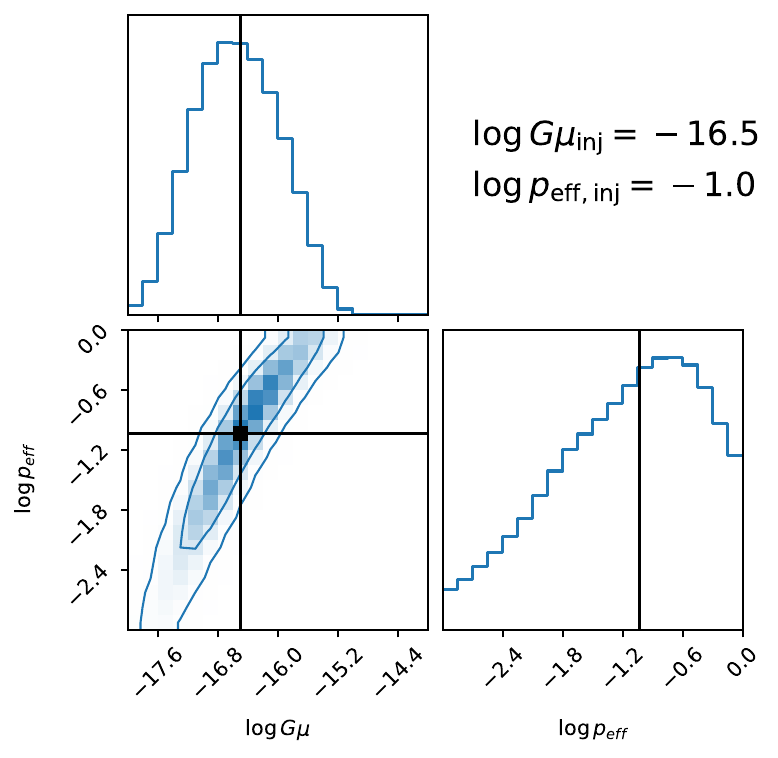}\\[-1em]
    \caption{
    Examples of the 2D reconstruction posterior in the $\{G\mu, p_{\rm eff}\}$ plane, for the template with small intercommutation probability ({\it i.e.}~for superstrings, Sect.~\ref{subsubsec:super_string}), where the outer and inner contour lines show the 68\% and 95\% CL regions. The intersection between the vertical and horizontal lines is the injected value for the true signal in each case.
    The marginalized 1D posteriors for both parameters are also in the adjacent insets. The left panel corresponds to the degenerate direction with large tension, $G\mu_{\rm inj} \gtrsim 10^{-15.5}$, while the right panel represents the case with a degenerate direction with small tension $G\mu_{\rm inj} \lesssim 10^{-15.5}$.
    }
    \label{fig:recon_posterior_superstring}\vspace{0.5em}
    {Reconstruction quality indicators ${\cal Q}$ for superstring templates} \\[0.25em]
    \includegraphics[width=1\textwidth]{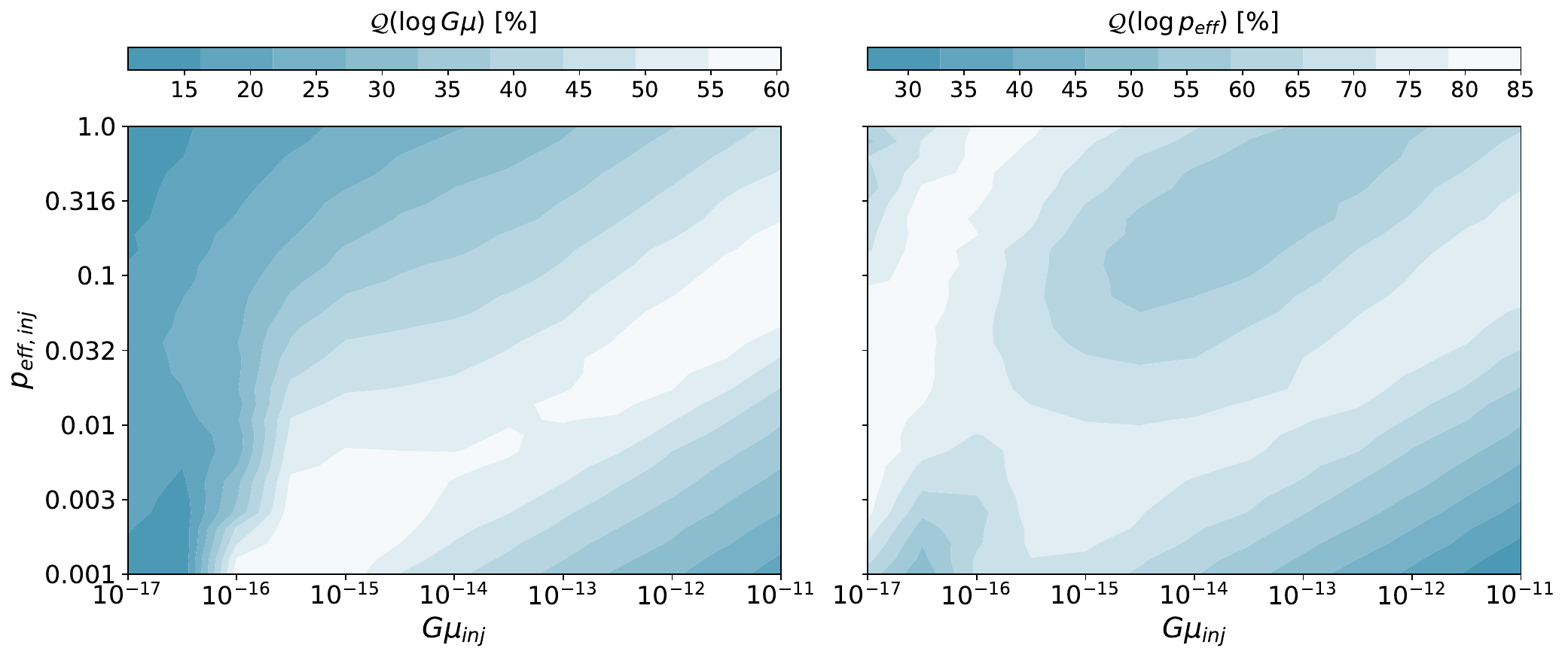}\\[-1.25em]
    \caption{The quality indicator ${\cal Q}$ from Eq.~(\ref{eq:quality_indicator}), for reconstructing $G\mu$ (left) and $p_{\rm eff}$ (right) when the mock data is built from the superstring template with $\{G\mu_{\rm inj},p_{\rm eff,inj}\}$.
    Examples of posteriors for two different choices of $\{G\mu_{\rm inj}, p_{\rm eff,inj}\}$ are shown in Fig.~\ref{fig:recon_posterior_superstring}.}
    \label{fig:recon_precision_superstring}
\end{figure}

\paragraph{Small intercommutation probability ({\it i.e.}~superstrings).}
Fig.~\ref{fig:recon_posterior_superstring} shows examples of the reconstruction posteriors for the superstring templates, when the mock signal is injected with $\{G\mu_{\rm inj}, p_{\rm eff,inj}\} = \{10^{-16.5}, 0.1\}$ and $\{10^{-11}, 0.1\}$.
We see that the reconstruction posterior in both cases is elongated diagonally due to the presence of the degeneracy between $G\mu$ and $p_{\rm eff}$.
These degenerate directions follow a different scaling in each case, as it is also evident from the constant-SNR contours in Fig.~\ref{fig:snr_beyond_1}-bottom-left. The degenerate direction for the case $\{G\mu_{\rm inj}, p_{\rm eff,inj}\} = \{10^{-11}, 0.1\}$ is present for $G\mu \gtrsim 10^{-15.5}$ with $p_{\rm eff} \propto \sqrt{G\mu}$, which is expected when the radiation-era contribution of GWB spectrum is sitting in LISA's window ($\Omega_{\rm GW} \propto \sqrt{G\mu}/p_{\rm eff}$).
This $p_{\rm eff} \propto \sqrt{G\mu}$ scaling explains the direction of the elongated posterior in Fig.~\ref{fig:recon_posterior_superstring}-left.
Another branch of degeneracy, for the case $\{G\mu_{\rm inj}, p_{\rm eff,inj}\} = \{10^{-16.5}, 0.1\}$, appears for $G\mu \lesssim 10^{-15.5}$ and follows $p_{\rm eff} \propto (G\mu)^2$, which is expected when the infrared tail falls within the LISA window, with $\Omega_{\rm GW} \propto (G\mu)^2/p_{\rm eff}$; see e.g., Fig.~3.4 of \cite{Gouttenoire:2019kij} for the $G\mu$ scaling. The elongated posterior in Fig.~\ref{fig:recon_posterior_superstring}-right follows roughly this $p_{\rm eff} \propto (G\mu)^2$ scaling.

As mentioned at the beginning of this section, while the reconstruction posteriors are well confined along the degenerate directions, the degeneracy degrades the reconstruction precisions for both $G\mu$ and $p_{\rm eff}$, in the absence of knowledge of one of the two parameters. As a consequence, their marginalized posteriors spread widely over the prior ranges, as seen in the adjacent insets of Fig.~\ref{fig:recon_posterior_superstring}. Due to large uncertainty in reconstructing $G\mu$ and $p_{\rm eff}$ simultaneously, Fig.~\ref{fig:recon_precision_superstring} presents, rather than the reconstruction precision, the quality indicators of $\log G\mu$ and $\log p_{\rm eff}$, scanning over the injected values of the mock signal. Both $100 \times \mathcal{Q}(\log G\mu)$ and $100 \times \mathcal{Q}(\log p_{\rm eff})$ are much larger than $10\%$ almost everywhere, implying that both parameters cannot be reconstructed well simultaneously. 
Furthermore, we observe two regimes where the quality indicators have different behaviors, arising from the two aforementioned degenerate directions.
In both panels of Fig.~\ref{fig:recon_precision_superstring}, constant-$\mathcal{Q}$ contours for $G\mu_{\rm inj} \gtrsim 10^{-15.5}$ follow roughly $p_{\rm eff} \propto \sqrt{G\mu}$ direction, while for $G\mu_{\rm inj} \lesssim 10^{-15.5}$ follow the $p_{\rm eff} \propto (G\mu)^2$ direction.
The existence of the constant-$\mathcal{Q}$ contours arises from the fact that all mock signals---having different $\{G\mu_{\rm inj},p_{\rm eff, inj}\}$ but sitting along the same degenerate direction---have the same posterior, and thus the same $\mathcal{Q}$. 

Furthermore, we observe a non-trivial trend for ${\cal Q}(\log G\mu)$ (left panel), which increases and then shrinks as $p_{\rm eff}$ decreases. This is mainly a prior effect, by which injected values  closer to their prior boundaries lead to narrower posteriors in general. We observe this effect more clearly for intermediate values of $G\mu_{\rm inj}$ as we move along the $p_{\rm eff}$ direction, but of course the same argument holds for the whole region of the parameter space. An analogous effect occurs for ${\cal Q}(\log p_{\rm eff})$ (right panel).

All in all, we see that the degeneracy between $G\mu$ and $p_{\rm eff}$ severely degrades the parameter reconstruction ability. 
This is just the first case where we start seeing difficulty in the parameter reconstruction of a multi-parameter template.
Potentially, there could be ways of improvement, as the posterior distribution of $G\mu$ and $p_{\rm eff}$ is confined along the aforementioned degenerate directions. Therefore, another independent probe of either $G\mu$ or $p_{\rm eff}$ would allow us to break this degeneracy and pin-down the true parameter.

\begin{figure}[t!]
    \centering
    {\sffamily LISA's reconstruction precision for metastable cosmic-string templates}\\[0.25em]
    \includegraphics[width=1\textwidth]{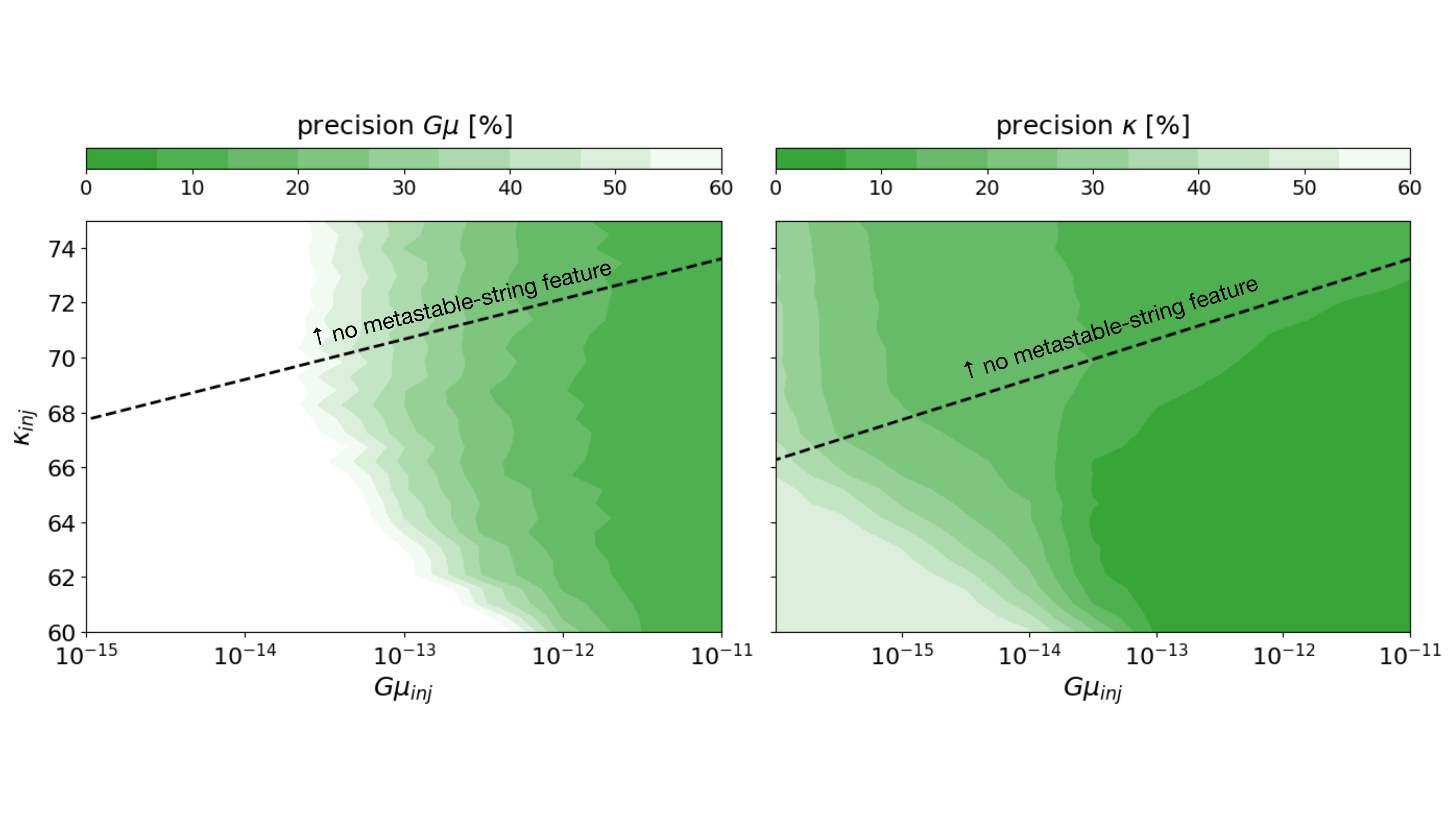}\\[-1em]
    \caption{The reconstruction precision for $G\mu$ (left) and $\kappa$ (right) when the mock data is built from the metastable-string template (Sect.~\ref{subsubsec:metastable_strings}) with $\{G\mu_{\rm inj},\kappa_{\rm inj}\}$.
    Above both black dashed lines, the GW spectrum of metastable strings has no low-frequency cutoff feature due to the condition~\eqref{eq:metastable_stable_network} and resembles the spectrum of the conventional template, i.e., the precision in $G \mu$ becomes independent of $\kappa$. Above the dashed line of the right panel, no mestable-string feature can be reconstructed, and its precision is dominated by the priors.
    }
    \label{fig:recon_precision_metastable}
\end{figure}

\paragraph{Metastable cosmic strings, {\it e.g.}~from GUT models.}
Fig.~\ref{fig:recon_precision_metastable} presents the reconstruction precision of $G\mu$ and $ \kappa$ of the metastable cosmic-string template discussed in Sect.~\ref{subsubsec:metastable_strings}. A true signal with $G\mu_{\rm inj} \gtrsim 10^{-13}$ can be reconstructed with $G\mu$-precision $\lesssim 40\%$ and $\kappa$-precision $\lesssim 5\%$.
Using Eqs.~\eqref{eq:tension} and \eqref{eq:metastable_kappa_def} and the propagation of uncertainty, we can translate the reconstruction uncertainties in both template parameters into the uncertainties of the two symmetry breaking scales, $m_M$ and $\eta$, by $\Delta \eta \approx \Delta G\mu/2$ and $\Delta m_M \approx \sqrt{4(\Delta \eta)^2 + (\Delta \kappa)^2/4} = \sqrt{(\Delta G\mu)^2 + (\Delta \kappa)^2/4} \sim \Delta G\mu$, where the last step uses $\Delta \kappa \ll \Delta G\mu$, as indicated from our results.
We see that LISA can reconstruct well the template parameters $\{G\mu, \kappa\}$ as well as the underlying particle-physics parameters $\{\eta, m_M\}$.

As expected, signals with larger SNR (shown in Fig.~\ref{fig:snr_beyond_1}-top-right) can be better reconstructed, {\it i.e.}~smaller $\Delta G\mu$ and $\Delta \kappa$.
However, a large SNR alone cannot lead to a better reconstruction precision of $\kappa$.
For signals of similar SNR, those with $\kappa_{\rm inj}$ below the black dashed line [determined by Eq.~\eqref{eq:metastable_stable_network}] have much smaller $\Delta \kappa$. 
This is because $\kappa$, which is associated with the infrared-cutoff feature shown in Fig.~\ref{fig:metastable_spectrum}, can be reconstructed only if the cutoff feature in the spectrum lieas within the LISA window. For large values of $\kappa_{\rm inj}$, the cutoff moves to lower frequencies, and the signal inside the LISA window tends to the conventional template's spectra. Since the conventional spectra are independent of $\kappa$, the reconstruction precision returns a flat $\kappa$-posterior for all $\kappa$ values extending from the black dashed line to the maximum $\kappa$ of the prior; see also the discussion at the beginning of this section. The reported $\kappa$-precision above the dashed-line in Fig.~\ref{fig:recon_precision_metastable}-right is therefore prior dependent and artificial. 
We shall see in Sect.~\ref{sec:model_comparison_beyond_conventional} that the ability to resolve $\kappa$ is crucial for discriminating between conventional and metastable-string templates.

\begin{figure}
    \centering
    {\sffamily LISA's reconstruction precision and quality indicators ${\cal Q}$ for current-carrying string templates}\\[0.25em]
    \includegraphics[width=\textwidth
    ]{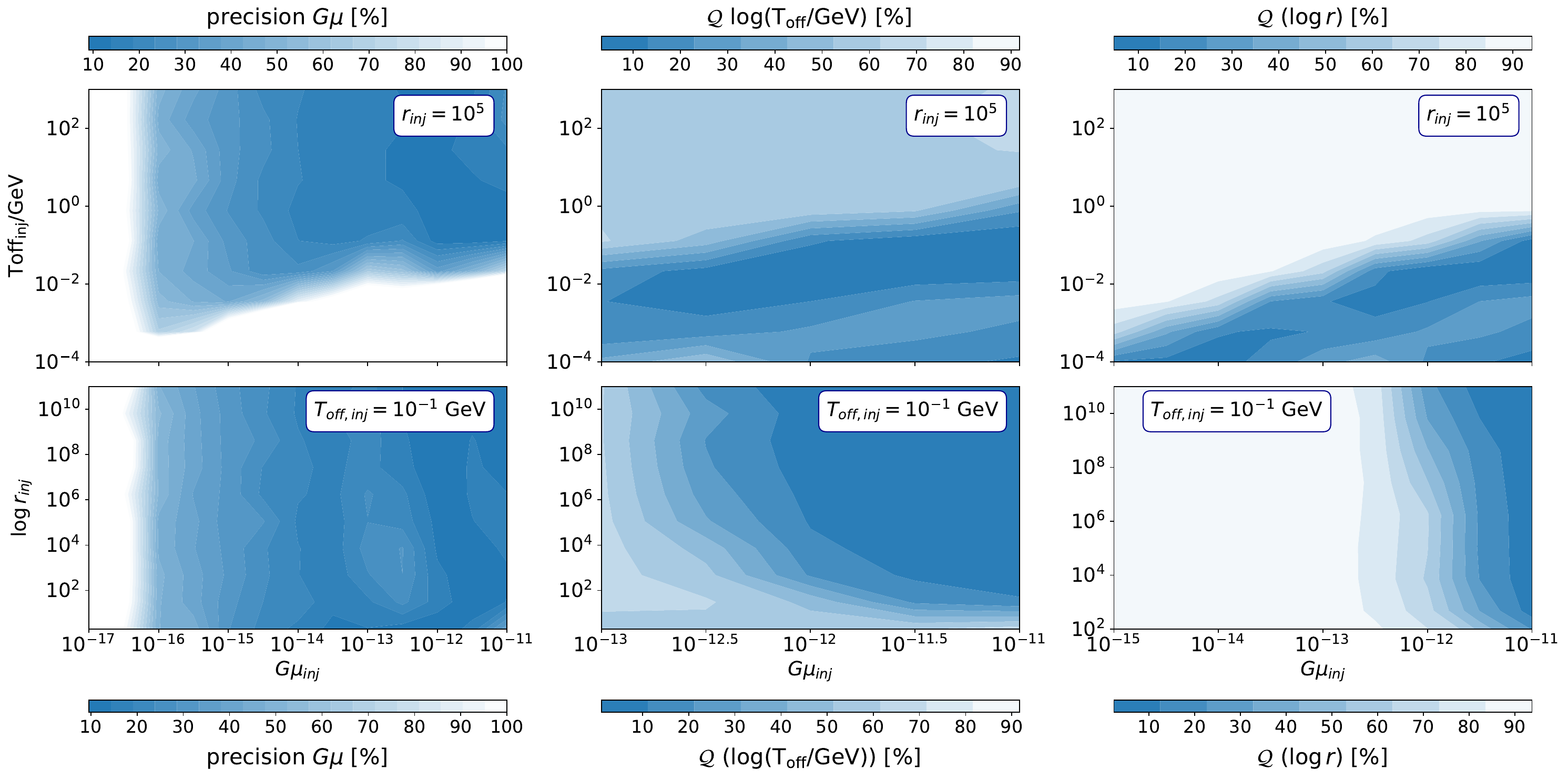}\\[-1em]
    \caption{
    The reconstruction precision for $G\mu$ (left) and the quality indicators for $\log (T_{\rm off}/{\rm GeV})$ (middle) and $\log r$ (right) when the mock signal is built from the current-carrying string template (Sect.~\ref{sec:current_carrying_string}) with $\{G\mu_{\rm inj},T_{\rm off, inj}, r_{\rm inj}\}$.
    The precision and the quality indicator of each parameter are calculated from Eqs.~\eqref{eq:precision_recon} and \eqref{eq:quality_indicator}, respectively.
    Due to the complexity of the template, we show only the case of $r_{\rm inj} = 10^5$ in the upper row and $T_{\rm off,inj} = 10^{-1} ~ {\rm GeV}$ in the lower row.
    }
    \label{fig:recon_precision_current_carrying}
\end{figure}

\paragraph{Current-carrying cosmic strings.}
Fig.~\ref{fig:recon_precision_current_carrying} shows the reconstruction precision in $G\mu$ and the quality indicators in $\log(T_{\rm off}/\rm GeV)$ and $\log r$ for the current-carrying string template from Sect.~\ref{sec:current_carrying_string}.
Due to the three-dimensional parameter space of the model, we consider a case with fixed value  $r_{\rm inj}= 10^{5}$ (top panels), where the true signal is built from different $\{G\mu_{\rm inj}, T_{\rm off,inj}\}$, and a case with fixed $T_{\rm off,inj} = 100 ~ {\rm MeV}$ (bottom panels), where the true signal is built from different $\{G\mu_{\rm inj}, r_{\rm inj}\}$. 
The analysis for other choices of injected parameters can be done straightforwardly and quickly, as we can also employ the neural network already used for these examples, pretrained for the whole range of the parameter priors.

\underline{\textit{Fixed $r_{\rm inj}$ (upper panel of Fig.~\ref{fig:recon_precision_current_carrying}).}}---The current-carrying effect considered in this work can induce a distinct peak feature in the GWB spectrum, as shown in Fig.~\ref{fig:current_carrying}.  For $r_{\rm inj}$ as large as $10^5$,  the spectrum can be estimated with $\Omega_{\rm GW} = \Omega_{\rm GW}^\Delta \times (f/f_\Delta)$ for $f\gtrsim f_\Delta$, with $f_\Delta$ being the lowest frequency at which the current-carrying effect starts imprinting the enhancement, and $\Omega_{\rm GW}^\Delta = \Omega_{\rm GW}(f_\Delta)$.
As it is evident in Fig.~\ref{fig:current_carrying}-top-right, the frequency $f_\Delta $ scales as $\propto T_{\rm off}$, and we recall that the GWB amplitude goes as $\Omega_{\rm GW}^\Delta \propto \sqrt{G\mu}$ for GWs emitted during RD.
Therefore, if a signal scaling as $\Omega_{\rm GW} \propto f$ is observed, we will not be able to reconstruct well $G\mu$ and $T_{\rm off}$, due to the degenerate direction $T_{\rm off} \propto \sqrt{G\mu}$ in the $\{G\mu, T_{\rm off}\}$ parameter space.

The upper panel of Fig.~\ref{fig:recon_precision_current_carrying} shows that when $T_{\rm off,inj}$ is sufficiently large, we cannot reconstruct $T_{\rm off}$ and $r$, as their quality factors are $\sim 90\%$. Instead, the precision in $G\mu$ becomes $T_{\rm off}$-independent and resembles that of the conventional template.
This is expected, since the spectral enhancement moves to higher frequency and evades LISA's window.
On the other hand, the $G\mu$-precision in Fig.~\ref{fig:recon_precision_current_carrying}-top-left shows that, once $T_{\rm off}$ is low enough, the $G\mu$-precision becomes substantially degraded.
This degradation arises from the aforementioned degeneracy between $T_{\rm off}$ and $G\mu$.
For $G\mu_{\rm inj} \simeq 10^{-11}$, this happens when $T_{\rm off,inj} \lesssim 10 ~ {\rm MeV}$. From the  degenerate direction $T_{\rm off} \propto \sqrt{G\mu}$ for $r= 10^5$, the condition for the degradation of reconstruction precision is $T_{\rm off} \lesssim 10 ~ {\rm MeV} (G\mu/10^{-11})^{1/2}$, as shown in Fig.~\ref{fig:recon_precision_current_carrying}-top-left.
Since LISA can probe the feature associated to $T_{\rm off}$ and $r$ in this small-$T_{\rm off,inj}$ region, their posteriors are smaller than those in the case of large $T_{\rm off}$. Their posteriors ought to be reconstructible along the aforementioned degenerate direction, as their quality indicators are $\ll 95\%$.
For example, for $G\mu_{\rm inj} \simeq 10^{-11}$ and $T_{\rm off,inj} \lesssim 0.1 ~ {\rm GeV}$, both quality indicators are up to 40\%.
We can also see that the contours of constant precision roughly follow this degenerate direction.
However, they are more complicated because of the prior effect (see similar effect in the superstring case, discussed previously).

\underline{\textit{Fixed $T_{\rm off,inj}$ (lower panel of Fig.~\ref{fig:recon_precision_current_carrying}).}}---The position of the current-carrying string feature is controlled by $T_{\rm off,inj}$, while the modulation of $r_{\rm inj}$ dictates both the size of spectral enhancement and its slope, as shown in Fig.~\ref{fig:current_carrying}-top-left.
Therefore, we do not expect a large degeneracy between $G\mu$ and $r$ that degrades the reconstruction precision, as in the case of $G\mu$ and $T_{\rm off}$.
Fig.~\ref{fig:recon_precision_current_carrying}-bottom-left shows that, for $T_{\rm off,inj} = 0.1 \, {\rm GeV}$, $G\mu$ can be reconstructed with a precision $\lesssim 50\%$ when $G\mu_{\rm inj} \gtrsim 10^{-16}$ and all values of $r_{\rm inj}$ (i.e., $r_{\rm inj} \gtrsim 0.5$).
Note that the degeneracy between $G\mu$ and $T_{\rm off}$ does not degrade the reconstruction precision here, because this degradation appears only when $T_{\rm off,inj} \lesssim 10 \, {\rm MeV}$ for $G\mu_{\rm inj} \lesssim 10^{-11}$, as  discussed earlier and shown in Fig.~\ref{fig:recon_precision_current_carrying}-top-left.

In the middle and right panels of Fig.~\ref{fig:recon_precision_current_carrying}-bottom, we indeed see that $\log T_{\rm off}/{\rm GeV}$ can be reconstructed with $\mathcal{Q}(\log T_{\rm off}/{\rm GeV}) < 10\%$ for $G\mu_{\rm inj} \gtrsim 10^{-12}$, and $\log r$ can be reconstructed with $\mathcal{Q}(r) < 20\%$ when $G\mu_{\rm inj} \gtrsim 3 \times 10^{-12}$.
A larger $r_{\rm inj}$ improves both reconstruction precisions, as the spectral enhancement becomes stronger.
Nonetheless, LISA cannot reconstruct both $\log T_{\rm off}$ and $\log r$ well for $G\mu \lesssim 10^{-13}$ (i.e., both quality indicators approach $\sim 90\%$), since the enhancement feature moves outside LISA's window.

\begin{figure}[t!]
    \centering
    {\sffamily Templates with alternative cosmic histories}\\[0.25em]
    \includegraphics[width=0.485\linewidth]{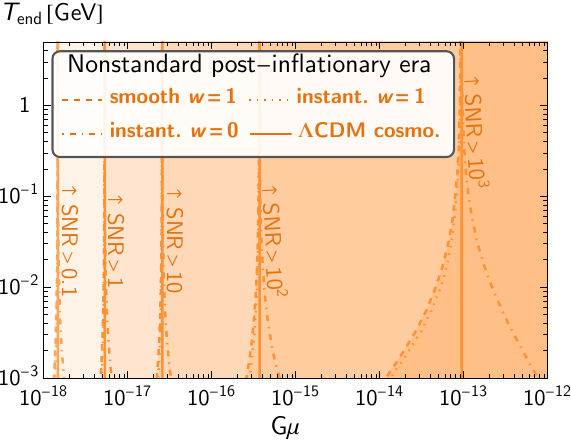}\hfill
    \includegraphics[width=0.485\linewidth]{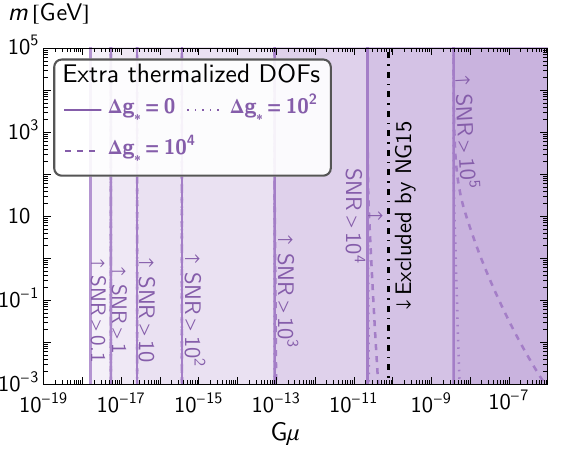}\\[-1em]
    \caption{SNR forecast [using Eq.~\eqref{eq:SNR_definition}] of GW signal observed at LISA in the beyond-conventional cosmic-string templates, where the cosmic history gets modified; cf. Sect.~\ref{sec:GWB_cosmic_history}. We do not show the case of the completely-secluded dofs (Sect.~\ref{subsec:nonst_cosmo_dark_DOFs}) because its signal is similar to the conventional case, as shown in  Fig.~\ref{fig:CS_dark_dof}.}
    \label{fig:snr_beyond_2}
\end{figure}

\begin{figure}[t!]
    \centering
    \sffamily LISA's reconstruction precision and quality indicators ${\cal Q}$ for templates with nonstandard era\\[0.25em]
\includegraphics[width=1\textwidth, height=0.3\textheight]{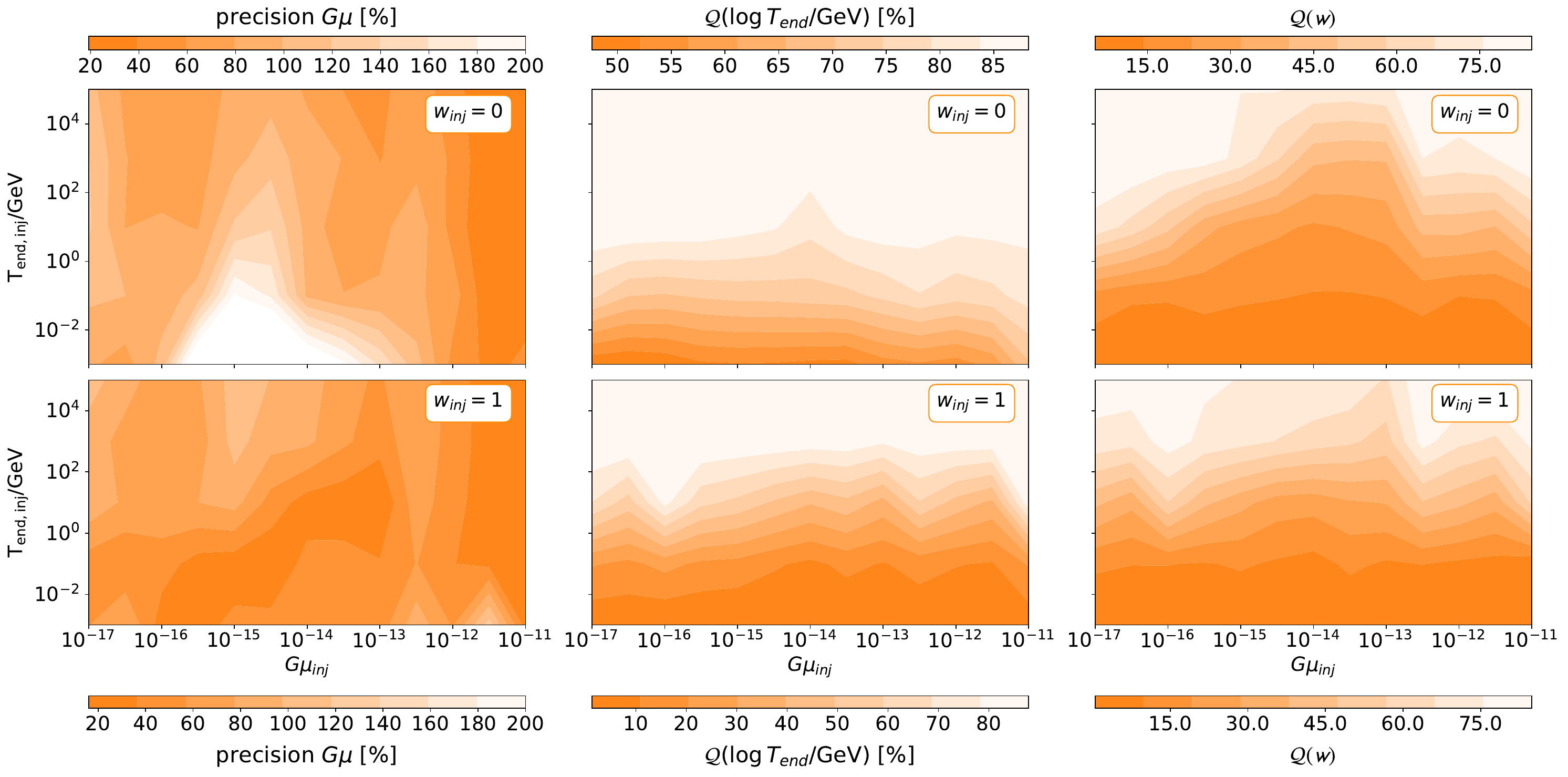}\\[-1em]
    \caption{
    The reconstruction precision for $G\mu$ (left) and the quality indicators $\mathcal{Q}$ for $\log (T_{\rm end}/{\rm GeV})$ (middle) and $w$ (right) when the mock signal is built from the template of the nonstandard era after inflation with an instantaneous change to radiation era (Sect.~\ref{subsec:nonst_cosmo_after_inflation}), assuming $\{G\mu_{\rm inj},T_{\rm end, inj}, w_{\rm inj}\}$.
    The precision is calculated from Eq.~\eqref{eq:precision_recon}, and the quality indicator follows Eq.~\eqref{eq:quality_indicator}. 
    The top row has $w_{\rm inj} = 0$ representing the matter-domination era case, while the bottom row has $w_{\rm inj} = 1$ corresponding to the kination era.
    }
    \label{fig:recon_precision_nonstandard_era}
\end{figure}

\subsection{Exploring cosmic histories}
\label{subsec:recon_explore_cosmic_hist}
Fig.~\ref{fig:snr_beyond_2} shows the SNR forecast for  GW signals from the templates with alternative cosmic histories (Sect.~\ref{sec:GWB_cosmic_history}).
We report only the cases of the nonstandard era after inflation (Sect.~\ref{subsec:nonst_cosmo_after_inflation}) and the extra dofs thermalized with SM particles (Sect.~\ref{subsec:nonst_cosmo_thermalized_DOFs}). 
The extra hidden-dof case (Sect.~\ref{subsec:nonst_cosmo_dark_DOFs}) has a SNR forecast similar to the conventional template, since these hidden DOFs only enhance the GW spectrum by few $\%$ in LISA's window; see Fig.~\ref{fig:CS_dark_dof}.

The features on the GWB spectrum are prominent in the LISA window when nonstandard era ends late [cf. Eq.~(121) of \cite{Gouttenoire:2019kij}] or the extra thermalized dofs have small  mass [Eq.~\eqref{eq:thermalized_dof_frequency}].
For example, Fig.~\ref{fig:snr_beyond_2} indicates that the extra thermalized dofs induces a substantial deviation from the conventional prediction (i.e., the constant SNR lines becomes sensitive to $m$) when $m < 1\,{\rm TeV}$ for $G\mu \lesssim 7.9\times 10^{-10}$, the value maximally allowed by PTA bounds. 
Nevertheless, this mass range is constrained by collider experiments \cite{ATLAS:2024fdw,CMS:2024zqs}.
As we shall see from the reconstruction results, the parameter $m$---when the true signal has mass as low as $10\, {\rm GeV}$---cannot even be reconstructed.

\paragraph{Nonstandard era after inflation.}
Fig.~\ref{fig:recon_precision_nonstandard_era} shows the reconstruction precision in $G\mu$, and the quality indicator  in $\log(T_{\rm end}/\rm GeV)$ and $w$, for signals built with the template with nonstandard era after inflation discussed in Sect.~\ref{subsec:nonst_cosmo_after_inflation}, depending on $\{G\mu_{\rm inj},T_{\rm end,inj}$ and $w_{\rm inj}\}$. As examples, we consider only the cases of $w_{\rm inj} = 0$ and $1$, which can be motivated by the matter-domination and kination eras, respectively.\footnote{
We note that~\cite{Blanco-Pillado:2024aca} presents the reconstruction posterior for the specific case of $G\mu_{\rm inj} = 10^{-10}, ~ T_{\rm end,inj} = 1 \, {\rm GeV},$ and $w_{\rm inj} = 1$. However, the authors also consider the variation of the initial loop size $\alpha$ and the exponent of GW emission power $q$ (which we will only discuss in Sect.~\ref{subsec:recons_loopProperties}), introducing further degeneracy with $G\mu$, so a direct comparison with our results is irrelevant.} 

First, we observe from the left column of Fig.~\ref{fig:recon_precision_nonstandard_era} that $G\mu$ can be reconstructed with precision $\Delta G\mu \lesssim 100\%$ almost everywhere, except when $T_{\rm end,inj} \lesssim 10^3 \, {\rm GeV}$ and $G\mu_{\rm inj} \in (10^{-16},10^{-13})$ for $w_{\rm inj} = 0$. 
We see obviously that, for smaller $T_{\rm end,inj}$ and in some range of $G\mu_{\rm inj}$, the reconstruction precision is  worsened for $w_{\rm inj} = 0$ but improved for $w_{\rm inj} = 1$.
This behavior can be explained by the spectral feature due to the nonstandard era that only enters LISA's sensitivity window for a low enough $T_{\rm end,inj}$. In other words, the turning-point frequency where the GW spectrum deviates from the conventional prediction follows $f_{\rm tp} \propto T_{\rm end}$; see {\it e.g.}~the frequency-temperature relation in~\cite{Gouttenoire:2019kij}. A postinflationary period with $w_{\rm inj} = 0$ ($w_{\rm inj} = 1$) suppresses (enhances) the high-frequency part of the GW spectrum compared to the conventional prediction, as shown in Fig.~\ref{fig:non-standard-era}. Therefore, the $G\mu$-reconstruction for $w_{\rm inj} = 0$ ($w_{\rm inj} = 1$) is degraded (improved) mildly by a slightly weaker (stronger) GW signal.

\begin{figure}[t!]
    \centering
    {\sffamily LISA's reconstruction precision for templates with}\\
    {\sffamily extra thermalized dofs (left) and completely-hidden dofs (right)} \\[0.25em]
    
    \includegraphics[width=0.47\textwidth]{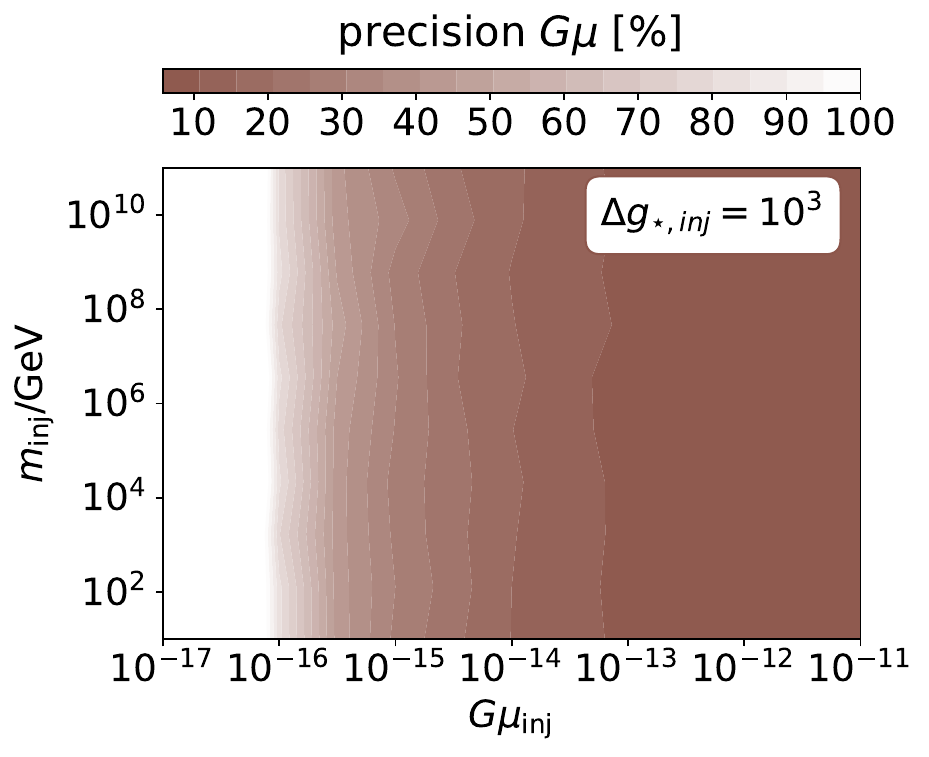}\hfill\includegraphics[width=0.48\textwidth]{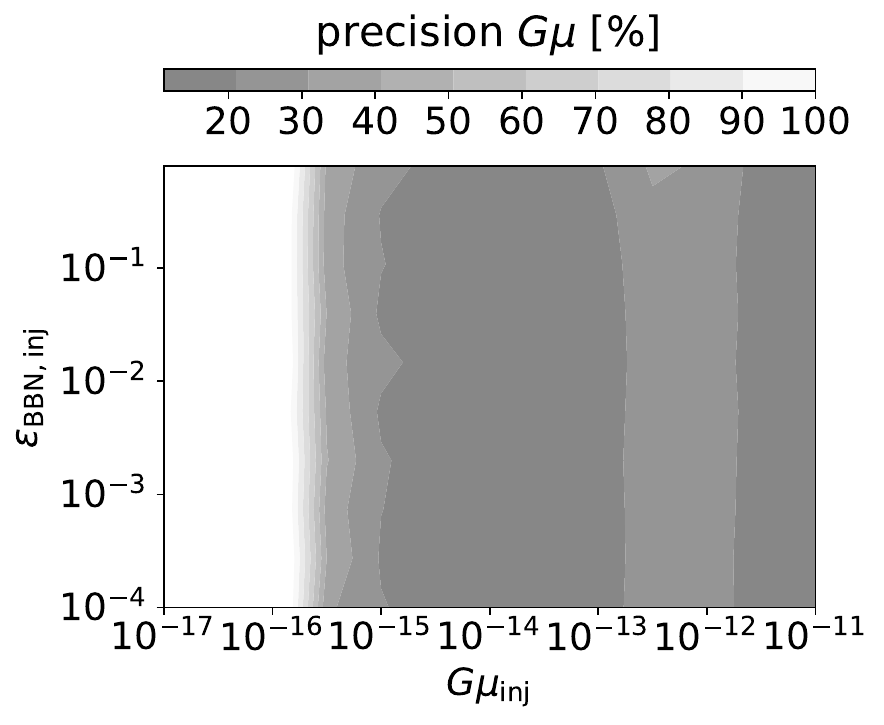}\\[-1em]
    \caption{
    (Left) The reconstruction precision for $\log G\mu$ when the mock signal is built from the template of extra thermalized dofs (Sect.~\ref{subsec:nonst_cosmo_thermalized_DOFs}) with $\{G\mu_{\rm inj},m_{\rm inj}\}$ and $ \Delta g_{*,{\rm inj}} = 10^3$.
    We do not show the precisions for $\log (m/{\rm GeV})$ and $\log \Delta g_*$, as we have checked that their quality indicators [Eq.~\eqref{eq:quality_indicator}] are roughly the maximal allowed value $\mathcal{Q} \approx 95\%$. 
    This indicates LISA's inability in probing the extra thermalized dofs of mass scale $m > 10 \,{\rm GeV}$. We also check that the result is independent of $\Delta g_{*,\rm inj}$.
    (Right) The reconstruction precision for $G\mu$  when the mock signal is built from the template of completely-hidden dofs (Sect.~\ref{subsec:nonst_cosmo_dark_DOFs}) with $\{G\mu_{\rm inj},\epsilon_{\rm BBN,inj}\}$.
    We omit to show the reconstruction quality of $\epsilon_{\rm BBN}$ as its quality indicator is $\sim 90-95\%$ everywhere in the shown parameter space.}
    \label{fig:recon_precision_extra_thermal_dof}
\end{figure}

Furthermore, the middle and right columns of Fig.~\ref{fig:recon_precision_nonstandard_era} show that the posteriors of $\log(T_{\rm end}/{\rm GeV})$ and $\log w$ are not well reconstructed  for $T_{\rm end,inj} \gtrsim 1$ GeV. The reason is that the feature associated to the nonstandard era is outside the LISA's windows.
For smaller $T_{\rm end,inj}$, their reconstructions are slightly improved (i.e., a narrower posterior and a smaller $\mathcal{Q}$).
The quality indicators for $\omega_{\rm inj} = 0$ are evidently larger than those with $\omega_{\rm inj} = 1$. This is because the enhancement effect from the kination era improves marginally the parameter reconstruction, unlike the suppression feature from the matter-domination era.
For kination case, it might be possible to reconstruct $T_{\rm end}$ and $w$ for $T_{\rm end} < 10 \, {\rm MeV}$.

\paragraph{Extra thermalized DOFs.}
Fig.~\ref{fig:recon_precision_extra_thermal_dof}-left provides LISA's reconstruction precision $G\mu$ for the extra thermalized-dof template, calculated from the marginalized posterior and Eq.~\eqref{eq:precision_recon}.
We omit showing the reconstruction precision for $m$ and $\Delta g_*$ because we have checked that their quality indicators [defined in Eq.~\eqref{eq:quality_indicator}] are roughly $\sim90$-$95\%$ everywhere in the shown $\{G\mu_{\rm inj}, m_{\rm inj}\}$ parameter space, indicating that posteriors of $m$ and $\Delta g_*$ flatten over the prior ranges, and hence cannot be reconstructed.
As examples, we consider the mock signal of different $\{G\mu_{\rm inj}, m_{\rm inj}\}$ values, while we fix  $\Delta g_{*,\rm inj} = 10^3$. 

We observe that the reconstruction precision of $G\mu$ is more or less independent of $m_{\rm inj}$. 
We also checked that the results remain similar for smaller values of $\Delta g_{*,\rm inj}$, up to the statistical fluctuation. 
Moreover, the values of the precision resemble the conventional prediction shown in Fig.~\ref{fig:keynote_accuracy_precision_reconstruction}.
This is the hint that LISA cannot probe any extra thermalized dofs of mass $m \geq 10~{\rm GeV}$ and $\Delta g_* \leq 10^{3}$.
The inability to reconstruct $m$ can be understood as the extra-DOFs feature in the GWB spectrum sits around the frequency~\eqref{eq:thermalized_dof_frequency}, shown in Fig.~\ref{fig:CS_dof}.
For $m = 10\,{\rm GeV}$ and $G\mu \simeq 10^{-11}$, the extra DOFs induce a $\lesssim 10-15\%$ deviation in $\Omega_{\rm GW}$ from the conventional template. Larger $m$ and smaller $G\mu$ values lead to even smaller deviations, as if there is no observable feature associated to $m$ and $\Delta g_*$. Note also that smaller $\Delta g_*$ leads to a weaker feature on the GWB spectrum, as shown in Fig.~\ref{fig:CS_dof}.
For $\Delta g_* > 10^3$, the feature would be slightly more prominent, but a large number of dofs might only be motivated in particular theories. All in all, our results, together with the collider constraints ($m \gtrsim {\rm TeV}$), suggest that LISA cannot probe any extra DOFs that thermalize with SM particles. The prospect of detecting such heavy particles could be realized however by GWB observatories operating at higher frequency ranges, such as Einstein Telescope (ET)~\cite{Hild:2010id,Punturo:2010zz,Abac:2025saz}, Cosmic Explorer (CE)~\cite{LIGOScientific:2016wof,Reitze:2019iox}, and Big Bang Observatory (BBO)~\cite{Yagi:2011wg}.

\paragraph{Completely hidden DOFs.} LISA's reconstruction precision in $G\mu$ for the template with completely-hidden dofs from Sect.~\ref{subsec:nonst_cosmo_dark_DOFs}, is shown in 
Fig.~\ref{fig:recon_precision_extra_thermal_dof}-right.
We see that the precision in $G\mu$ is approximately independent of $\epsilon_{\rm BBN,inj}$ because $\epsilon_{\rm BBN}$ induces a small enhancement on the GWB spectrum, as shown in Fig.~\ref{fig:CS_dark_dof}.
The reconstruction precision of $\log \epsilon_{\rm BBN}$ is omitted as we have checked that the quality indicator of $\log \epsilon_{\rm BBN}$ is about $\sim 90\%$ everywhere in the parameter space, implying that LISA cannot pin down $\epsilon_{\rm BBN}$ of the true signal. This result suggests that LISA cannot probe a completely hidden relativistic DOF sector. Other GW observatories with higher frequency windows (e.g., ET, CE, and BBO) are actually needed to search for the signature of such a sector, as the enhancement in the GWB spectrum is slightly larger at higher frequencies~\cite{servantSpectro}; see Fig.~\ref{fig:CS_dark_dof}.
We also observe that, although the tendency of the $G\mu$-precision is similar to the conventional prediction in Fig.~\ref{fig:keynote_accuracy_precision_reconstruction}, the actual uncertainty in this case is slightly larger.
This could be explained by a mild degeneracy between $G\mu$ and $\epsilon_{\rm BBN}$, since both of them control the amplitude $\Omega_{\rm GW}$.

\begin{figure}[t!]
    \centering
    {\sffamily Templates with varying loop's properties}\\[0.25em]
    \includegraphics[width=0.485\linewidth]{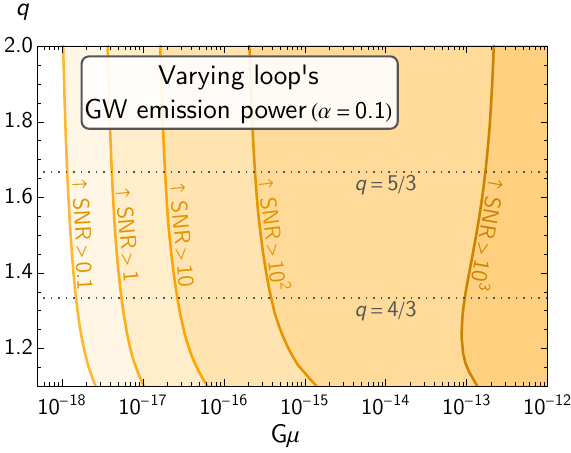}
    \includegraphics[width=0.485\linewidth]{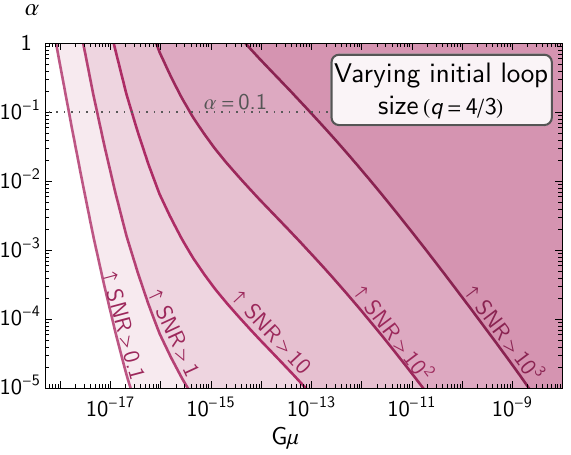}\\[-1em]
    \caption{SNR forecast [using Eq.~\eqref{eq:SNR_definition}] of GW signal observed at LISA in the beyond-conventional cosmic-string templates, where the loop number properties get modified; cf. Sect.~\ref{subsec:loopProperties}.}
    \label{fig:snr_beyond_3}
\end{figure}

\begin{figure}[t!]
    \centering
    {\sffamily Examples of reconstruction posteriors for template with varying loop properties} \\[0.25em]
    \includegraphics[width=\linewidth]{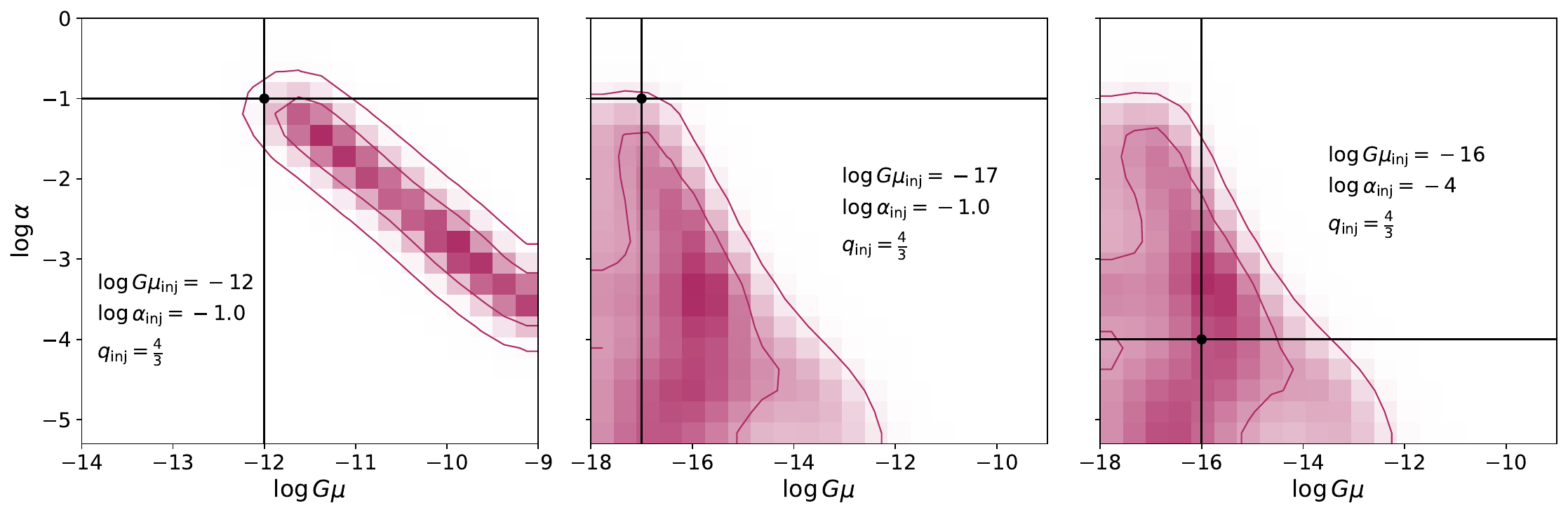}\\[-1em]
    \caption{Examples of the 2D reconstruction posterior in the $\{G\mu, \alpha\}$ plane, where the outer and inner contour lines show the 68\% and 95\% CL regions.
    With $q_{\rm inj} =4/3$, the intersection between the vertical and horizontal lines is the injected value of $\{G\mu_{\rm inj},\alpha_{\rm inj}\}$ for the true signal in each case.
    The left panel shows the degenerate direction in the strong signal regime, while the posteriors in middle and right panels are similar as expected from the weak signal regime.
    We provide the scans of reconstruction quality over $\{G\mu_{\rm inj},q_{\rm inj}\}$ when $\alpha_{\rm inj} = 0.1$ and $\{G\mu_{\rm inj},\alpha_{\rm inj}\}$ when $q_{\rm inj}= 4/3$ in Figs.~\ref{fig:recon_precision_loop_qa_1} and \ref{fig:recon_precision_loop_qa_2}, respectively.
    }
    \label{fig:recon_posterior_alpha_q}
\end{figure}

\subsection{Changing loop properties}
\label{subsec:recons_loopProperties}
The SNR forecast of the GW signals for the template with varying loop properties (Sect.~\ref{subsec:loopProperties}) are shown in Fig.~\ref{fig:snr_beyond_3}, where the left panel has a fixed initial loop size $\alpha = 0.1$ and the right panel has a fixed loop's emission power (or shape parameter) $q= 4/3$.
We see that the variation over $q$ does not impact strongly the SNR  of the signal, as one can expect from the small $q$-dependency of the GW spectrum shown in Fig.~\ref{fig:spec_q_plot}.
On the other hand, the variation of the initial loop size $\alpha$ impacts the GW spectrual amplitude strongly, as shown in Fig.~\ref{fig:spec_alpha_plot}.

We see that the line of constant SNR for $G\mu \gtrsim 10^{-16}$ in Fig.~\ref{fig:snr_beyond_3}-right follows the degenerate direction $\alpha \propto (G\mu)^{-1}$, which corresponds to $\Omega_{\rm GW} \propto \sqrt{\alpha G\mu}$ for GW emitted during radiation era (see e.g., \cite{Auclair:2019wcv,Gouttenoire:2019kij}).
As discussed at the beginning of this section and shown in Fig.~\ref{fig:recon_posterior_alpha_q}-left, the posterior of the parameter reconstruction elongates along this degenerate direction, degrading its parameter reconstruction's quality.
On the other hand, we expect that any signal with $\{G\mu,\alpha\}$ below the line of ${\rm SNR} = 10$ in Fig.~\ref{fig:snr_beyond_3}-right is barely observable and reconstructible, leading to the degradation of reconstruction quality which is due to no observable signal, instead of the degeneracy; see the middle and right panels of Fig.~\ref{fig:recon_posterior_alpha_q}.

We consider now the reconstruction quality of this template; we will show only examples where either $\alpha_{\rm inj} = 0.1$, representing the conventional initial loop size, or $q_{\rm inj}= 4/3$, representing the conventional (cusp) shape.\footnote{A similar analysis with $\alpha_{\rm inj} = 0.1$ is also performed in \cite{Blanco-Pillado:2024aca} (Figs.~5-6), using \texttt{SGWBinner}~\cite{Caprini:2019pxz}.}
An analogous analysis for other values of $\alpha_{\rm inj}$ and $q_{\rm inj}$ can be done straightforward and fast with the neural network that is already employed for the cases presented here.

\begin{figure}[t!]
    \centering
    {\sffamily LISA's reconstruction quality indicators ${\cal Q}$ for templates with varying loop properties (I)}
    \\[0.25em]
    \includegraphics[width=\textwidth]{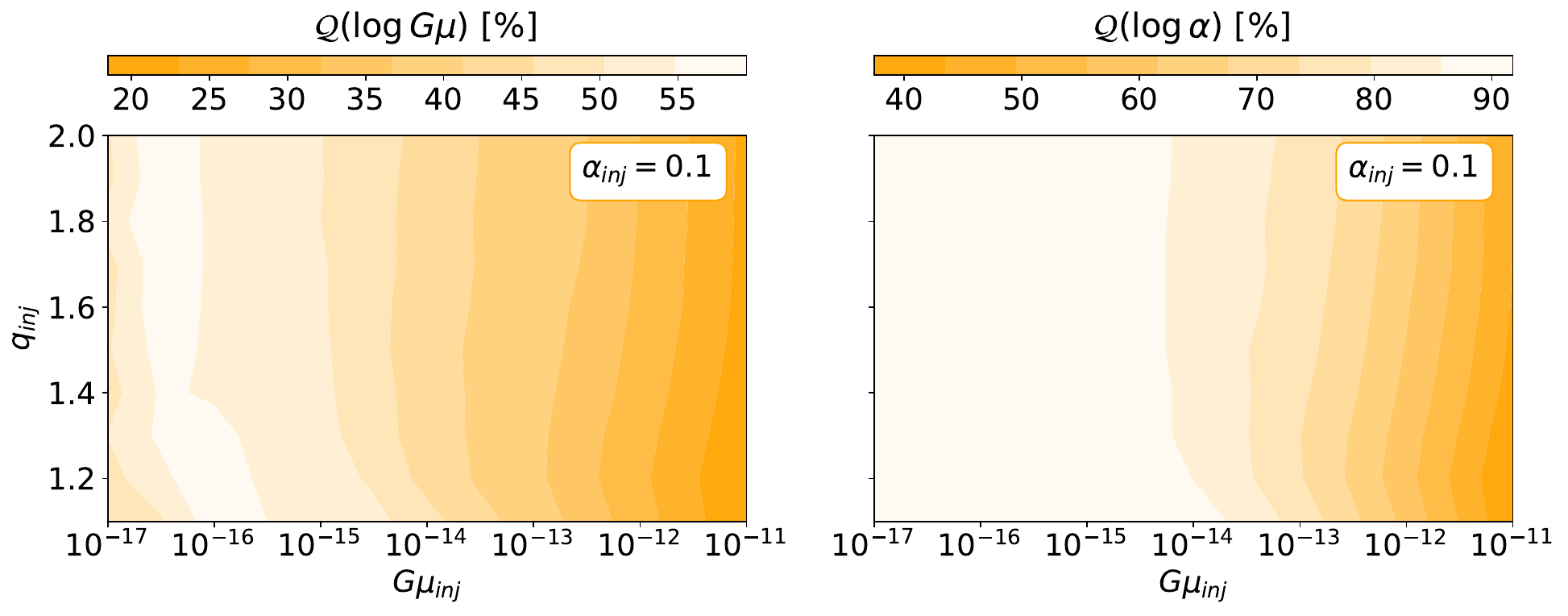}\\[-1em]
    \caption{
    The quality indicators for reconstructing $\log G\mu$ (left) and $\log \alpha$ (right) when the mock signal is built from the template of non-conventional loop properties (Sect.~\ref{subsec:loopProperties}) with $\{G\mu_{\rm inj},q_{\rm inj}\}$ and a fixed $\alpha_{\rm inj} = 0.1$. The reconstruction of $G\mu$ and $\alpha$ cannot be precise due to the degeneracy between them.
    The reconstruction result of $q$ is omitted as $\mathcal{Q}(q) \sim 90\%$ everywhere across the $\{G\mu_{\rm inj},q_{\rm inj}\}$ parameter space, meaning its posterior is flat and $q$ is not reconstructible.
    }\vspace*{1cm}
    \label{fig:recon_precision_loop_qa_1}
    {\sffamily {LISA's reconstruction quality indicators ${\cal Q}$ for templates with varying loop properties (II)}}
    \\[0.25em]
    
    \includegraphics[width=\textwidth]{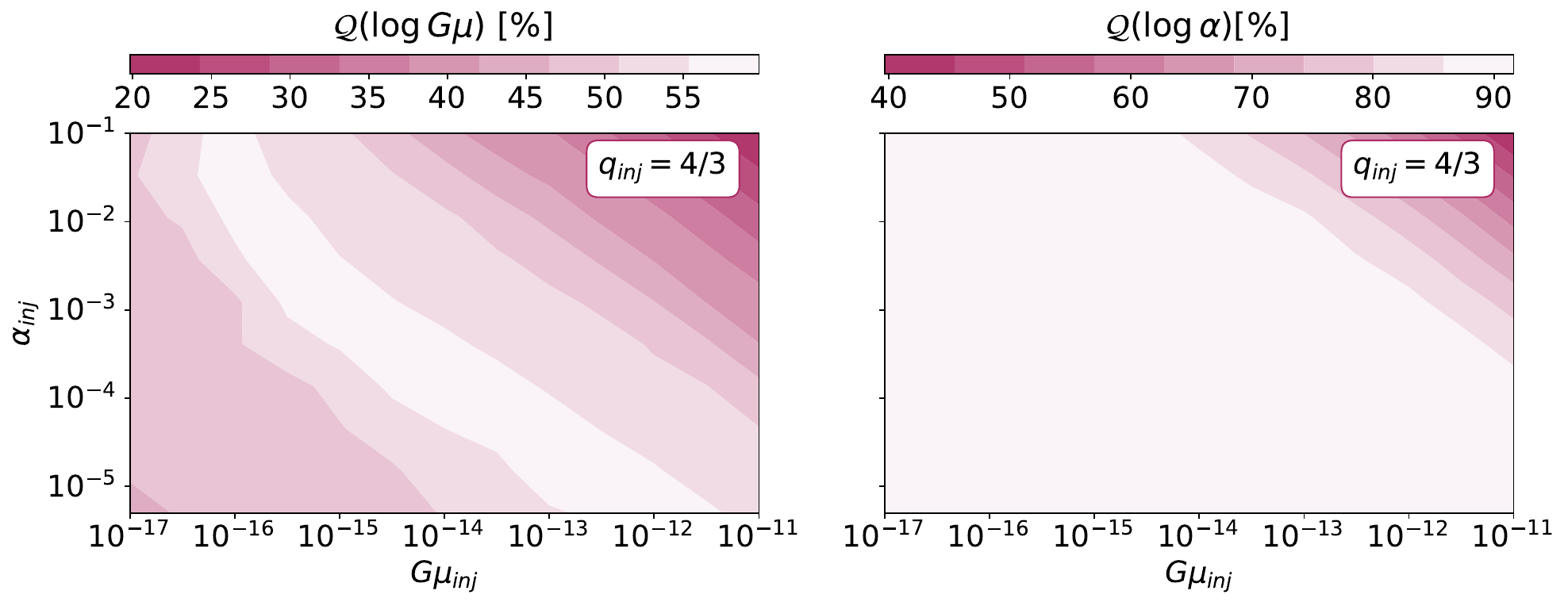}\\[-1em]
    \caption{
    The quality indicators for reconstructing $\log G\mu$ (left) and $\log \alpha$ (right) when the mock signal is built from the template of non-conventional loop properties (Sect.~\ref{subsec:loopProperties}) with $\{G\mu_{\rm inj},\alpha_{\rm inj}\}$ and a fixed $q = 4/3$.
    The reconstruction result of $q$ is omitted again as $\mathcal{Q}(q) \sim 90\%$ everywhere.
    Unlike Fig.~\ref{fig:recon_precision_loop_qa_1}, the degenerate direction can be seen clearly in the $\{G\mu_{\rm inj},\alpha_{\rm inj}\}$ plane, as the lines of constant $\mathcal{Q}$.}
\label{fig:recon_precision_loop_qa_2}
\end{figure}

\paragraph{Loop's GW emission power.}
Fig.~\ref{fig:recon_precision_loop_qa_1} shows the quality indicators for $\log G\mu$ and $\log \alpha$ in the parameter space of $\{G\mu_{\rm inj},q_{\rm inj}\}$ when we fix $\alpha_{\rm inj} = 0.1$ for the injected signal.
The quality indicator of $q$ is omitted because it is roughly $\sim 90\%$ everywhere in the shown parameter space, reflecting that its posterior flattens over the prior range and its reconstruction is not possible.
This result is expected, since $q$ only modifies the GW spectrum mildly; see Fig.~\ref{fig:spec_q_plot}.
Although the quality indicators of $\log G\mu$ and $\log \alpha$ can be smaller than $90\%$, they are still larger than 10\% everywhere, indicating a degradation of parameter reconstruction.

The easiest way to understand the trend of $\mathcal{Q}(\log G\mu)$ and $\mathcal{Q}(\log \alpha)$ is to look at $G\mu-\alpha$ degeneracy. By picking $\alpha_{\rm inj} = 0.1$ and $G\mu_{\rm inj} \gtrsim 10^{-16}$, the joint posterior distribution for each true signal's choice will be elongated along $\alpha \propto (G\mu)^{-1}$ degenerate direction in $\{G\mu,\alpha\}$ parameter space, from $G\mu \sim 10^{-16}$ to larger $G\mu$ value, as an example shown in Fig.~\ref{fig:recon_posterior_alpha_q}-left.
For larger $G\mu_{\rm inj}$, the posterior moves closer to the prior boundary and gets truncated (see the superstring case for a similar argument), reducing (improving) the quality indicators of both $G\mu$ and $\alpha$.
On the other hand, for signals with $G\mu_{\rm inj} \lesssim 10^{-16}$ (${\rm SNR} < 10$) that  have difficult reconstruction, their posteriors would be similar to each other as shown in Fig.~\ref{fig:recon_posterior_alpha_q}-middle and right. I.e., they span the region in $\{G\mu,\alpha\}$ parameter space below the line of ${\rm SNR} = 10$ in Fig.~\ref{fig:snr_beyond_3}-right.
Within the entire range of $\alpha$-prior, the marginalized posterior of $G\mu$ cannot extend up to too large $G\mu$ value, reducing its posterior width.
This explains why $\mathcal{Q}(\log G\mu) < 90\%$, while $\mathcal{Q}(\log \alpha) \sim 90\%$ in the weak signal regime.

\paragraph{Initial loop size.}
Fig.~\ref{fig:recon_precision_loop_qa_2} shows the quality indicators for $\log G\mu$ and $\log \alpha$ in the parameter space of $\{G\mu_{\rm inj},\alpha_{\rm inj}\}$ when we fix $q_{\rm inj} = 4/3$ for the injected signal.
Similarly to the previous case, the quality indicator of $q$ is again omitted, since its posterior mostly flattens over the prior range for all $\{G\mu_{\rm inj},\alpha_{\rm inj}\}$ choices.
The reconstructions of $G\mu$ and $\alpha$ are better than $q$ in some regions ($\mathcal{Q}(\log G\mu)$, $\mathcal{Q}(\log \alpha) \ll 95\%$); still, their uncertainties are too large to say the reconstruction is precise.
We see that $\mathcal{Q}(\log G\mu)$ and $\mathcal{Q}(\log \alpha)$ are larger than $\sim 10\%$ and also exhibit a non-trivial behavior, caused by the degeneracy between $G\mu$ and $\alpha$.

As discussed earlier in this subsection, there exists a degenerate direction $\alpha \propto (G\mu)^{-1}$ in the strong signal regime, as shown in Fig.~\ref{fig:recon_posterior_alpha_q}-left.
This explains the presence of the lines of constant $\mathcal{Q}(\log G\mu)$ and $\mathcal{Q}(\log \alpha)$ in the $\{ G\mu_{\rm inj}, \alpha_{\rm inj}\}$ plane of Fig.~\ref{fig:recon_precision_loop_qa_2}, {\it i.e.}~all points of $\{ G\mu_{\rm inj}, \alpha_{\rm inj}\}$ sitting along the same degenerate direction will have the same posterior (populating the parameter space for $G\mu \gtrsim 10^{-16}$ up to the prior boundary\footnote{The posterior does not extend towards lower $G\mu$, as the GWB spectrum in this regime has different shape and is not degenerate with the GW spectrum in this large $G\mu$ regime, {\it i.e.}~this is when the infrared tail of the spectrum enters the LISA window.}) and hence the same $\mathcal{Q}$.
We also see that from the top-right corner of the parameter space the quality indicators start increasing when $G\mu_{\rm inj}$ or $\alpha_{\rm inj}$ decrease. 
For a particular set of $\{ G\mu_{\rm inj}, \alpha_{\rm inj}\}$ along the same degenerate direction, its posterior can traverse the longest distance across the parameter space that is allowed by the prior, such that it has the largest $\mathcal{Q}$. This results in the white stripe in Fig.~\ref{fig:recon_precision_loop_qa_2}-left.
For weak signals (e.g., ${\rm SNR < 10}$), their posteriors do not follow the degenerate direction but now span the region in $\{G\mu,\alpha\}$ parameter space below the ${\rm SNR} = 10$ line in Fig.~\ref{fig:snr_beyond_3}-right, as shown in Fig.~\ref{fig:recon_posterior_alpha_q}-middle and right.
Similar to the previous case of varying $q_{\rm inj}$, this leads to $\mathcal{Q}(\log G\mu) < 90\%$, while $\mathcal{Q}(\log \alpha) \sim 90\%$.

\subsection{Model Comparison Examples}
\label{sec:model_comparison_beyond_conventional}
We have seen that LISA can not always reconstruct beyond-conventional templates with good precision. In some scenarios, template parameters cannot be well reconstructed, as their associated spectral features reside outside the LISA window or there are degeneracies among the model parameters. Despite these difficulties, it is however still possible to address the following key question: \emph{can LISA distinguish beyond-conventional signals from conventional backgrounds?} 
In this section, we focus on a couple of beyond-conventional templates, metastable strings and superstrings. In particular, we obtain the regions of parameter space where LISA can distinguish between these templates and connventional signals, using the methodology shown in Fig.~\ref{fig:flowchart}-right, which we recall is explained in Appendix~\ref{app:model_comparison}). 
Similar to the results in  Sect.~\ref{subsec:ModelComparisonVOSvsBOSvsDOF}, we shall see again that the usual fixed-SNR criterion does not directly quantify the ability to distinguish one model spectrum from others.

Before presenting our results, an important comment is in order. As we will see below, it can happen that one model, despite yielding sharper posteriors than  another---{\it i.e.}~more precise parameter reconstruction---, is disfavored relative to the latter according to the Bayes factor. This outcome may appear counterintuitive, but it reflects a fundamental principle of Bayesian model comparison: the marginal likelihood (or evidence) favors models that achieve higher average likelihood across the prior parameter space, rather than merely attaining a higher likelihood at a single parameter value. This mechanism embodies a trade-off between goodness-of-fit and model complexity. Here, complexity arises not only from the number of parameters but also from the extent of prior support. On the one hand, models with larger number of parameters get penalized as long as the goodness-of-fit does not improve significantly. On the other hand,
a more concentrated (informative) prior reduces the effective complexity and can be favored, provided the likelihood is sufficiently aligned with the prior. Conversely, a broad prior relative to the likelihood implies that much of the prior mass lies in regions of low likelihood, thereby reducing the evidence. In our analysis below, although the two competing models we compare share the same prior range for the tension, the first model (conventional template) has a narrower likelihood. As a result, the prior is effectively less informative in the region where the data constrain the parameters, which leads to a stronger Occam penalty and lower evidence compared to the non-conventional template, even if the latter contains more than one parameter.

\begin{figure}
    \centering
    {\sffamily LISA's model-comparison ability for metastable cosmic-string templates}\\[0.25em]
    \includegraphics[width=0.6\linewidth]{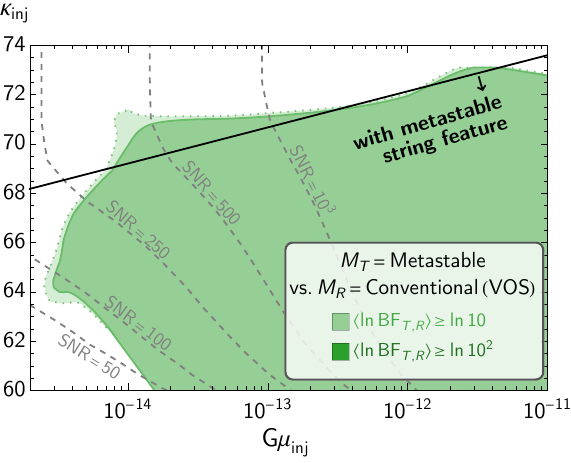}\\[-1em]
    \caption{LISA's ability to discriminate the metastable-string template in Sect.~\ref{subsubsec:metastable_strings} from the conventional VOS template in Sect.~\ref{sec:GW_semi_analytic}, assuming the signal from the true model $M_T$ is of the metastable strings with parameters $\{G\mu_{\rm inj}, \kappa_{\rm inj}\}$. The colored regions correspond to the parameter space where LISA can confidently distinguish the two models (favouring the true model), and each color denotes different discriminating power, expressed by the averaged $\langle{\ln \rm BF}_{T,R}\rangle$ in Eq.~\eqref{eq:average_BF} of Sect.~\ref{subsec:ModelComparisonVOSvsBOSvsDOF}. The region below the black solid line [Eq.~\eqref{eq:metastable_stable_network}] is where the low-frequency cutoff from the metastable strings is present on the spectrum and allows us to distinguish the metastable template from the conventional scenario. The gray dashed contours correspond to SNR of the metastable-string GWB, which are also shown in Fig.~\ref{fig:snr_beyond_1}-top-right.}
    \label{fig:bayes_metastable}
\end{figure}

\paragraph{Example I. Metastable strings.}
The colored regions in Fig.~\ref{fig:bayes_metastable} show the parameter space $\{G\mu_{\rm inj},\kappa_{\rm inj}\}$ where LISA can distinguish decisively the signal---built from the metastable-string template as the true model $M_T$---from the conventional template as its rival model $M_R$, {\it i.e.}~the averaged logarithmic Bayes factor [defined in Eq.~\eqref{eq:average_BF}] $\langle \ln {\rm BF}_{T,R}\rangle \geq \ln 10^2$.
We choose the VOS template from Sect.~\ref{sec:GW_semi_analytic} to represent the conventional case, as our metastable-string template is built from the VOS loop-number density; see Eq.~\eqref{eq:loop_number_density_metastable}.

\begin{figure}
    \centering
    {\sffamily LISA's model-comparison ability for superstring templates}\\[0.25em]
    \includegraphics[width=0.6\linewidth]{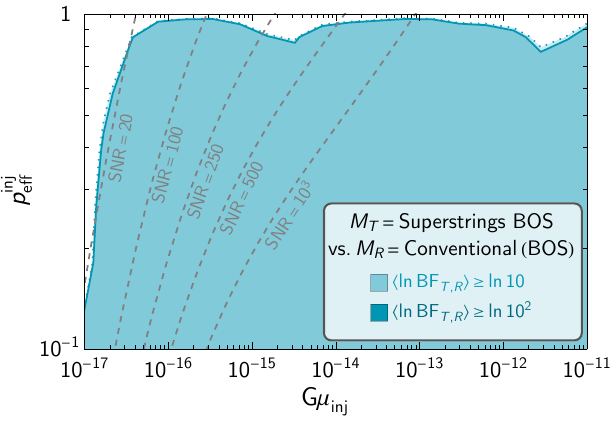}\\[-1em]
    \caption{LISA's ability to discriminate the superstring-BOS template in Sect.~\ref{subsubsec:metastable_strings} from the conventional BOS template in Sect.~\ref{sec:GW_semi_analytic}, assuming the signal from the true model $M_T$ is of the superstring template with parameters $\{G\mu_{\rm inj}, p_{\rm eff,inj}\}$. The colored regions correspond to the parameter space where LISA can confidently distinguish the two models, and each color denotes different discriminating power, expressed by the averaged $\langle{\ln \rm BF}_{T,R}\rangle$ in Eq.~\eqref{eq:average_BF} of Sect.~\ref{subsec:ModelComparisonVOSvsBOSvsDOF}. The gray dashed contours correspond to SNR of the superstring GWB, which are also shown in Fig.~\ref{fig:snr_beyond_1}-bottom-left.}
    \label{fig:bayes_superstring}
\end{figure}

As expected, LISA can only distinguish the metastable string template from the conventional template below the black solid line in Fig.~\ref{fig:bayes_metastable} [defined by Eq.~\eqref{eq:metastable_stable_network}] where the low-frequency cutoff feature is within the LISA frequency window. 
We show the dashed-gray contour for each SNR value, and note that the usual ${\rm SNR} = 10$ criterion cannot identify the metastable strings. One would need ${\rm SNR} > 50$ for $\kappa_{\rm inj} > 60$; however, the region with a confident detection of metastable-string template does not correspond to a fixed value of SNR.
Instead, the shape of this region is similar to areas of the $\kappa$-precision in Fig.~\ref{fig:recon_precision_metastable}-right.
This suggests that the ability to distinguish any beyond-conventional templates lies in reconstructing the BSM parameters that control the GWB spectral features.
In the current case, LISA requires a precision of $\Delta \kappa \lesssim 20\%$ to discriminate the metastable-string signal from the conventional one.

\paragraph{Example II. Small intercommutation probability.}
Fig.~\ref{fig:bayes_superstring} shows the region of parameter space $\{G\mu_{\rm inj}, p_{\rm eff,inj}\}$ where the superstring template from Sect.~\ref{subsubsec:super_string}---assumed to be the true signal---can be distinguished from the conventional template using LISA.
We focus on the case where both superstring and conventional templates are built from the BOS model (though, of course, we expect a similar result for the VOS model). For $p_{\rm eff} < 1$ and large $G\mu$, Fig.~\ref{fig:bayes_superstring} shows that the superstring spectrum can be distinguished well from the conventional template.
Nonetheless, we have seen in Sect.~\ref{sec:reconstruction} that conventional signals with $G\mu < 10^{-16}$ cannot be reconstructed with high precision due to its small amplitude. This explains why the model-comparison ability is lost for small $G\mu$ values, unless the signal amplitude is boosted via decreasing $p_{\rm eff}$; the superstring signal can then be reconstructed with a good precision in that case, making it distinguishable from the conventional template. Looking at the SNR values shown by the gray dashed lines in the figure, a signal with ${\rm SNR} \gtrsim 20$ can be robustly claimed to be originated from superstrings, as opposed to from a conventional template.

\section{Conclusion and outlook}
\label{sec:conclude}

This work shows LISA’s ability to explore early-Universe physics by assessing how effectively it can detect and reconstruct GWB signals from cosmic string scenarios, as well as distinguish among the various modelings. We focus on GWBs from Nambu-Goto cosmic-strings, for which state-of-the-art predictions of the background spectra are available. The first part of this work presents these templates in a survey style, accompanied by the publicly available \href{https://github.com/peerasima/cosmic-strings-GWB}{\faGithub\,repository} where we make their GWB spectrum templates available. As shown in Fig.~\ref{fig:template_summary}, we categorize signals into conventional templates that do not require any other BSM physics besides the existence of cosmic strings following standard scaling arguments, and beyond-conventional templates where either extra scaling assumptions (one case) or extra BSM physics (all other cases) are involved, leading to richer structures in the GWB spectra as compared to standard scaling cosmic strings. With a well defined recipe for calculating GWB spectra discussed in Sect.~\ref{sec:GWB_recap}, obtaining GWB templates from any modeling is straightforward once all ingredients—i.e., loop number density, GW emission from a single loop, and the cosmic history—are known.

We employ the SBI technique, implemented in \href{https://github.com/AndronikiDimitriou/GWBackFinder}{\faGithub\,{\tt GWBackFinder}}, to scrutinize LISA's ability to detect and reconstruct the cosmic string GWB signals, following the methodology outlined in Fig.~\ref{fig:flowchart}. 
With its pre-trained advantage of performing Bayesian predictions without requiring new simulations in the inference stage, this technique enables us to evaluate LISA's ability for any possibility of the true signal across the model's parameter space and address the key questions posed in the introduction:

\textit{\underline{I) Figures of merit \& II) Parameter space exploration:}}
With our pipeline, we can successfully reconstruct the parameters of both conventional and beyond-conventional templates, as shown by {\it e.g.}~posteriors in Figs.~\ref{fig:recon_bos_potato} and \ref{fig:recon_posterior_superstring}.
Using {\it precision} [Eq.~\eqref{eq:precision_recon}] as a measure of uncertainty from parameter reconstruction, Fig.~\ref{fig:keynote_accuracy_precision_reconstruction} shows the resulting precision as a function of the injected value of the string tension $G\mu$ for  conventional templates. 
For a signal with $G\mu \sim 3 \times 10^{-12}$, LISA can detect cosmic strings with the best precision in the tension, $\Delta G\mu \sim 2-3 \%$, while it can only reconstruct $G\mu$ at best up to an order of magnitude for $G\mu \lesssim 3 \times 10^{-17}$.
This information about reconstruction uncertainty cannot be obtained from the commonly-used SNR calculation.
Another interesting finding is the precision degradation for $G\mu \gtrsim 10^{-12}$, which is expected to come from the irreducible signal's variance.
Since this noise would serve as a limitation for signal reconstruction, we intend to investigate further this aspect in future work.

For beyond-conventional templates, the situation differs immensely as they involve multiple parameters (with the exception of the LRS model). Obviously, LISA cannot reconstruct the parameter when the feature associated with it in the spectrum lies outside the LISA frequency window. For example, our results show that LISA cannot probe extra DOFs, neither thermalized nor completely decoupled from SM particles, unless experimental and observational bounds are violated. When a beyond-conventional template feature is within LISA's sensitivity, we have seen that in many cases a realistic reconstruction is prone to degeneracies among the model parameters ---{\it e.g.}~in superstrings, current-carrying strings, or strings with modified GW emission ($\alpha$ and $q$)---which can degrade substantially the reconstruction precision of a single signal parameter. The ability to perform parameter reconstruction then becomes prior dependent. In these cases, we chart the LISA's reconstruction quality across the whole parameter space in terms of another quantity, the \textit{quality indicator} $Q$ [Eq.~(\ref{eq:quality_indicator})], which for a given parameter captures how wide the posterior is with respect to the prior. Future work should be dedicated to developing a systematic approach to analyzing such degeneracy problems and improve the reconstruction method, {\it e.g.}~using a combination of model parameters as a new variable for the analysis, using the information from the synergy with other detectors, or from some other multi-messenger astronomy input.

\underline{\textit{III) Model comparison:}} Employing for the first time the model-comparison methodology via the SBI technique, we see in Fig.~\ref{fig:bayes_vos_bos} that LISA is capable of decisively distinguishing between the conventional VOS and BOS templates for the same tension, for values $G\mu \gtrsim 5 \times 10^{-13}$. On the other hand, LISA cannot tell apart different SM DOFs evolutions for any $G\mu$.
Regarding the beyond conventional templates, we focus on examples of metastable strings and superstrings. Their results show large regions of parameter space where LISA can distinguish them from conventional templates and therefore claim the detection of extra BSM physics.

The next step for our work is to include the astrophysical foregrounds---i.e., galactic and extragalactic binaries---which will be guaranteed within the LISA sensitivity band, but were currently ignored in the analysis presented here. We will analyze in the next papers of our planned series of works for cosmic-string signals at LISA (see discussion about this at the end of Sect.~\ref{sec:intro}) how these foregrounds affect and degrade parameter reconstructions and model comparison abilities. Additionally, it will be interesting to see if LISA can distinguish between cosmological backgrounds and astrophysical foregrounds.
As a future direction, we also highlight that cosmic-string GWB signals span a broad frequency range, and thus they can reside within several detectors' windows. The analysis of a multi-detector synergy will involve signal and many detector noise parameters, and the SBI technique used here would be a natural approach to address it.

Last but not least, we note that the extra energy loss due to particle production by network loops, evident in field-theory string simulations (see  discussion in Sect.~\ref{subsec:FieldTheoryNetworks}), would not only scale down the GWB amplitude (for given tension), but also shift the spectrum frequency profile, as it affects how string loops shrink. While a frequency template for such field-theory strings is still missing, if it becomes eventually available, the methodology for parameter reconstruction and model comparison shown in this work will be straightforwardly applicable for analysis of this background within the LISA window.

\section*{Acknowledgement}
We thank Mark Hindmarsh, Mauro Pieroni, Ivan Rybak, Kai Schmitz,  G{\'e}raldine Servant, Lara Sousa, and Daniele Steer, as well as participants of the 2024 workshop ``\href{https://indico.ific.uv.es/event/7656/}{Towards a realistic forecast detection of Primordial Gravitational Wave Backgrounds}" and of the 2025 workshop ``\href{https://www.benasque.org/2025gwc/}{Dawn of gravitational wave cosmology}", for comments and insightful discussions.
We also acknowledge the LISA cosmology working group for interaction with many of their members.
This work is supported by the grants CIPROM/2022/69, EUR2022-134028, by the Generalitat Valenciana Grant PROMETEO/2021/083, and by Spanish Ministerio de
Ciencia e Innovaci\'on grant PID2023-148162NB-C22. AD and BZ acknowledge the support from Generalitat Valenciana through the “GenT
program”, ref.: CIDEGENT/2020/055. The authors gratefully acknowledge
the computer resources of ARTEMISA (ASFAE/2022/024) and the technical support provided by the Instituto de Fisica Corpuscular, IFIC(CSIC-UV). 

\appendix

\section{Derivation of cosmic-string GWB}
\label{app:GWB_derivation}
The local cosmic-string network is a long-lasting GW source which produces GW in a two-step process. First, the network produces loops along the cosmic history; later, these loops oscillate and lose their energy via GW emission.
We will start reviewing briefly each step of the calculations and later combining them to obtain the master formula~\eqref{eq:master_formula_GWB_strings} for the total GWB from the cosmic strings network.
The technicality for efficiently mass-producing templates of GWB spectra is discussed in appendix~\ref{app:mode_summation}.

Starting from a loop population of length within range $[l,l+dl)$ at the GW-emission time $t_e$ with a loop number density, $\#_{\rm loop}(t_e) = {\tt n}(l,t_e) dl$, we express the length $l$ in terms of GW frequency $f_e$ using Eq.\,\eqref{eq:GW_frequency_loop_length},
\begin{align}
    \#_{\rm loop}^j(t_e) = \frac{2j}{f_e}{\tt n}\left(\frac{2j}{f_e^2},t_e\right) (- d f_e).
\end{align}
Each loop oscillates with $j^{\rm th}$-harmonic and emits GW with the power $dE/dt_e = G\mu^2 P_j$ discussed in Sect.~\ref{subsec:singleLoopGWemission}.
Within time $[t_e,t_e+dt_e)$, the loops of mode $j$ thus emit GW signal of frequency $f_e$ with energy density
\begin{align}
    d \rho_{{\rm GW},e}^j =   \#_{\rm loop}^j(t_e) \times \frac{dE_j}{dt_e} \, dt_e = G\mu^2 P_j \left(\frac{2j}{f_e^2}\right) {\tt n}\left(\frac{2j}{f_e},t_e\right) (- d f_e) dt_e, 
\end{align}
which red-shifts as radiation until today,
\begin{align}
    d \rho_{\rm GW,0}^j=d \rho_{{\rm GW},e}^j \left[\frac{a(t_e)}{a(t_0)}\right]^4 = G\mu^2 P_j \left(\frac{2j}{f^2}\right) \frac{{\tt n}\left[\frac{2j}{(1+z)f},t_e(z)\right]}{H(z)(1+z)^6} d f dz.
\end{align}
The today's energy density spectrum from the mode $j^{\rm th}$ oscillation reads,
\begin{align}
    \frac{d \rho^j_{\rm GW,0}}{d \log f} &= G \mu^2 P_j \frac{2 j}{f} \int_{z_1}^{z_2}\frac{dz}{H(z)(1+z)^6} \, {\tt n}\left[\frac{2j}{(1+z)f},t(z)\right],\\
    &= G \mu^2 P_j \frac{2 j}{f} \int_{a_2}^{a_1}\frac{da}{H(a)}\left(\frac{a}{a_0}\right)^4 \, {\tt n}\left[\frac{2j}{f} \cdot \frac{a}{a_0},t(a)\right],
\end{align}
where the second line expresses the red-shift $z$ in terms of the scale factor $a$
\footnote{To deal with error from the numerical integrator, switching between the redshift $z$ and the scale factor $a$ might be needed.}.
We usually express the energy density spectrum in terms of the fraction of the total energy density of the Universe today, i.e., $\Omega_{\rm GW}^j(f) = \rho_{c,0}^{-1}(d \rho_{\rm GW,0}^j/d \log f)$. Overall, the GWB spectrum from the cosmic-string network is the sum of all modes\footnote{In fact, it is not possible to sum to infinitely large $k$-mode, as these fast oscillations can produce massive particles (when $f > \eta$); cf. e.g., \cite{Gouttenoire:2019kij}. However, the upper limit on $k$ is still larger than the sufficiently large $k$ requires for saturating the GW spectrum, e.g., $10^8$ modes as shown in Fig.~\ref{fig:jmax}} ; that is the master formula in Eq.\,\eqref{eq:master_formula_GWB_strings}.

\subsection{Mode summation}
\label{app:mode_summation}
The GWB spectrum in Eq.\,\eqref{eq:master_formula_GWB_strings} includes GWB contributions from many modes, which can be computationally expensive due to many integrations. 
However, there exists a trick \cite{Cui:2017ufi,Cui:2018rwi,Gouttenoire:2019kij} where the integration is required to be evaluated only once.
By observing the master equation~\eqref{eq:master_formula_GWB_strings}, we can rewrite it as 
\begin{align}
    \Omega_{\rm GW}(f) = \sum_{j=1}^{j_{\rm max}} P_j \tilde{\Omega}_{\rm GW}^{(j)}(f) = \sum_{j=1}^{j_{\rm max}} P_j \tilde{\Omega}_{\rm GW}^{(1)}(f/j),
    \label{eq:summation_trick}
\end{align}
where $(j)$ denotes the $j^{\rm th}$ mode, $j_{\rm max}$ is the maximum mode number which can be regarded as infinity, and the normalized GW spectrum of mode $j^{\rm th}$ is
\begin{align}
    \tilde{\Omega}_{\rm GW}^{(j)}(f) = \left(\frac{2j}{f}\right) \frac{G \mu^2}{3 H_0^2 m_{\rm Pl}^2} \int_{a_2}^{a_1} da \, \frac{1}{H(a)}\left(\frac{a}{a_0}\right)^4 \, {\tt n}\left[\frac{2j}{f} \cdot \frac{a}{a_0},t(a)\right] = \tilde{\Omega}_{\rm GW}^{(1)}\left(\frac{f}{j}\right).
    \label{eq:single_mode_simplify}
\end{align}
This means that we require only the GW spectrum of the fundamental mode $\tilde{\Omega}_{\rm GW}^{(1)}(f)$, and the spectra of any $j^{\rm th}$ can be retrieved via simply rescaling its frequency. The form Eq.\,\eqref{eq:summation_trick} works for both the approximated and numerical $P_j$ discussed in Sect.~\ref{subsec:singleLoopGWemission}; this differs slightly from the form originally used in \cite{Cui:2017ufi,Cui:2018rwi,Gouttenoire:2019kij} which applies only to the approximated $P_j$.

\subsection{Further technicality}
The above trick enables us to replace the GWB spectra of higher-$j^{\rm th}$ modes with the fundamental mode. Still, the summation of these spectra is a bottleneck for mass-producing the spectra.
We will discuss now the further technical details which allows us to calculate the total GWB spectrum faster.

\begin{figure}[t]
    \centering
    {\sffamily Effect of including more modes}\\
    \includegraphics[width=0.65\textwidth]{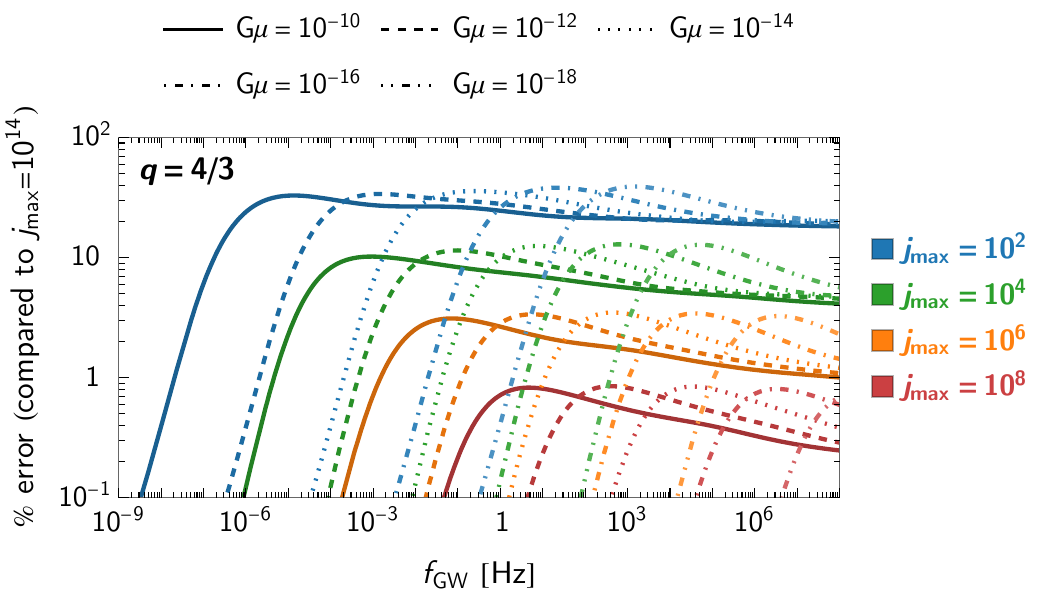}\\[-1em]
    \caption{The relative error---defined by $\%~ {\rm error} = 100 \times \left|{\Omega_{\rm GW}(f;j_{\rm max})/\Omega_{\rm GW}^{\rm ref}} -1\right|$---between spectra $\Omega_{\rm GW}(f;j_{\rm max})$ with different $j_{\rm max}$ in Eq.\,\eqref{eq:summation_trick} and the reference spectra $\Omega_{\rm GW}^{\rm ref} = \Omega_{\rm GW}(f;j_d = 10^{6},j_{\rm max}=10^{14})$.}
    \label{fig:jmax}
\end{figure}

\begin{figure}[t!]
    \centering
    {\sffamily Effect of the integration trick}\\
    \includegraphics[width=0.485\textwidth]{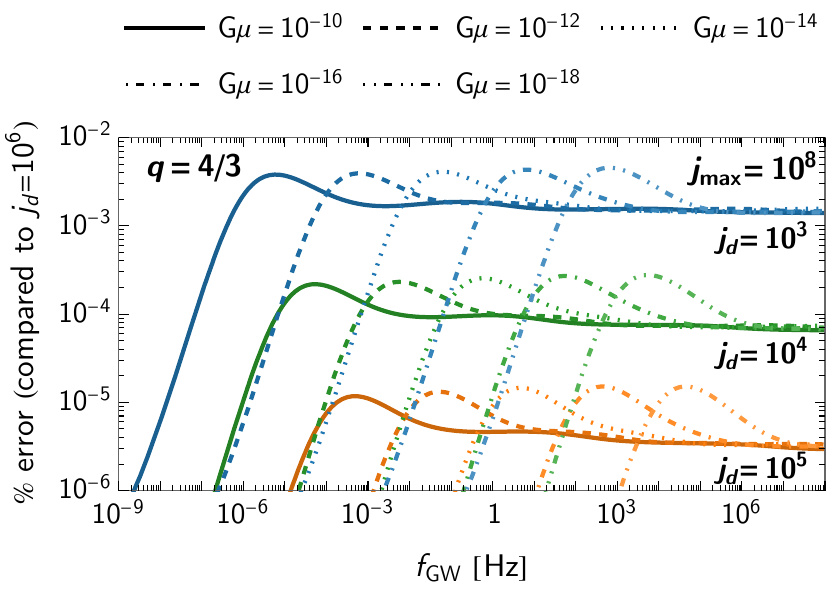}\hfill
    \includegraphics[width=0.485\textwidth]{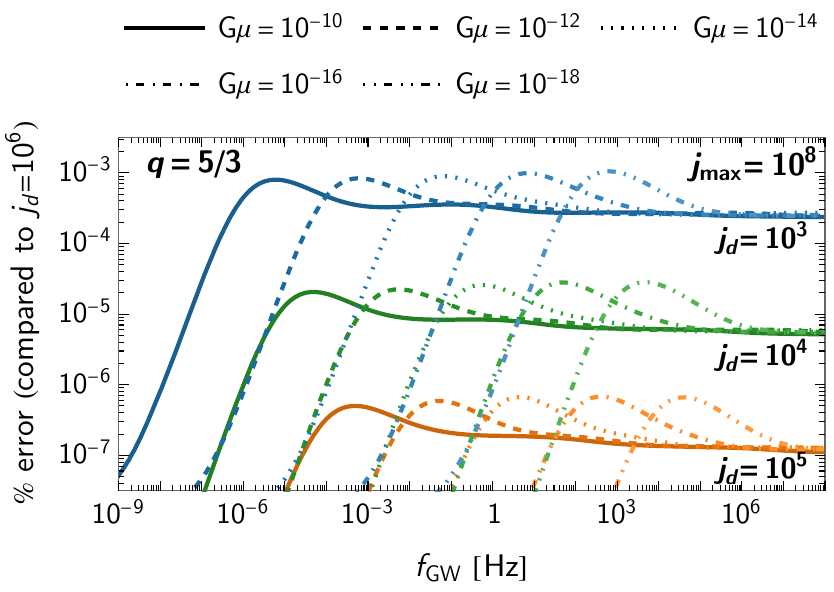}\\[-1em]
    \caption{The relative error---$\%~ {\rm error} = 100 \times \left|{\Omega_{\rm GW}(f;j_d,j_{\rm max})/\Omega_{\rm GW}^{\rm ref}}-1\right|$---between spectra with different discrete summed modes $\Omega_{\rm GW}(f;j_d,j_{\rm max})$ in Eq.\,\eqref{eq:discrete_sum_integration} and the reference spectra $\Omega_{\rm GW}^{\rm ref} = \Omega_{\rm GW}(f;j_d = 10^{6},j_{\rm max}=10^8)$. We show results for $q=4/3$ (left) and $q=5/3$ (right), which match well the analytic estimate of the error $\sim j_d^{-q}/\zeta(q)$.}
    \label{fig:summation_and_integration}
\end{figure}

\paragraph{Maximum modes.}
As discussed in Sect.~\ref{subsec:singleLoopGWemission}, the GW emission from higher-$j^{\rm th}$ mode is suppressed by a factor $j^{-q}$, so the GWB contribution from $(j \gg 1)$ mode is much smaller than fundamental-mode contribution and can be treated as a correction to the overall GWB spectrum.
Hence, the inclusion of higher modes have negligible effects, and one can truncate the summation to some large $j_{\rm max}$.
Fig.\,\ref{fig:jmax} shows the GWB spectra from the summation with different $j_{\rm max}$, where we use the integration tricks to sum the modes with $j > 10^6$. Note that the error coming from the integration trick in Fig.~\ref{fig:summation_and_integration} is much smaller than the error from the mode truncation.
We use $j_{\rm max} = 10^8$ in this work, as the inclusion of higher modes contribute to less than 1\% in $\Omega_{\rm GW}$.
The $\mathcal{O}(0.1\%)$ correction can be obtained by summing up to $j_{\rm max} = 10^{12}$; however, it modifies the GW spectrum outside LISA frequency window.
Moreover, the reconstruction precision of LISA is at best $\mathcal{O}(2-3\%)$ in $G\mu$, which can be translated to the precision $\mathcal{O}(1\%)$ in $\Omega_{\rm GW}$, as shown in Fig.~\ref{fig:keynote_accuracy_precision_reconstruction} of Sect.~\ref{subsec:quality_recon_conven}.
This justifies our choice of $j_{\rm max} = 10^{8}$.

\paragraph{Summation and integration.}
Since the contributions for larger $j^{\rm th}$ modes can be treated as corrections, it is possible to replace the summation for very high $j$ by the integration,
\begin{align}
    \Omega_{\rm GW}(f) = \sum_{j=1}^{j_{\rm max}} P_j \tilde{\Omega}_{\rm GW}^{(j)}(f) \simeq \sum_{j=1}^{j_{d}} P_j \tilde{\Omega}_{\rm GW}^{(j)}(f) + \int_{j_d}^{j_{\rm max}} P_j \tilde{\Omega}_{\rm GW}^{(j)}(f) \equiv \Omega_{\rm GW}(f;j_d,j_{\rm max}).
    \label{eq:discrete_sum_integration}
\end{align}
The error occurs from this integration trick is $\sim j_{d}^{-q}/\zeta(q)$, which is estimated from the leading order of the Euler-Maclaurin formula\footnote{The leading-order correction is $P_{j_d} \tilde{\Omega}_{\rm GW}^{(j_d)}(f) = P_{j_d} \tilde{\Omega}_{\rm GW}^{(1)}(f/j_d) = \Gamma \, j_d^{-q} \tilde{\Omega}_{\rm GW}^{(1)}(f/j_d)/\zeta(q)$ where the last step assumes the approximated $P_j$ in Eq.\,\eqref{eq:grav_emission_power_loop_j}. For the scale-invariant part of the $(j=1)$ spectrum with amplitude $\tilde{\Omega}_{\rm GW}^{(1)}(f/j_d)=\tilde{\Omega}_{\rm GW}^{(1)}$, the actual summation yields $\Omega_{\rm GW} = \Gamma \tilde{\Omega}_{\rm GW}^{(1)}$. Hence, the ratio of error to the amplitude from the actual summation is $j_d^{-q}/\zeta(q)$.} that relates a discrete sum to a integral \cite{Blanco-Pillado:2024aca}.
Fig.\,\ref{fig:summation_and_integration} shows the relative errors between the spectra with several $j_D$ values and the reference spectra with large $j_D = 10^6$.
To optimize the computational resources, we calculate our spectra using $j_D = 10^{4}$, i.e., the estimation's error is less than $\mathcal{O}(10^{-4}\%)$ and $\mathcal{O}(10^{-5}\%)$ of the actual summation results for $q=4/3$ and $5/3$, which is much smaller than the precision from the reconstruction method.

\subsection{GWB from different loop populations}
\label{app:pieces_CS_GWB}
\begin{figure}[t!]
    \centering
    \includegraphics[width=0.65\textwidth]{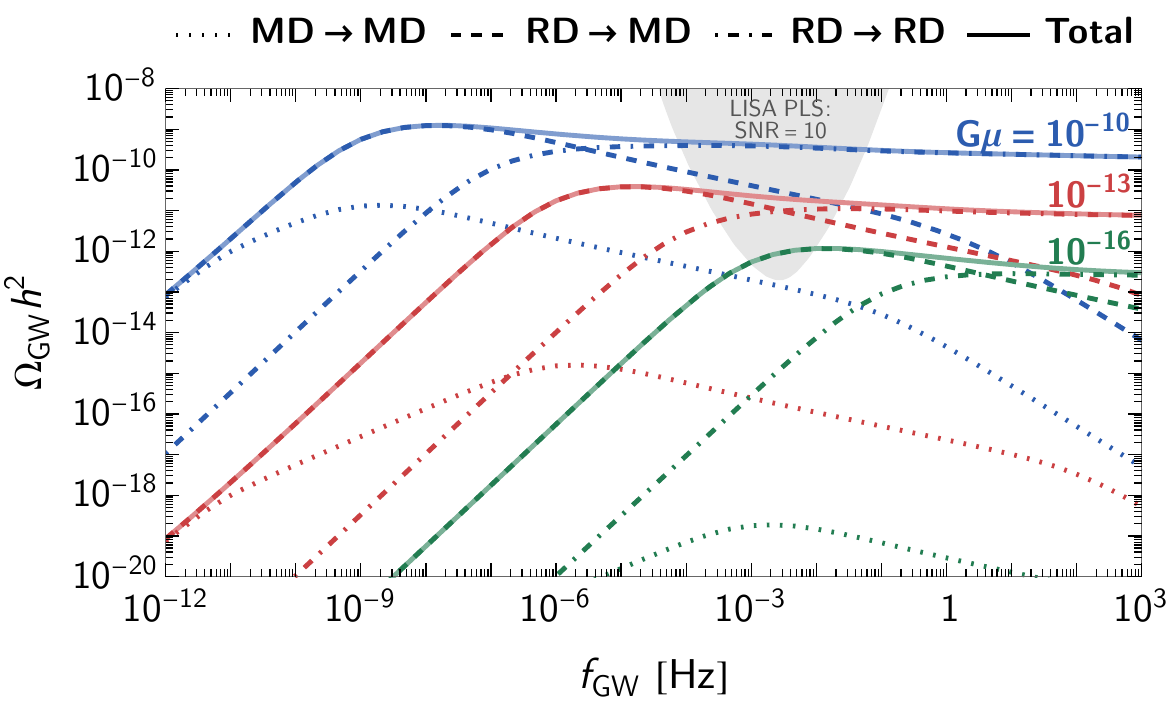}\\[-1em]
        \caption{Assuming BOS model, we show three GW components---corresponding to the eras when loops are formation and when they emit GW---for different $G\mu$ values. Loops formed and emitting GW during RD era lead to RD\,$\rightarrow$\,RD contribution; loops formed during RD but emitting GW in MD era are denoted by RD\,$\rightarrow$\,MD; loops formed in MD are labeled with MD\,$\rightarrow$\,MD. We see that, for $G\mu \leq 10^{-10}$, the MD\,$\rightarrow$\,MD contribution is suppressed within LISA window, compared to the contributions from loops formed during the RD era, independent of when these loops decay. The solid colored line shows the sum of all contributions. The gray region is the LISA PLS of ${\rm SNR}\,=\,10$.}
    \label{fig:BOS_each_contribution}
\end{figure}

A GW spectrum from cosmic strings is composed of  contributions from many loop populations.
Although each loop population produced GW with a broad frequency range due to loops' length shrinking and the redshift of GW, the GW signal today has a maximal amplitude at frequency estimated by the frequency-temperature relation in \cite{Cui:2017ufi,Cui:2018rwi,Gouttenoire:2019kij}.
Using BOS template in Sect~\ref{sec:GW_simulated}, Fig.\,\ref{fig:BOS_each_contribution} shows GW contributions from three loop populations classified by the loop production time and the GW emission  happening before or after the matter-radiation equality $t_{\rm eq}$: \emph{i)} loops produced and emitting GW in radiation era (denoted RD$\rightarrow$RD) \emph{ii)} loops produced in radiation era and emitting  GW in matter era (denoted RD$\rightarrow$MD) \emph{iii)} loops produced and emitting GW in matter era (denoted MD$\rightarrow$MD).
Earlier versions of this plot can be found e.g., in \cite{Blanco-Pillado:2017oxo,Cui:2018rwi,Gouttenoire:2019kij}. We show it again just for the completion.
GW from loops produced during radiation era (RD$\rightarrow$RD and RD$\rightarrow$MD) dominates the GW spectrum in the LISA frequency window.
With $G\mu$ as large as $10^{-10}$, the MD$\rightarrow$MD population contributes to $\Omega_{\rm GW}$ as small as 0.1\% of that from radiation-era loop populations.
Note that even a change in the cosmic expansion history at $t_{\rm eq}$ results in three different loop populations, which have distinct spectral features. Various cases of nonstandard cosmic histories (e.g., \cite{Cui:2017ufi,Cui:2018rwi,Gouttenoire:2019kij,Gouttenoire:2019rtn,Gouttenoire:2021jhk,Co:2021lkc,Ghoshal:2023sfa}) lead to more loop populations and many more interesting GWB features.

\section{Generalized VOS equations}
\label{app:generalized_VOS_current}

The loop number density from the VOS model (section~\ref{sec:GW_semi_analytic}) depends on the string network parameters, which evolve according to the system of equations called Velocity-dependent One-Scale (VOS) model.
We consider the generalized VOS equations \cite{Martins:2020jbq}, which has been extended to the case of current-carrying strings and describe three variables: the correlation length of long strings $L$, the long string's root-mean-squared velocity $\bar{v}$, and the current strength $Y$ of the long strings.
For the standard VOS model \cite{Martins:1995tg,Martins:1996jp, Martins:2000cs, Sousa:2013aaa, Sousa:2014gka} for cosmic strings without current ($Y=0$), we can neglect the evolution equation of $Y$.
The generalized VOS equations read \cite{Martins:2020jbq},

\begin{align}
    \frac{dL}{dt} &= \frac{H L}{1+ Y}(1+ \bar{v}^2 + 2Y) + \frac{b \tilde{c} \bar{v}}{2 \sqrt{1+Y}},
    \label{eq:generalized_VOS_equations_1}\\
    \frac{d\bar{v}}{dt} &= \frac{1-\bar{v}^2}{1+Y}\left[\frac{k(\bar{v})(1-Y)}{L \sqrt{1+Y}}-2 H \bar{v}\right],
    \label{eq:generalized_VOS_equations_2}\\
    \frac{dY}{dt} &= 2 Y \left[\frac{k(\bar{v})\bar{v}}{L\sqrt{1+Y}} - H\right] - \frac{\tilde{c} \bar{v}}{L}(b-1)\sqrt{1 + Y},
    \label{eq:generalized_VOS_equations_3}
\end{align}
where $\tilde{c} \simeq 0.23$ \cite{Martins:2000cs} is the loop chopping coefficient\footnote{When the current is present, the efficiency of loop chopping from long strings could differ from the current-less case, such that $\tilde{c}$ deviates from 0.23. Nonetheless, its precise value should be determined from the numerical simulation, and we will use $\tilde{c} \simeq 0.23$ throughout this work.}, the momentum variable is defined by
\begin{align}
    k(\bar{v}) = \frac{2\sqrt{2}}{\pi}\left(\frac{1 - 8\bar{v}^6}{1 + 8\bar{v}^6}\right) \left(1- \bar{v}^2\right) \left(1 + 2\sqrt{2}\bar{v}^3\right),
\end{align}
and $b = \sqrt{1+ Y}$ is chosen such that the loop production (the terms with $\tilde{c}$) is the same as the $Y=0$ case \cite{Martins:2020jbq}.

\section{Different implementation of BOS models}
\label{app:diff_model_calculation}
\begin{figure}[t!]
    \centering
    \includegraphics[width=0.55\textwidth]{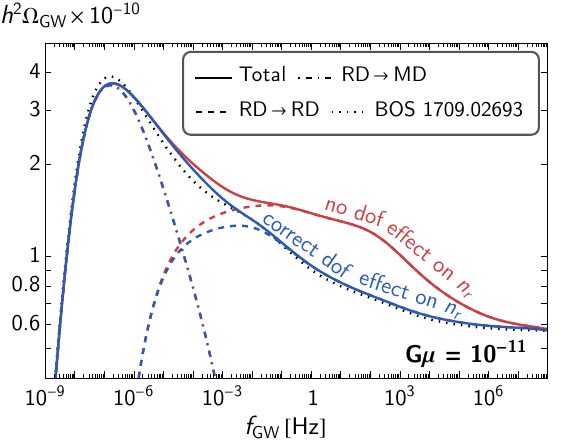}\\[-1em]
    \caption{GW spectra from different implementation of BOS templates, for $G\mu = 10^{-11}$.
    The dashed line is the RD contribution [GW emitted when $z > z_{\rm eq}$], the dot-dashed line is the RD$\to$MD contribution [loops produced when $z > z_{\rm eq}$ but GW emitted when $z <z_{\rm eq}$]. The solid lines are the sum of both contributions. The contribution from loops produced during the MD era is negligible. The blue curves include the dof effect on the dilution of loop number density [Eq.~\eqref{eq:BOS_loop_number_density}], while the red curve uses the DOF effect on the GW emission only [obtained from ${\tt n}_{r, \rm approx}(l, t) \simeq 0.18 t^{-3/2}(l + \Gamma G \mu t)^{-5/2}$]. The black-dotted curve is the GWB spectrum provided by Ref.~\cite{Blanco-Pillado:2017oxo}.}
    \label{fig:diff_calculation}
\end{figure}

As discussed in Sect.~\ref{sec:GW_simulated}, the usual  formula of the loop number density of loop formed during radiation era---i.e. ${\tt n}_{r, \rm approx}(l, t) \simeq 0.18 t^{-3/2}(l + \Gamma G \mu t)^{-5/2}$---neglects the effect of dof evolution.
Figure~\ref{fig:diff_calculation} show a comparison between the results when the dof effect is omitted in red and when the dof effect is correctly implemented in the loop number density in blue.
The reference spectrum generated by Ref.~\cite{Blanco-Pillado:2017oxo} is also shown as black dotted line.
Note that a slight difference between our result in the solid blue line and that of Ref.~\cite{Blanco-Pillado:2017oxo} is caused by different $g_*$--$g_{*s}$ modelings. Nonetheless, according to the result in Sect.~\ref{subsec:ModelComparisonVOSvsBOSvsDOF} we do not expect such slight deviation to be distinguishable by LISA.

\section{LISA noise model and data generation}
\label{app:LISA_noise}

\paragraph{LISA noise:}
By adopting time-delay-interferometry (TDI)~\cite{Tinto:2020fcc}, most of the instrumental LISA noises can be optimally cleaned, except for two effective contributions: the {\it interferometry metrology system} (IMS) and {\it acceleration} noises. The analytical approximations of their power spectral densities are given by~\cite{LISA:2017pwj,LISAnoise}
\begin{align}
    P_{\rm IMS} =&~ 1.6 \times 10^{-43} A_P^2 \left[1 + \left({2\, {\rm mHz}}/{f}\right)^4\right] x_f^2 ~ {\rm Hz}^{-1},\\
    P_{\rm acc} =&~ 7.7374 \times 10^{-46} A_{\rm acc}^2 \left[1 + \left({0.4\, {\rm mHz}}/{f}\right)^2\right] \left[1 + \left({f}/{8\, {\rm mHz}}\right)^4\right] x_f^{-2} ~ {\rm Hz}^{-1},
\end{align}
where $x_f \equiv {2\pi f L}/{c}$, with $L = 2.5 \times 10^{9} \, {\rm m}$ LISA's arm length (assuming a perfect equilateral triangular configuration) and $c$ the speed of light. We assume uniform priors for the noise parameters, centered at $A_P = 15$ and $A_{\rm acc} = 3$ with 20\% margin~\cite{LISAnoise,SciRD},  motivated by the range between the allocated noise budget and the
current best estimate, see Fig.~7.1 of Ref.~\cite{LISA:2024hlh}. 

To separate noise from signal, we consider the three uncorrelated data streams, known as A, E, T \cite{Hogan:2001jn,Adams:2010vc}, which are transformed from the original TDI channels.
The power spectrum densities of the LISA noise in the AET basis are diagonalizable, with $N_{\alpha \beta} = 0 $ for $\alpha \neq \beta$, where $\alpha,\beta \in \{{\rm A,\,E,\,T\}}$. Following~\cite{Flauger:2020qyi}, we write
\begin{align}
 N_{\rm AA}(f;A_{P},A_{\rm acc}) = & N_{\rm EE}(f;A_{P},A_{\rm acc})\nonumber \\
 =8\sin^{2} & x_f \left[ 4(1+\cos x_f +\cos^{2} x_f)P_{\rm acc}(f;A_{\rm acc}) + \left(2+\cos x_f \right)P_{\rm IMS}(f;A_{P}) \right],\\
N_{\rm TT}(f;A_{P},A_{\rm acc}) =&~16\sin^{2} x_f \left[2(1-\cos x_f )^{2}P_{\rm acc}(f;A_{\rm acc}) + (1-\cos x_f )P_{\rm IMS}(f;A_{P})\right].
\end{align}
These detector noises can be translated into an equivalent energy-density power spectrum through 
\begin{align}
\Omega_{\rm noise}^{\alpha\beta}(f; A_{P},A_{\rm acc})\equiv \frac{4\pi^{2}f^{3}}{3H_{0}^{2}}\frac{N_{\alpha\beta}(f;A_{P},A_{\rm acc})   }{16 x_f^2 \sin^2\left(x_f\right) \tilde{R}_{\alpha\beta}(x_f)} \,,~~~\alpha, \beta = {\rm A, E, T}\,,
\label{eq:Omeganoise}
\end{align}
where $\tilde{R}_{\alpha\beta}(x_f)$ are the (geometry-dependent part of the) response functions of LISA, which we take from Ref.~\cite{Flauger:2020qyi}, where a detailed derivation can be found in their Appendix A.3. While their exact functional form depend
on (not very illuminating) expressions, we note that simple
analytic approximations can still be written down~\cite{Robson:2018ifk,Flauger:2020qyi} as $\tilde{R}_{\rm AA}(f) = \tilde{R}_{\rm EE}(f) \approx {9}/{(20 + 14 x_f^2)}$ and $\tilde{R}_{\rm TT}(f) \approx 9x_f^6/(3.6 \times 10^{4} + 14 x_f^8)$. We remark that in our present work we use, in any case, the exact functional form, as given in Ref.~\cite{Flauger:2020qyi}.

\paragraph{Mock data generation:}
Following~\cite{Flauger:2020qyi} we consider that the data collected during the science run of LISA spans over $\sim$ 3 years\footnote{While the nominal mission time is 4 years, only $75\%$ of the time will correspond to the science run.}, and will be segmented into 94 chunks of 11.5 days each, which are labeled as $j=1,2,...,94$. For each chunk, the data spans within the range $[3 \times 10^{-5}$, $5 \times 10^{-1}]$ Hz with a frequency spacing of $\Delta f=\textrm{(11.5 days)}^{-1} \simeq 1 \, \mu{\rm Hz}$, i.e.~data are stored in $\sim 5 \cdot 10^5$  bins located at the frequencies $f_i \equiv i\cdot \Delta f\,,~i = 1, 2, 3, ...$.

Following \cite{Dimitriou:2023knw}, for each frequency $f_i$ of each chunk $j$, within each channel $(\alpha, \beta)$, we generate data ($D$) which are the sum of random realizations of a GWB signal ($S$) and of the detector noise ($\mathcal{N}$), 
\begin{align}
  D_{i,j}^{\alpha\beta} = S_{i,j} + \mathcal{N}_{i,j}^{\alpha\beta} ,
  \label{eq:data}
\end{align}
where 
\begin{equation}
    S_{i,j}=\frac{1}{2}\left|{G_{1}\left(0,\sqrt{\Omega_{\rm GW}(f_{i})}\right)+iG_{2}\left(0,\sqrt{\Omega_{\rm GW}(f_{i})}\right)}\right|^{2}
\end{equation}
and
\begin{equation}
\mathcal{N}_{i,j}^{\alpha\beta}=\frac{1}{2}\left|{G_{3}\left(0,\sqrt{\Omega_{\rm noise}^{\alpha\beta}(f_{i})}\right)+iG_{4}\left(0,\sqrt{\Omega_{\rm noise}^{\alpha\beta}(f_{i})}\right)}\right|^{2}.
\end{equation}
with $G_k(0,\sigma)\in\mathbb{R}$ a random number generator from a Gaussian distribution of zero mean and standard deviation $\sigma$.
The data averaged over time chunks at each frequency $f_i$ is
\begin{equation}
    \bar{D}_i^{\alpha \beta} = \frac{1}{N_c}\sum_{j=1}^{N_c} D_{i,j}^{\alpha \beta} \, ,
\end{equation}
where $N_c=94$ is the total number of chunks.

\paragraph{Coarse-graining of data:} Without losing much information in the high-frequency bins where $\Delta f/f \ll 1$, we consider a coarse-graining of the data to reduce the computational load. 
We {\it re-bin} the data within the frequency range of $[10^{-3},0.5]$ Hz into 1000 log-spacing `macro' bins with  $\Delta\log(f/{\rm Hz}) \simeq 2.7\times 10^{-3}$, while we keep the original finer bins in the $[3\times 10^{-5},10^{-3}]$ Hz range. The entire LISA window for each data chunk contains then 1970 bins in total.
The re-binned data is represented by coarse-grained frequencies and data $(\tilde f, \tilde D)$, as~\cite{Giese:2021dnw}
\begin{equation}
\label{eq:weightedFreqs}
{\tilde f}_k \equiv \sum_{i\in{\rm bin}\,k} w_i f_i ~ ~ {\rm and} ~ ~ \tilde{D}_k \equiv \sum_{i\in{\rm bin}\,k} w_i \bar{D}_i~.
\end{equation}
where we omitted the noise indices $\{\alpha,\beta\}$ for clarity, $k$ enumerates the macro-bins, and the weight of the re-binned data, which depends on the noise parameters ${\bf n} \equiv \{A_P,A_{\rm acc}\}$,  is defined\footnote{While we follow previous analyses for this coarse-graining strategy, we note that the resulting binning depends on the a-priori chosen values of the noise parameters. Consequently, when applied to an observed dataset, it is not guaranteed that the procedure is optimal. } by $
w_i \equiv w(f_i; {\bf n}) = {[\Omega_{\rm noise}(f_i; {\bf n})]^{-1}}/{ \sum_{l\in{\rm bin}\,k} [\Omega_{\rm noise}(f_l; {\bf n})]^{-1} }$.

\section{Summary of our SBI technique}
\label{app:SBI_summary}

We use an improved version of the SBI methodology implemented recently in \cite{Dimitriou:2023knw} by some of us. The most important improvement concerns the procedure of inferring the signal parameters which, contrary to the previous implementation using a 2-step  procedure, it is done now simultaneously together with the inference of the noise parameters, in a single step. 
A qualitative picture of our procedures is shown in the flow charts of Fig. \ref{fig:flowchart}. We use the Neural Posterior Estimation (NPE) type of SBI, on the one hand, for the signal reconstruction of a given model (see left panel). We generate mock data from the latter and the noise spectrum, and obtain posterior distributions of the signal parameters with NPE. On the other hand, for model comparison purposes (see right panel of the same figure) we adopt a Neural Likelihood Estimation (NLE) procedure. 
Upon building mock data from a given ``true" model, we estimate via NLE the {\it evidence} of both the true model and a ``rival'' model. We confront these by computing the expected Bayes factor, where the average is performed across mock data generated from the true model.

\section{MCMC likelihood}
\label{app:mcmc}
In this appendix we specify the likelihood we assumed when implementing the MCMC procedure, which we compare with our SBI method in sect. \ref{subsec:sbi_vs_MCMC}.  We follow the literature~\cite{Verde_2003} and assume the following likelihood
\begin{equation}
    \log {\cal L}_{\rm G+LN} = \frac{1}{3}\log{\cal L}_{\rm G}+
\frac{2}{3}\log{\cal L}_{\rm LN}~,
\label{eq:likelihood}
\end{equation}
where the Gaussian and log-normal contributions read,
\begin{align}
\log{\cal L}_{\rm G}(G\mu, A_{P},A_{\rm acc}) =& -\frac{N_c}{2}
\sum_{\alpha, \beta}\sum_k n_{\alpha \beta}^{(k)} \left[
  \frac{D^{\rm th}_{\alpha \beta}(f_{\alpha \beta}^{(k)},G\mu, A_{P},A_{\rm acc})-\bar D_{\alpha \beta}^{(k)}}
  {D^{\rm th}_{\alpha \beta}(f_{\alpha \beta}^{(k)},G\mu, A_{P},A_{\rm acc})}
\right]^2~,
\label{eq:Gaussian}\\
\log{\cal L}_{\rm LN}(G\mu, A_{P},A_{\rm acc}) =& -\frac{N_c}{2}
\sum_{\alpha, \beta}\sum_k n_{\alpha \beta}^{(k)}
\log^2 \left[ \frac{D^{\rm th}_{\alpha \beta}(f_{\alpha \beta}^{(k)},G\mu, A_{P},A_{\rm acc})}{D_{\alpha \beta}^{(k)}}\right]~.
\label{eq:lognormal}
\end{align}
Here, the theoretical prediction is,
\[D^{\rm th}_{\alpha \beta}(f_{\alpha \beta}^{(k)},G\mu, A_{P},A_{\rm acc})=h^{2} \Omega_{\rm GW}(f_{\alpha \beta}^{(k)},G\mu, A_{P},A_{\rm acc})+h^{2} \Omega_{\rm noise}^{\alpha \beta}(f_{\alpha \beta}^{(k)},G\mu, A_{P},A_{\rm acc})~,\]
where the indices $\alpha, \beta$ run over the channel combinations, the index $k$ denotes the coarse-grained data points, and $n_{\alpha \beta}^{(k)}$ is defined as the number of points within the bin-$k$ for the cross-spectrum of channels $\alpha$ and $\beta$. In the AET basis, $\log \mathcal{L}_{\rm G+LN}$ reduces, in any case, to the sum of diagonal elements $\alpha = \beta$. 

Note that this likelihood, Eq.~\eqref{eq:likelihood}, typically adopted in the MCMC analyses, is not the true likelihood of the adopted simulation process, although it turns out to be a good approximation.

\section{Bayesian model comparison}
\label{app:model_comparison}

In this appendix we present the standard formalism for performing Bayesian model comparison, whose results are presented in sect.\ref{subsec:ModelComparisonVOSvsBOSvsDOF}.

Let us recall first that the probability that a model $M_i$ could be behind certain observed data $D$, is equivalent to the posterior of the model itself, $\mathbb{P}(M_i|D) \propto \mathbb{P}(M_i) \mathbb{P}(D|M_i)$.
Given a priori no preference on any model ({\it i.e.}~equal model priors $\mathbb{P}(M_i)$), the ratio between two models' posteriors (given the same data) is then simplified to the so-called \emph{Bayes factor} (BF),
\begin{equation}
    \frac{\mathbb{P}(M_i|D)}{\mathbb{P}(M_j|D)} = \frac{\mathbb{P}(D|M_i)}{\mathbb{P}(D|M_j)} \equiv {\rm BF}_{i,j}(D),
    \label{eq:bayes_factor_definition}
\end{equation}
where $\mathbb{P}(D|M_i)$ is the \emph{evidence of model} $M_i$, which corresponds to the likelihood of the data $\mathbb{P}(D|\vec{\theta},M_i)$, marginalized over the set of all parameters $\vec{\theta}$ of $M_i$,
\begin{equation}
    \mathbb{P}(D|M_i)=\int \mathbb{P}(D|\vec{\theta},M_i) \mathbb{P}(\vec{\theta}|M_i) d\vec{\theta} \, ,
\end{equation}
with $\mathbb{P}(\vec{\theta}|M_i)$ the prior over the model parameters.

To answer question {\it I)} in sect.\ref{subsec:ModelComparisonVOSvsBOSvsDOF}, we follow the methodology outlined in Fig.~\ref{fig:flowchart}.
Focusing on data $D(\vec\theta_{\rm inj})$ which is built from a (true) template $M_{T}$ with a fixed set of parameters $\vec\theta_{\rm inj}$, we calculate the evidences of the true model and of a rival model $M_{R}$, and determine the corresponding Bayes factor.
Sticking to the SBI approach, we pre-train an approximation $\hat{\mathbb{P}}(D|\vec{\theta},M_i)$ of the true likelihood, which does not require further simulations to be evaluated. We adopt the Neural Likelihood Estimation (NLE) SBI method \cite{2018arXiv180507226P} to obtain $\hat{\mathbb{P}}(D|\vec{\theta},M_i)$, which is similar in spirit to the NPE approach we adopted to estimate the posterior distributions of the model parameters.
In practice, we approximate the evidence of model $M_i$ as
\begin{equation}
    \mathbb{P}(D|M_i) \approx \int \hat{\mathbb{P}}(D|\vec{\theta},M_i) \mathbb{P}(\vec{\theta}|M_i) d\vec{\theta} \, .
\end{equation}
Since the data $D$ is prone to the statistical fluctuations from the data generation, we consider the average of the (log) Bayes factor over the sampled datasets, defined by
\begin{equation}
    \left\langle \ln {\rm BF}_{T,R} \right\rangle = \int dD~ \mathbb{P}(D|M_{T}) ~\ln {\rm BF}_{T,R}(D) 
    \approx
    \frac{1}{N}
    \sum_{j=1}^N
    \ln {\rm BF}_{T,R}(D_j)
    \,,
    \label{eq:average_BF}
\end{equation}
where
$D_{j}$ are $N$ dataset samples from $\mathbb{P}(D|M_{T})$, which are nothing but the simulated datasets from $M_{T}$. In the last step, the above integral is approximated with the Monte Carlo method.
Note that, in general, ${\rm BF}_{T,R}(D)$ could be smaller than unity ({\it i.e.}~$\ln {\rm BF}_{T,R} < 0$) for a given dataset $D$, which means the wrong model $M_{R}$ is favored by the data. However, on average, $M_{T}$ will always be favored, since $\langle\ln {\rm BF}_{T,R}\rangle$ in Eq.~\eqref{eq:average_BF} is positive definite\footnote{This follows from the mathematical properties of Eq.~\eqref{eq:average_BF}, which is actually a {\it Kullback-Leibler} (KL) divergence, describing the similarity between the two distributions, and it is positive definite by construction~\cite{bishop2006pattern}. The more similar the two distributions are, the smaller the KL is, such that KL vanishes only if the two distributions are identical.}. 
\newline\newline\noindent
Addressing now question {\it II)} in sect.\ref{subsec:ModelComparisonVOSvsBOSvsDOF}, we focus on the Bayes factor of two models $M_1$ and $M_2$, when none of which coincide with the true -unknown- data-generating model $M_{T}$. This is most likely the situation we shall encounter when LISA observations become available. In this case, Eq.\,\eqref{eq:average_BF} is modified as,
\begin{align}
\left\langle \ln {\rm BF}_{1,2} \right\rangle &= \int dD~ \mathbb{P}(D|M_{T}) \ln {\rm BF}_{1,2}(D) \nonumber\\
 &= \int dD~ \mathbb{P}(D|M_{T}) \ln {\rm BF}_{T,2}(D) - \int dD~ \mathbb{P}(D|M_{T}) \ln {\rm BF}_{T,1}(D)~,
 \label{eq:average_BF2}
\end{align}
where the second equality is obtained by multiplying and dividing by $P(D|M_{T})$ inside the logarithm. By looking at Eq.~\eqref{eq:average_BF2}, we can draw two conclusions: {\it a)} $\left\langle \ln {\rm BF}_{1,2} \right\rangle$ is no longer positive definite, and {\it b)} positive (negative) $\left\langle \ln {\rm BF}_{1,2} \right\rangle$ 
implies that model $M_1$ ($M_2$) is favored, i.e., the evidence which is more similar to the evidence of the true model $M_{T}$ yields a smaller integral value.

The standard criterion for the model-comparison ability can be referred to the Jeffreys scale \cite{Jeffreys:1939xee}, which classifies the value of Bayes factor into several regimes:
\begin{itemize}
\item ${\rm BF}_{i,j}(D) = 1$, suggests that the data favors models $M_i$ and $M_j$ equally,
\item ${\rm BF}_{i,j}(D) \in  [1,3], \,[3,10], \,[10,30],$ and $[30,10^{2}]$, implies weak, 
substantial, strong, and very strong evidence favoring the model $M_i$, respectively,
\item ${\rm BF}_{i,j}(D) > 10^2$ corresponds to a  decisive evidence favoring the model $M_i$.
\end{itemize}
We consider that this Jeffreys criterion can also be applied to the average of logarithmic BF in Eq.~\eqref{eq:average_BF}.
For \emph{decisive evidence} (${\rm BF}_{T,R} \gtrsim 10^2$) for discriminating between $M_{T}$ and $M_{R}$, we require $\langle \ln {\rm BF}_{T,R} \rangle > \ln 10^2 \simeq 4.6$.

\bibliographystyle{JHEP}
\bibliography{auto.bib,manual.bib}

\end{document}